\documentclass[showpacs,preprintnumbers,amsmath,amssymb,nofootinbib]{revtex4}

\usepackage{graphicx}% Include figure files
\usepackage{bm}
\usepackage{epstopdf}
\usepackage{float}
\usepackage{dcolumn}% Align table columns on decimal point
\usepackage{bm}% bold math
\usepackage{mathrsfs}
\usepackage{amsmath}
\usepackage{footnote}
\usepackage{accents}
\usepackage{tensor}
\usepackage[titletoc]{appendix}
\usepackage{color}

\begin{document}

%Sergio: here comes a general list
%1. remove 1st unnecessary text which is anyhow commented
%2. use small p for pressure. replace capitla P by p (Done)
%3. no Hehl name should appear (Done)
%4. contortion or contorsion? I think teh 1st. check and replace (Done)
%5. everywhere constently we should write <...> for averaging (Done... though we should recheck)
%6. a_0=1 every where (did we define a as R/R_0? (YES)) (Done... but how about the appendix)
%7. sigma and sigma_0 to tilde{sigma} and tilde{sigma_0) on page 23 to 25 everywhere (DONE recheck)
%8. check the new cosmology results
%9. fix the gap before eq 201 (Temporarily solved)
%10 refernces in tehri order as they apepar in text (Mostly done - still uncited references)
%11 check for missing refrence sin the text (MISSING)
%12. mention all figures in the text (MISSING)
%13. check other ``Sergios'' in the text below
%14. appendix C not clear at all
%15. can we say something about the uniqueness of the Ricci tensor from Riemann?
%15. dont drink too much beer!
%\nofiles

%\preprint{APS/123-QED}

\title{Einstein--Cartan Cosmologies}
%with
% cosmological constant}% Force line breaks with \\
\author{Sergio Bravo Medina}
\email{s.bravo58@uniandes.edu.co}
\affiliation{
Departamento de F\'isica,\\ Universidad de los Andes, Cra.1E
No.18A-10, Bogot\'a, Colombia
}
\author{Davide Batic}
\email{davide.batic@ku.ac.ae}
\affiliation{
Department of Mathematics,\\ 
Khalifa University of Science and Technology, Main Campus
Abu Dhabi, United Arab Emirates}
\author{Marek Nowakowski}
\email{mnowakos@uniandes.edu.co}
\affiliation{
Departamento de F\'isica,\\ Universidad de los Andes, Cra.1E
No.18A-10, Bogot\'a, Colombia
}

\date{\today}% It is always \today, today,
             %  but any date may be explicitly specified

\begin{abstract}
Cosmologies based on General Relativity encompassing an anti-symmetric
connection (torsion) can display nice desirable features as the absence of the 
initial singularity and the possibility of inflation in the
early stage of the universe. After briefly reviewing the standard approach to
the cosmology with torsion, we generalize it to demonstrate
that several theories of torsion gravity are possible using
different choices of the diffeomorphic invariants in the Lagrangians.
As a result, distinct cosmologies emerge. In all of them it is possible that the universe 
avoids the initial singularity and passes through an initial accelerated expansion.
Differences between these theories are highlighted. 
\end{abstract}

\pacs{ }% PACS, the Physics and Astronomy
                             % Classification Scheme.
\keywords{Torsion, Einstein-Cartan, Spin Fluid, Torsion Gravity}%Use showkeys class option if keyword
                              %display desired
\maketitle

\section{Introduction}
Our understanding of the universe has undergone a revolution in the last decades. This is
mainly due to important observational discoveries like
the current accelerated stage of the expansion \cite{Peebles}, precise 
measurements of the Cosmic Background Radiation \cite{Penzias} and other crucial
cosmological parameters \cite{Planck}. 
%Sergio: some more?
Yet a few cosmological puzzles are
still escaping a full explanation. Among them is the problem of the 
initial singularity (best revealed by the Kretschman invariant
${\cal K=R_{\mu \nu \alpha \beta}R^{\mu \nu \alpha \beta}}=12 (\ddot{a}/a)^2 +(\dot{a}/a)^4$
which tends to $\infty$ for the scale factor going to zero, $a \to 0$)
and the nature and mechanism of early inflation which, in turn, seems to be a necessary ingredient of the early
cosmology to explain, for instance,  the homogeneity and isotropy of the universe. Whereas it is certain
that we have several candidates for the inflationary scenario as well as for the
avoidance of the initial singularity, none of these theories seems to be fully accepted as 
observational data, which could discriminate between these theories, are lacking and might continue to be missing for some time.
It is therefore of interest to explore different scenarios which provide us with
a solution. In choosing one of them a helpful guiding principle
might be the unifying wisdom of Occam's razor: we consider a theory more
powerful and elegant the more phenomena it is capable to explain by
invoking a minimal number of new assumption and ingredients. A theory
like Einstein--Cartan could be the harbinger of such a unifying scheme.
It is a logical extension of Einstein's General Relativity based on a new
connection which allows the former to have a symmetric (the standard Christoffel part) and
a new anti-symmetric part called contortion. As such the roots of this particular extension are 
purely geometrical modifying, e.g., the geometrical part (say, the Einstein tensor as an example)
of the Einstein equations (this would be then unlike extensions which invoke new fields apart from the metric). 
It is a physically consistent theory (this would be unlike some 
theories with non-symmetric metric which do have deficiencies) which so far, is neither confirmed nor falsified.  Indeed, it is fully unconstrained as
no data exist which could tell us something about its validity. Its  appeal lies in the cosmological aspects or, to be more exact,
in the aspects which have to do with early cosmology. 
It has been known for some time that the Einstein--Cartan cosmology leads to a singularity
free universe which at the same time can undergo an inflationary early expansion
thanks to the introduction of an anti-symmetric connection. The standard particular
Einstein--Cartan theory which has these desirable features is based on two pillars.
The first one is the Lagrangian chosen to be
$\sqrt{-g}R[\Gamma]$ where $R[\Gamma]$ is the full Ricci scalar
obtained by using the connection $\Gamma=\accentset{\circ}{\Gamma} -K$, where $\accentset{\circ}{\Gamma}$ is
the Christoffel connection and $K$ the anti-symmetric part called contortion.
The second basic ingredient is the choice of the physical source of the modified Einstein equations.
The Euler-Lagrange equations following from ${\cal L}=\sqrt{g}R$ are not of the Einstein form 
which coined on the general relativistic result would be `Einstein tensor being proportional to the source'.
Insisting on this form one adds a term on both sides of the equation which promotes
the left-hand-side to become the Einstein tensor (using again the full connection) whereas the right-hand-side
being the sum of the metric energy-momentum tensor and the newly added term is identified as the physical source
called customarily the canonical energy-momentum tensor. Both assumptions, the choice of the Lagrangian and the identification
of the canonical energy-momentum tensor as the physical source, are not stringent and, in view of the desirable features
of the Einstein--Cartan cosmology, it appears worthwhile to probe into other combinations and possibilities. 
In a general setup of the Einstein--Cartan gravity there exist, apart from $R$,  several independent
diffeomorphic invariants  which, in linear combination, can yield different suitable Lagrangians.  
One can also identify the metric energy-momentum tensor inclusive new spin contributions as the physical source without running into
contradictions. Taken all together, new Einstein--Cartan cosmologies emerge which we will study in this paper.
We will demonstrate how, subject to the choice of parameters, they solve the initial singularity problem and can
display  an initial accelerated stage of expansion. Three solutions of the former are logically possible: (i) the universe
is in a sense `eternal' as it contracts to a minimum non-zero length (which we identify as the Big-Bang) from which it
expands \cite{Bounce}, (ii) the universe is born spontaneously, but at a non-zero value of the scale factor and (iii)   
the scale factor approaches its zero value only asymptotically at $t \to -\infty$. We will probe into the questions
which one of these possibilities is realized in the Einstein--Cartan cosmologies.

Usually, the avoidance of the initial singularity is attributed to quantum effects (or a quantum gravity theory). 
A question then arises what could be the input of quantum mechanics, if at all, in Einstein--Cartan theory. By inspecting the choice
of the source of the torsion part, this query can be answered. The aforementioned source is taken to be
$s_{\mu \nu}u^{\alpha}$ where $u^{\alpha}$ is the four-velocity and the tensor $s_{\mu \nu}$ can be either
connected to the spin identifying it as one of the generators of the Poincar\'e--Lie algebra or as a classical angular momentum tensor
referring to extended bodies, see for details \cite{P1, P2}. In the early universe it appears more appropriate to refer
to spin rather than angular momenta.
In such a case $s_{\mu \nu}$ is connected to the spin four vector by $s_{\alpha}=\epsilon_{\alpha \mu \nu \rho}s^{\mu \nu}u^{\rho}/\sqrt{g}$.
Here, the spin vector itself is usually associated with the expectation value of the spin operators 
being proportional to the Planck constant $ \hbar$. Hence, we could say that the source of the
torsion is of quantum mechanical origin. Indeed, the source could be a sum of two terms: one which can be traced back to the
spin and is phenomenologically taken as above and a classical angular momentum contribution which we will not consider here.
The spin tensor $s_{\mu \nu}$ and the four-velocity $u^{\alpha}$ are determined by the Mathisson--Papapetrou--Dixon (MPD) equations \cite{Mathisson,Papapetrou,Dixon}
which depend also on the metric. Therefore, in principle, the modified Einstein equations and the MPD equations are coupled.
By using an averaging of the spin we can reach at a simpler system as we will explain in the main text.

There exist many reviews and books which expose the Einstein--Cartan gravity in detail \cite{Hehl,Trautmann,TrautmanNature,Hammond,PoplawskiInflation,Odinstov,ArcosPereira,Boehmer,Barrow}. Here we take the occasion to review
the main aspects of the Einstein--Cartan cosmology to be able to make the comparison with the extension of these theories
made possible by starting with more general Lagrangians.
 
The paper is organized as follows. To make it complete we briefly review in the next section the conventions and
identities which enter the Einstein--Cartan theory, in general, and its cosmology, in particular. Section three
is devoted to the equations of motion derived from the Lagrangian $\sqrt{g}R$. We present them in two different
forms, first in the standard way and secondly in a novel way after splitting off a total divergence.
We dwell upon this point because of a subtlety which follows and which is the identification of the
physical source. The original Euler-Lagrange equations are not of the usual Einstein $G=\kappa \Sigma$ form as mentioned already above.
These equations
have the metric energy-momentum as the physical source. Only after algebraic manipulation it is
possible to arrive at the Einstein form and choose the so-called canonical energy-momentum tensor as a physical source.
In the next section we discuss in detail the standard cosmology which emerges from the choice $G=\kappa \Sigma$.
We identify the term in the canonical energy-momentum tensor which is responsible for the nice features of this cosmological model.  Next we turn to alternative Lagrangians and their cosmological implications.    
We present a simple, and a more general model and compare them with the standard one discussed in the preceding section.
As we will show in both models, it is possible to maintain the avoidance of the initial singularity and
have an accelerated initial condition on the expansion. All results depend on an averaging or the spin tensor. 
Roughly, we assume that spin quantities entering our expression average to zero if we have to do with non-contracted open indices.
In section six we discuss briefly an alternative approach to this averaging for all models presented before.
In the final section we summarize our conclusions.

\section{Conventions and Identities}%Generalities of Einstein-Cartan}
\subsection{Conventions}
Throughout the text we use the metric signature $(+,-,-,-)$.
When torsion is introduced into spacetime, the standard identities and symmetries can be lost and makes sense to mention these differences explicitly before embarking
on the equations of motion and cosmology.  
We derive our results following the notation and conventions introduced in \cite{Hehl}. 
We shall start with the definition of torsion as the antisymmetric part of the connection and thus a proper tensor
\begin{equation}
S\indices{_{\mu\nu}^{\alpha}}=\Gamma_{[\mu\nu]}^{\alpha}=\frac{1}{2}(\Gamma^{\alpha}_{\mu \nu}-\Gamma^{\alpha}_{\nu \mu}).
\end{equation}

Apart from this definition demanding metricity with respect to the new covariant derivative, i.e. $\nabla_{\mu}g_{\alpha\beta}=0$, the Einstein--Cartan (EC) connection is obtained
(see Appendix A for details)
\begin{equation}\label{ECConnection}
\Gamma_{\mu\nu}^{\alpha}=\accentset{\circ}{\Gamma}_{\mu\nu}^{\alpha}-K\indices{_{\mu\nu}^{\alpha}},
\end{equation}

where $\accentset{\circ}{\Gamma}_{\mu\nu}^{\alpha}$ is the standard Christoffel connection
\begin{equation}
\accentset{\circ}{\Gamma}_{\mu\nu}^{\alpha}=\frac{1}{2}g^{\alpha\lambda}\left(\partial_{\mu}g_{\nu\lambda}+\partial_{\nu}g_{\mu\lambda}-\partial_{\lambda}g_{\mu\nu} \right),
\end{equation}

and $K\indices{_{\mu\nu}^{\alpha}}$ is the contortion tensor, related to torsion by
\begin{equation}
K\indices{_{\mu\nu}^{\alpha}}=-S\indices{_{\mu\nu}^{\alpha}}+S\indices{_{\nu}^{\alpha}_{\mu}}-S\indices{^{\alpha}_{\mu\nu}}=-K\indices{_{\mu}^{\alpha}_{\nu}},
\end{equation}
\begin{equation}
S\indices{_{\mu\nu}^{\alpha}}=-K\indices{_{[\mu\nu]}^{\alpha}}=-\frac{1}{2}\left(K\indices{_{\mu\nu}^{\alpha}}-K\indices{_{\nu\mu}^{\alpha}}\right).
\end{equation}

For the sake of convention we shall use terms with an over-circle, e.g. $\accentset{\circ}{R}_{\mu\nu}, \accentset{\circ}{\nabla}$, 
to refer to objects defined with respect to the Christoffel connection. For the Riemann tensor and the covariant derivative we use the same convention as in \cite{Hehl}. 
The covariant derivative of a mixed tensor is defined as
\begin{equation}
\begin{split}\label{DerivativeHehl}
\nabla_{\mu}V\indices{_{\nu\alpha\ldots}^{\beta\gamma\ldots}}=& \partial_{\mu}V\indices{_{\nu\alpha\ldots}^{\beta\gamma\ldots}}+\Gamma_{\mu\lambda}^{\beta}V\indices{_{\nu\alpha\ldots}^{\lambda\gamma\ldots}}+\Gamma_{\mu\lambda}^{\gamma}V\indices{_{\nu\alpha}^{\beta\lambda\ldots}}+\ldots \\
& -\Gamma_{\mu\nu}^{\lambda}V\indices{_{\lambda\alpha\ldots}^{\beta\gamma\ldots}}-\Gamma_{\mu\alpha}^{\lambda}V\indices{_{\nu\lambda\ldots}^{\beta\gamma\ldots}}-\ldots.
\end{split}
\end{equation}
The Riemann tensor is given by \footnote{D'Inverno \cite{Inverno} has a different convention for the Riemann tensor and the covariant derivative, since one can always 
think of a different order of the indices on this tensor, or even the order in the commutation relation \cite{MTW}.}

\begin{equation}\label{RiemannHehl}
R\indices{_{\mu\nu\alpha}^{\beta}}=\partial_{\mu}\Gamma_{\nu\alpha}^{\beta}-\partial_{\nu}\Gamma_{\mu\alpha}^{\beta}+\Gamma_{\mu\lambda}^{\beta}\Gamma_{\nu\alpha}^{\lambda}-\Gamma_{\nu\lambda}^{\beta}\Gamma_{\mu\alpha}^{\lambda}.
\end{equation}
The signature of the metric is chosen to be $(-,+,+,+)$. This sets our conventions. 

\subsection{General Results and Identities}

Of interest are also general identities independent of the the equations of motion. This includes splitting the tensors into
the standard component defined over the
Christoffel connection plus a part with torsion (or contortion), Bianchi identities and other new identities. 
For instance, by using the EC connection as defined in equation (\ref{ECConnection}) we can write the Riemann tensor as 
\footnote{In order to establish antisymmetry only on certain indices we use the
 \[
A_{[\mu|\nu|\alpha]}=\frac{1}{2}\left(A_{\mu\nu\alpha}-A_{\alpha\nu\mu} \right).
\]}
\begin{equation}\label{RCartan}
\begin{split}
R\indices{_{\mu\nu\alpha}^{\beta}}=& \accentset{\circ}{R}\indices{_{\mu\nu\alpha}^{\beta}}-\nabla_{\mu}K\indices{_{\nu\alpha}^{\beta}}+\nabla_{\nu}K\indices{_{\mu\alpha}^{\beta}}-K\indices{_{\nu\mu}^{\lambda}}K\indices{_{\lambda\alpha}^{\beta}}\\ & +K\indices{_{\mu\nu}^{\lambda}}K\indices{_{\lambda\alpha}^{\beta}}-K\indices{_{\nu\alpha}^{\lambda}}K\indices{_{\mu\lambda}^{\beta}}+K\indices{_{\mu\alpha}^{\lambda}}K\indices{_{\nu\lambda}^{\beta}}\\
= & \accentset{\circ}{R}\indices{_{\mu\nu\alpha}^{\beta}}-2\nabla_{[\mu}K\indices{_{\nu]\alpha}^{\beta}}+2K\indices{_{[\mu\nu]}^{\lambda}}K\indices{_{\lambda\alpha}^{\beta}}+2K\indices{_{[\mu| \alpha}^{\lambda}}K\indices{_{|\nu]\lambda}^{\beta}} .
\end{split}
\end{equation}

We may contract the Riemann tensor above to obtain the Ricci tensor. 
On this convention the Ricci tensor is obtained through the contraction suggested in \cite{Hehl}, i.e.  $R\indices{_{\rho\mu\nu}^{\rho}}=R_{\mu\nu}$. This yields
\begin{equation}\label{Ricci}
R_{\mu\nu}=\accentset{\circ}{R}_{\mu\nu}-\nabla_{\lambda}K\indices{_{\mu\nu}^{\lambda}}+\nabla_{\mu}K\indices{_{\lambda\nu}^{\lambda}}+K\indices{_{\rho\mu}^{\lambda}}K\indices{_{\lambda\nu}^{\rho}}-K\indices{_{\mu\nu}^{\lambda}}K\indices{_{\rho\lambda}^{\rho}}.
\end{equation}
There is only one non-trivial contraction possible when going from the Riemann to the Ricci tensor in Einstein--Cartan theory, which can be seen from equations (\ref{RiemannHehl}) and (\ref{RCartan}). Namely we see
\begin{equation}
R\indices{_{\alpha}^{\alpha}_{\mu\nu}}=0, \quad R\indices{_{\mu\alpha\nu}^{\alpha}}=-R\indices{_{\alpha\mu\nu}^{\alpha}}=-R_{\mu\nu} , \quad R\indices{_{\mu\nu\alpha}^{\alpha}}=0.
\end{equation}
%...Sergio.  Should I show it? (YES)
Note that in the case of torsion the Ricci tensor is non-symmetric, unlike $\accentset{\circ}{R}_{\mu\nu}$ which is an essential part of the Einstein field equations in General Relativity. 
We may write its antisymmetric part in the following way
\begin{equation}
R_{[\mu\nu]}=\left(\nabla_{\lambda}+2S\indices{_{\lambda\beta}^{\beta}} \right)S\indices{_{\mu\nu}^{\lambda}}+\nabla_{\mu}S\indices{_{\nu\beta}^{\beta}}-\nabla_{\nu}S\indices{_{\mu\beta}^{\beta}}=\accentset{\star}{\nabla}_{\lambda}T\indices{_{\mu\nu}^{\lambda}},
\end{equation}
where we have defined the \emph{modified torsion} tensor 
\begin{equation}
T\indices{_{\mu\nu}^{\alpha}}\equiv S\indices{_{\mu\nu}^{\alpha}}+\delta_{\mu}^{\alpha}S\indices{_{\nu\lambda}^{\lambda}}-\delta_{\nu}^{\alpha}S\indices{_{\mu\lambda}^{\lambda}}
\end{equation}
as well as the star derivative\footnote{Note that the star derivative has the following property when acting upon a vector
\[
\sqrt{-g}\accentset{\star}{\nabla}_{\lambda}V^{\lambda}=\partial_{\lambda}\left(\sqrt{-g}V^{\lambda} \right).
\].} 
\begin{equation}
\accentset{\star}{\nabla}_{\lambda}\equiv \nabla_{\lambda}+2S\indices{_{\lambda\alpha}^{\alpha}}. 
\end{equation}
We can contract the full Ricci tensor to obtain the Ricci scalar in terms of torsion or contortion
\begin{equation}\label{Ricci-Full}
\begin{split}
R & = \accentset{\circ}{R}+2\nabla_{\lambda}K\indices{_{\alpha}^{\lambda\alpha}}+K\indices{_{\rho}^{\alpha\lambda}}K\indices{_{\lambda\alpha}^{\rho}}-K\indices{_{\alpha}^{\alpha\lambda}}K\indices{_{\rho\lambda}^{\rho}}, \\
%\end{equation}
%or in terms of torsion,
%\begin{equation}
 & = \accentset{\circ}{R}-4(\nabla_{\lambda}-S\indices{_{\alpha\lambda}^{\alpha}})S\indices{_{\rho}^{\lambda\rho}}+2S_{\alpha\beta\gamma}S^{\gamma\beta\alpha}+S_{\alpha\beta\gamma}S^{\alpha\beta\gamma}.
\end{split}
\end{equation}

It is worth noting that for any vector \cite{Weinberg}
\begin{equation*}
\accentset{\circ}{\nabla}_{\lambda}V^{\lambda}=\frac{1}{\sqrt{-g}}\partial_{\lambda}\left(\sqrt{-g}V^{\lambda} \right) \mbox{ }\mbox{ } \mbox{ } \mbox{ }\mbox{ and } \mbox{ }\mbox{ }\mbox{ } \mbox{ } \mbox{ }
\nabla_{\lambda}V^{\lambda}=\frac{1}{\sqrt{-g}}\partial_{\lambda}\left(\sqrt{-g}V^{\lambda} \right)-K\indices{_{\lambda\alpha}^{\lambda}}V^{\alpha}.
\end{equation*}

It can be easily shown that $K\indices{_{\lambda\alpha}^{\lambda}}=-2S_{\alpha}$ with $S_{\alpha}=S\indices{_{\lambda \alpha}^{\lambda}}$. Then the Ricci scalar in terms of torsion and contortion yields
%\begin{equation}
%\nabla_{\lambda}V^{\lambda}=\frac{1}{\sqrt{-g}}\partial_{\lambda}\left(\sqrt{-g}V^{\lambda}\right)+2S_{\alpha}V^{\alpha}
%\end{equation}
%. Finally the Ricci scalar is:
%
\begin{eqnarray}\label{RicciTorsion}
R&=&\accentset{\circ}{R}-\frac{4}{\sqrt{-g}}\partial_{\lambda}\left(\sqrt{-g}S^{\lambda} \right)-4S_{\lambda}S^{\lambda}+2S_{\alpha\beta\gamma}S^{\gamma\beta\alpha}+S_{\alpha\beta\gamma}S^{\alpha\beta\gamma}, \\
\label{RicciContorsion}
&=&\accentset{\circ}{R}+\frac{2}{\sqrt{-g}}\partial_{\lambda}\left(\sqrt{-g}K\indices{_{\alpha}^{\lambda\alpha}}\right)+K\indices{_{\rho}^{\alpha\lambda}}K\indices{_{\lambda\alpha}^{\rho}}+K\indices{_{\alpha}^{\alpha\lambda}}K\indices{_{\rho\lambda}^{\rho}}.
\end{eqnarray}
All terms in (\ref{RicciContorsion}) are invariants and as such they are candidates for a Lagrangian built up by a linear
superposition of these terms, the Ricci scalar being only one possibility. We will examine such theories and apply them
to cosmology.
%\subsection{Conservation Laws and Bianchi Identities}
Apart from the relations found above, it is often useful to have some analogue to the standard Bianchi identities and
conservation laws in Einstein--Cartan Theory \cite{Schouten, PoplawskiFields}. Since we are indeed looking for identities similar to the Bianchi identities, these must give the
standard Bianchi identities when the torsion is put to zero.
To derive them we start with the commutator of the covariant derivative
acting on an arbitrary  vector. Given the connection in Einstein--Cartan
we obtain

\begin{equation}\label{CommutatorNabla}
%\begin{split}
[\nabla_{\gamma},\nabla_{\alpha}]B^{\beta}%&=2\nabla_{[\gamma}\nabla_{\alpha]}B^{\beta}=2\partial_{[\gamma}\nabla_{\alpha]}B^{\beta}-2\Gamma_{[\gamma\alpha]}^{\lambda}\nabla_{\lambda}B^{\beta}+2\Gamma_{[\gamma|\lambda}^{\beta}\nabla_{\alpha]}B^{\lambda}\\
%&=2\partial_{[\gamma}\left(\Gamma_{\alpha]\lambda}^{\beta}B^{\lambda} \right)+S\indices{_{\gamma\alpha}^{\lambda}}\nabla_{\lambda}B^{\beta}+2\Gamma_{[\gamma|\lambda}^{\beta}\left(\partial_{\alpha]}B^{\lambda}+\Gamma^{\lambda}_{\alpha]\rho}B^{\rho} \right)\\
%&= 2S\indices{_{\gamma\alpha}^{\lambda}}\nabla_{\lambda}B^{\beta}+2(\partial_{[\gamma}\Gamma_{\alpha]\rho}^{\beta}+\Gamma_{[\gamma|\lambda}^{\beta}\Gamma_{\alpha]\rho}^{\lambda})B^{\rho}\\
%&
=2S\indices{_{\alpha\gamma}^{\lambda}}\nabla_{\lambda}B^{\beta}+R\indices{_{\gamma\alpha\lambda}^{\beta}}B^{\lambda},
%\end{split}
\end{equation}
%Since we have the following field equations
%
%\begin{equation}
%G_{\mu\nu}\equiv R_{\mu\nu}-\frac{1}{2}g_{\mu\nu}R=\kappa \Sigma_{\mu\nu}
%\end{equation}
%
%it is useful to calculate what the covariant derivative of the `new' Einstein tensor is, namely $\nabla_{\nu}G^{\mu\nu}$. To obtain this is not so straightforward, we start by noting:
%
which means that we are entitled to  write
\begin{equation}\label{NablaCommutator}
%\begin{split}
\nabla_{\mu}\nabla_{[\nu}\nabla_{\alpha]}B^{\beta}
%&=\nabla_{\mu}\left[\nabla_{[\nu}\left(\partial_{\alpha]}B^{\beta}+\Gamma_{\alpha]\lambda}^{\beta}B^{\lambda} \right) \right]\\ &=\nabla_{\mu}\left[-\Gamma_{[\nu\alpha]}^{\lambda}\partial_{\lambda}B^{\beta}+\Gamma_{[\nu|\lambda}^{\beta}\partial_{\alpha]}B^{\lambda}+\partial_{[\nu}(\Gamma_{\alpha]\lambda}^{\beta}B^{\lambda})-\Gamma_{[\nu\alpha]}^{\gamma}\Gamma_{\gamma\lambda}^{\beta}B^{\lambda}+\Gamma_{[\nu|\gamma}^{\beta}\Gamma_{\alpha]\lambda}^{\gamma}B^{\lambda} \right]\\
%&=\nabla_{\mu}\left[(\partial_{[\nu}\Gamma_{\alpha]\lambda}^{\beta}+\Gamma_{[\nu|\gamma}^{\beta}\Gamma_{\alpha]\lambda}^{\gamma})B^{\lambda}-\Gamma_{[\nu\alpha]}^{\lambda}\left(\partial_{\lambda}B^{\beta}+\Gamma_{\gamma\lambda}^{\beta}B^{\lambda} \right) \right] \\
%&
= \frac{1}{2}\nabla_{\mu}\left(R\indices{_{\nu\alpha\lambda}^{\beta}}B^{\lambda} \right)+\nabla_{\mu}(S\indices{_{\alpha\nu}^{\lambda}}\nabla_{\lambda}B^{\beta}).
%\end{split}
\end{equation}
We also obtain the following antisymmetric condition on the covariant derivatives
\begin{equation}
%\begin{split}
\nabla_{[\mu}\nabla_{\nu]}\nabla_{\alpha}B^{\beta}
%=&\nabla_{[\mu}\left[ \partial_{\nu]}(\nabla_{\alpha}B^{\beta})-\Gamma_{\nu]\alpha}^{\lambda}\nabla_{\lambda}B^{\beta}+\Gamma_{\nu]\lambda}^{\beta}\nabla_{\alpha}B^{\lambda}\right]\\
%=& \left[\partial_{[\mu}(-\Gamma_{\nu]\alpha}^{\lambda}\nabla_{\lambda}B^{\beta}+\Gamma_{\nu]\lambda}^{\beta}\nabla_{\alpha}B^{\lambda})-\Gamma_{[\mu\nu]}^{\gamma}\left(\partial_{\gamma}\nabla_{\alpha}B^{\beta}-\Gamma_{\gamma\alpha}^{\lambda}\nabla_{\lambda}B^{\beta}+\Gamma_{\gamma\lambda}^{\beta}\nabla_{\alpha}B^{\lambda}\right) \right. \\
%& \left. -\Gamma_{[\mu|\alpha}^{\gamma}\left(\partial_{\nu]}\nabla_{\gamma}B^{\beta}-\Gamma_{\nu]\gamma}^{\lambda}\nabla_{\lambda}B^{\beta}+\Gamma_{\nu]\lambda}^{\beta}\nabla_{\gamma}B^{\lambda} \right)+\Gamma^{\beta}_{[\mu|\gamma}\left(\partial_{\nu]}\nabla_{\alpha}B^{\gamma}-\Gamma_{\nu]\alpha}^{\lambda}\nabla_{\lambda}B^{\gamma}+\Gamma_{\nu]\lambda}^{\gamma}\nabla_{\alpha}B^{\lambda} \right)\right]\\
%=& \left[S\indices{_{\nu\mu}^{\gamma}}\nabla_{\gamma}\nabla_{\alpha}B^{\beta}-\left(\partial_{[\mu}\Gamma_{\nu]\alpha}^{\lambda}+\Gamma_{[\mu|\alpha}^{\gamma}\Gamma_{\nu]\gamma}^{\lambda} \right)\nabla_{\lambda}B^{\beta}+\left(\partial_{[\mu}\Gamma_{\nu]\lambda}^{\beta}+\Gamma_{[\mu|\gamma}^{\beta}\Gamma_{\nu]\lambda}^{\gamma} \right)\nabla_{\alpha}B^{\lambda} \right]\\
%&
=S\indices{_{\nu\mu}^{\gamma}}\nabla_{\gamma}\nabla_{\alpha}B^{\beta}-\frac{1}{2}R\indices{_{\mu\nu\alpha}^{\lambda}}\nabla_{\lambda}B^{\beta}+\frac{1}{2}R\indices{_{\mu\nu\lambda}^{\beta}}\nabla_{\alpha}B^{\lambda}.
%\end{split}
\end{equation}
%This last expression can be reorganized using the general result
This can be re-organized using the commutator on equation (\ref{CommutatorNabla}) as follows
%\begin{equation}
%\begin{split}
%[\nabla_{\gamma},\nabla_{\alpha}]B^{\beta}%&=2\nabla_{[\gamma}\nabla_{\alpha]}B^{\beta}=2\partial_{[\gamma}\nabla_{\alpha]}B^{\beta}-2\Gamma_{[\gamma\alpha]}^{\lambda}\nabla_{\lambda}B^{\beta}+2\Gamma_{[\gamma|\lambda}^{\beta}\nabla_{\alpha]}B^{\lambda}\\
%&=2\partial_{[\gamma}\left(\Gamma_{\alpha]\lambda}^{\beta}B^{\lambda} \right)+S\indices{_{\gamma\alpha}^{\lambda}}\nabla_{\lambda}B^{\beta}+2\Gamma_{[\gamma|\lambda}^{\beta}\left(\partial_{\alpha]}B^{\lambda}+\Gamma^{\lambda}_{\alpha]\rho}B^{\rho} \right)\\
%&= 2S\indices{_{\gamma\alpha}^{\lambda}}\nabla_{\lambda}B^{\beta}+2(\partial_{[\gamma}\Gamma_{\alpha]\rho}^{\beta}+\Gamma_{[\gamma|\lambda}^{\beta}\Gamma_{\alpha]\rho}^{\lambda})B^{\rho}\\
%&
%=2S\indices{_{\alpha\gamma}^{\lambda}}\nabla_{\lambda}B^{\beta}+R\indices{_{\gamma\alpha\lambda}^{\beta}}B^{\lambda},
%\end{split}
%\end{equation}
%
%the combination of this results yields
\begin{equation}\label{NablaCommutator1}
\begin{split}
\nabla_{[\mu}\nabla_{\nu]}\nabla_{\alpha}B^{\beta}=&\frac{1}{2}R\indices{_{\mu\nu\lambda}^{\beta}}\nabla_{\alpha}B^{\lambda}-\frac{1}{2}R\indices{_{\mu\nu\alpha}^{\gamma}}\nabla_{\gamma}B^{\beta}+S\indices{_{\nu\mu}^{\gamma}}\nabla_{\alpha}\nabla_{\gamma}B^{\beta}\\&+2S\indices{_{\nu\mu}^{\gamma}}S\indices{_{\alpha\gamma}^{\lambda}}\nabla_{\lambda}B^{\beta}+S\indices{_{\nu\mu}^{\gamma}}R\indices{_{\gamma\alpha\lambda}^{\beta}}B^{\lambda},
\end{split}
\end{equation}
%If we now antisymmetrize in the indices $\mu$, $\nu$ and $\alpha$ for the expressions above these give
Finally antisymmetrizing in the three indices we get from equation (\ref{NablaCommutator})
\begin{equation}
\nabla_{[\mu}\nabla_{\nu}\nabla_{\alpha]}B^{\beta}=\frac{1}{2}\nabla_{[\mu}R\indices{_{\nu\alpha]\lambda}^{\beta}}B^{\lambda}+\frac{1}{2}R\indices{_{[\nu\alpha|\lambda}^{\beta}}\nabla_{|\mu]}B^{\lambda}+\nabla_{[\mu}S\indices{_{\alpha\nu]}^{\lambda}}\nabla_{\lambda}B^{\beta}+S\indices{_{[\alpha\nu}^{\lambda}}\nabla_{\mu]}\nabla_{\lambda}B^{\beta},
\end{equation}
while equation (\ref{NablaCommutator1}) gives
\begin{equation}
\begin{split}
\nabla_{[\mu}\nabla_{\nu}\nabla_{\alpha]}B^{\beta}=&-\frac{1}{2}R\indices{_{[\mu\nu\alpha]}^{\gamma}}\nabla_{\gamma}B^{\beta}+\frac{1}{2}R\indices{_{[\mu\nu|\lambda}^{\beta}}\nabla_{|\alpha]}B^{\lambda}+S\indices{_{[\nu\mu}^{\gamma}}\nabla_{\alpha]}\nabla_{\gamma}B^{\beta}\\&+2S\indices{_{[\nu\mu}^{\gamma}}S\indices{_{\alpha]\gamma}^{\lambda}}\nabla_{\lambda}B^{\beta}+S\indices{_{[\nu\mu|}^{\gamma}}R\indices{_{\gamma|\alpha]\lambda}^{\beta}}B^{\lambda}.
\end{split}
\end{equation}
By equating these equivalent expressions we see that the following identity holds 
\begin{equation*}
\frac{1}{2}\nabla_{[\mu}R\indices{_{\nu\alpha]\lambda}^{\beta}}B^{\lambda}+\nabla_{[\mu}S\indices{_{\alpha\nu]}^{\lambda}}\nabla_{\lambda}B^{\beta}=-\frac{1}{2}R\indices{_{[\mu\nu\alpha]}^{\gamma}}\nabla_{\gamma}B^{\beta}+2S\indices{_{[\nu\mu}^{\gamma}}S\indices{_{\alpha]\gamma}^{\lambda}}\nabla_{\lambda}B^{\beta}+S\indices{_{[\nu\mu}^{\gamma}}R\indices{_{\gamma|\alpha]\lambda}^{\beta}}B^{\lambda},
\end{equation*}
Comparing in the next step  the $B^{\lambda}$ and $\nabla_{\lambda}B^{\beta}$ coefficients we obtain the equivalent to the first Bianchi identity
\begin{equation}
\nabla_{[\mu}R\indices{_{\nu\alpha]\lambda}^{\beta}}=2S\indices{_{[\nu\mu}^{\gamma}}R\indices{_{\gamma|\alpha]\lambda}^{\beta}},
\end{equation}
and the second Bianchi identity which involves only torsion
\begin{equation}
\begin{split}
\nabla_{[\mu}S\indices{_{\alpha\nu]}^{\lambda}}&=-\frac{1}{2}R\indices{_{[\mu\nu\alpha]}^{\lambda}}+2S\indices{_{[\nu\mu}^{\gamma}}S\indices{_{\alpha]\gamma}^{\lambda}},\\
R\indices{_{[\mu\nu\alpha]}^{\lambda}}&=-2\nabla_{[\mu}S\indices{_{\alpha\nu]}^{\lambda}}+4S\indices{_{[\nu\mu}^{\gamma}}S\indices{_{\alpha]\gamma}^{\lambda}}.
\end{split}
\end{equation}
The first Bianchi identity may be contracted to yield
%\begin{equation}
%\nabla_{\mu}R\indices{_{\nu\alpha\lambda}^{\beta}}+\nabla_{\nu}R\indices{_{\alpha\mu\lambda}^{\beta}}+\nabla_{\alpha}R\indices{_{\mu\nu\lambda}^{\beta}}=2S\indices{_{\nu\mu}^{\gamma}}R\indices{_{\gamma\alpha\lambda}^{\beta}}+2S\indices{_{\alpha\nu}^{\gamma}}R\indices{_{\gamma\mu\lambda}^{\beta}}+2S\indices{_{\mu\alpha}^{\gamma}}R\indices{_{\gamma\nu\lambda}^{\beta}}
%\end{equation}
%
%setting $\beta=\nu$ to `contract' we get:
%
%\begin{equation}
%\nabla_{\mu}R_{\alpha\lambda}+\nabla_{\nu}R\indices{_{\alpha\mu\lambda}^{\nu}}+\nabla_{\alpha}R\indices{_{\mu\nu\lambda}^{\nu}}=2S\indices{_{\nu\mu}^{\gamma}}R\indices{_{\gamma\alpha\lambda}^{\nu}}+2S\indices{_{\alpha\nu}^{\gamma}}R\indices{_{\gamma\mu\lambda}^{\nu}}+2S\indices{_{\mu\alpha}^{\gamma}}R\indices{_{\gamma\nu\lambda}^{\nu}}.
%\end{equation}
%
%Further contracting with $g^{\alpha\lambda}$ this yields
%
\begin{eqnarray}
\nabla_{\mu}R+\nabla_{\nu}R\indices{^{\lambda}_{\mu\lambda}^{\nu}}+\nabla_{\lambda}R\indices{_{\mu\nu}^{\lambda\nu}}&=&2S\indices{_{\nu\mu}^{\gamma}}R\indices{_{\gamma\lambda}^{\lambda\nu}}+2S\indices{_{\lambda\nu}^{\gamma}}R\indices{_{\gamma\mu}^{\lambda\nu}}+2S\indices{_{\mu\lambda}^{\gamma}}R\indices{_{\gamma\nu}^{\lambda\nu}}\\
\nabla_{\mu}R-2\nabla_{\lambda}R\indices{_{\mu}^{\lambda}}&=&4S\indices{_{\nu\mu}^{\gamma}}R\indices{_{\gamma}^{\nu}}+2S\indices{_{\lambda\nu}^{\gamma}}R\indices{_{\gamma\mu}^{\lambda\nu}}.
\end{eqnarray}
Defining a new Einstein tensor over the full connection
\begin{equation}
G_{\mu\nu}\equiv R_{\mu\nu}-\frac{1}{2}g_{\mu\nu}R
\end{equation}
we can appreciate how the last result helps us to obtain the covariant divergence of the Einstein tensor 
%can 
%finally obtain a condition upon  , namely
\begin{equation} \label{G}
\nabla_{\nu}G\indices{_{\mu}^{\nu}}=\nabla_{\nu}R\indices{_{\mu}^{\nu}}-\frac{1}{2}\nabla_{\mu}R=-2S\indices{_{\nu\mu}^{\gamma}}R\indices{_{\gamma}^{\nu}}-S\indices{_{\lambda\nu}^{\gamma}}R\indices{_{\gamma\mu}^{\lambda\nu}}.
\end{equation}
So far, our results were independent of the equation of motion. We can apply them also
directly to the equations of motion.
There will be two sources in the theory, one for the modified Einstein equations which for now
we write as $G=\kappa \Sigma$ and for the torsion which we call $\tau\indices{_{\lambda\nu}^{\gamma}}$.  
%To write this result the way Hehl obtains it, we use:
By using the $\accentset{\star}{\nabla}$ derivative we can recast the result (\ref{G}) as 
\begin{equation}
\accentset{\star}{\nabla}_{\nu}G\indices{_{\mu}^{\nu}}=-2S\indices{_{\nu\mu}^{\gamma}}R\indices{_{\gamma}^{\nu}}-S\indices{_{\lambda\nu}^{\gamma}}R\indices{_{\gamma\mu}^{\lambda\nu}}+2S\indices{_{\nu\alpha}^{\alpha}}R\indices{_{\mu}^{\nu}}-S\indices{_{\mu\alpha}^{\alpha}}=\kappa \accentset{\star}{\nabla}_{\nu}\Sigma\indices{_{\mu}^{\nu}}
\end{equation}
Some straightforward manipulation gives
\begin{equation}
\accentset{\star}{\nabla}_{\nu}\Sigma\indices{_{\mu}^{\nu}}=\frac{1}{\kappa}\left(-S\indices{_{\lambda\nu}^{\gamma}}-\delta_{\lambda}^{\gamma}S\indices{_{\nu\alpha}^{\alpha}}+\delta_{\nu}^{\gamma}S\indices{_{\lambda\alpha}^{\alpha}} \right)R\indices{_{\gamma\mu}^{\lambda\nu}}-2\Sigma\indices{_{\gamma}^{\nu}}S\indices{_{\nu\mu}^{\gamma}}.
\end{equation}
Anticipating part of the results derived explicitly later i.e., using equation (\ref{TorsionHehl})
which yields $\tau\indices{_{\mu\nu}^{\alpha}}$ in terms of torsion we can write
\begin{equation}\label{BianchiTorsion}
\accentset{\star}{\nabla}_{\nu}\Sigma\indices{_{\mu}^{\nu}}=\tau\indices{_{\nu\lambda}^{\gamma}}R\indices{_{\mu\gamma}^{\nu\lambda}}+2\Sigma\indices{_{\gamma}^{\nu}}S\indices{_{\mu\nu}^{\gamma}}.
\end{equation}
In case our equation of motion is of the form $G=\kappa \Sigma$ this equation is the new 'conservation law'.
For completeness we mention that in \cite{HehlPart2} this `conservation law' is written defining the following derivative
\begin{equation}
\accentset{+}{\nabla}\psi_{\nu}\equiv \nabla_{\mu}\psi_{\nu}+4S\indices{_{\mu(\nu}^{\alpha}}\psi_{\alpha)}=\accentset{\star}{\nabla}_{\mu}\psi_{\nu}+2S\indices{_{\mu\nu}^{\alpha}}\psi_{\alpha},
\end{equation}
valid for only covariant indices in the form
\begin{equation}
\accentset{+}{\nabla}_{\nu}\Sigma\indices{_{\mu}^{\nu}}=\tau\indices{_{\nu\lambda}^{\gamma}}R\indices{_{\mu\gamma}^{\nu\lambda}}.
\end{equation}

Finally, we are aiming at an identity relating the antisymmetric part of $\Sigma$ to the source of the torsion.
The expanded second Bianchi identity is
\begin{equation}
R\indices{_{[\alpha\mu\nu]}^{\alpha}}=\frac{2}{3}\left[\nabla_{\alpha}S\indices{_{\mu\nu}^{\alpha}}+\nabla_{\mu}S\indices{_{\nu\alpha}^{\alpha}}-\nabla_{\nu}S\indices{_{\mu\alpha}^{\alpha}} \right]+\frac{4}{3}\left[S\indices{_{\nu\mu}^{\gamma}}S\indices{_{\alpha\gamma}^{\alpha}} \right].
\end{equation}
On the other hand, one can also show
\begin{equation}
\accentset{\star}{\nabla}_{\alpha}T\indices{_{\mu\nu}^{\alpha}}=\nabla_{\lambda}S\indices{_{\mu\nu}^{\lambda}}+2S\indices{_{\lambda\beta}^{\beta}}S\indices{_{\mu\nu}^{\lambda}}+\nabla_{\mu}S\indices{_{\nu\beta}^{\beta}}-\nabla_{\nu}S\indices{_{\mu\beta}^{\beta}}.
\end{equation}
Combining the two equations above appropriately, we obtain
\begin{equation}\label{AntisymmetricRicciG}
R_{[\mu \nu]}=G_{[\mu \nu]}=\frac{3}{2}R\indices{_{[\alpha\mu\nu]}^{\alpha}}=\accentset{\star}{\nabla}_{\alpha}T\indices{_{\mu\nu}^{\alpha}}
\end{equation}
which still lacks an explicit physical interpretation. However, we are able to derive an identity 
which we will need in the next section
 \begin{equation}
 \accentset{\star}{\nabla}_{\lambda}\tau\indices{_{\mu\nu}^{\lambda}}-\Sigma_{[\mu\nu]}=0.
 \end{equation}

\section{Lagrangian and Equations of Motion}
In Einstein--Cartan Theory it is generally assumed that the Lagrangian be taken proportional to the Ricci scalar in a
space-time with torsion. From its 
variation the equations of motion are obtained, as in any other theory in which a Lagrangian formalism may be applied \cite{Variations-GR}.
However, there seems to be
a preferred approach to these equations of motion, which lead - quite nicely - to
an equivalent expression to the Einstein Field Equations (EFE) defined over the full
connection. We will show in
this section that this is not the only approach in which one can obtain the field equations and, moreover,
the alternative way of approaching them leads to equations of motion which are in principle different
but can formally be transformed into the 
standard result. Given these equivalent expressions, the question arises what is the physical interpretation of the energy--momentum tensor
 in each one of these approaches.
 \subsection{Standard Approach}
 What we shall call the standard approach to the variational principle is discussed in 
 \cite{Hehl} and some details can be found in \cite{HehlPart1,HehlPart2,HehlPRD}.
We 
will see that the key to this approach is to work directly with the quantities of the Riemann--Cartan geometry without having to go back to the usual General Relativity geometrical
quantities defined over the Christoffel symbol.
%We will use Hehl's conventions as given by equations (\ref{RiemannHehl}) and (\ref{DerivativeHehl}), and recover his results to get the field equations as coming from the variational principle. We will do this to check that the convention reproduces his results.
%First we use Hehl's definition of the `new' Einstein tensor in this theory which he directly writes as
%\begin{equation}
%{}^{(H)}G_{\mu\nu}\equiv R_{\mu\nu}-\frac{1}{2}g_{\mu\nu}R.
%\end{equation}
%Next he sets up a Lagrangian for the gravitational sector, as well as one for the matter sector 
We start from a gravitational Lagrangian given by $\mathfrak{R}\equiv\sqrt{-g}g^{\mu\nu}R_{\mu\nu}$, so that the full action, including a 
matter Lagrangian, is
\begin{equation}
S=\frac{1}{2\kappa} \int d^{4}x \mathfrak{R}+\int d^{4}x \mathcal{L}_{m}.
\end{equation} 
Varying the gravitational portion of the action with respect to the inverse metric $g^{\mu\nu}$ yields
\begin{equation} 
\delta_{g} \mathfrak{R}=R_{\mu\nu}\delta_{g} \mathfrak{g}^{\mu\nu}+\mathfrak{g}^{\mu\nu}\delta_{g} R_{\mu\nu},
\end{equation}
where we have $\mathfrak{g}^{\mu\nu}=\sqrt{-g}g^{\mu\nu}$. Now, $\delta_{g} R_{\mu\nu}$ can be obtained from the contracted Riemann tensor
%\begin{equation}
%\delta_{g} R\indices{_{\rho\mu\nu}^{\rho}}=\partial_{\rho}(\delta_{g} \Gamma^{\rho}_{\mu\nu})-\partial_{\mu}(\delta_{g} \Gamma^{\rho}_{\rho\nu})+\Gamma_{\mu\nu}^{\lambda}(\delta_{g} \Gamma_{\rho\lambda}^{\rho})+\Gamma_{\rho\lambda}^{\rho}(\delta_{g} \Gamma^{\lambda}_{\mu\nu})-\Gamma_{\rho\nu}^{\lambda}(\delta_{g} \Gamma_{\mu\lambda}^{\rho})-\Gamma_{\mu\lambda}^{\rho}(\delta_{g} \Gamma_{\rho\nu}^{\lambda}).
%\end{equation}
%here we have used the commutation of the variation with respect to the partial derivative. 
which, using the definition of the covariant derivative given above, yields
\begin{equation}
\begin{split}
\delta_{g} R_{\mu\nu} & =\nabla_{\rho}(\delta_{g} \Gamma^{\rho}_{\mu\nu})-\nabla_{\mu}(\delta_{g} \Gamma_{\rho\nu}^{\rho})+\Gamma_{\rho\mu}^{\lambda}(\delta_{g} \Gamma_{\lambda\nu}^{\rho})-\Gamma^{\lambda}_{\mu\rho}(\delta_{g}\Gamma^{\rho}_{\lambda\nu}),\\
%& =2\nabla_{[\rho}(\delta\Gamma_{\mu]\nu}^{\rho})+2\Gamma^{\lambda}_{[\rho\mu]}(\delta \Gamma_{\lambda\nu}^{\rho})\\ 
& =2\nabla_{[\rho}(\delta_{g} \Gamma_{\mu]\nu}^{\rho})+2S\indices{_{\rho\mu}^{\lambda}}(\delta_{g} \Gamma_{\lambda\nu}^{\rho}).
\end{split}
\end{equation}
%\begin{equation}
%\delta R_{\mu\nu}=2\nabla_{[\lambda}\delta \Gamma^{\lambda}_{\mu]\nu}+2S\indices{_{\lambda\mu}^{\alpha}}\delta \Gamma^{\lambda}_{\alpha\nu}	
%\end{equation}
Contracting the first term with $g^{\mu\nu}$ reveals an explicit total divergence
\begin{equation}
\begin{split}
\nabla_{\rho}(\delta_{g} \Gamma_{\mu\nu}^{\rho}g^{\mu\nu}-g^{\rho\nu}\delta_{g} \Gamma_{\lambda\nu}^{\lambda})=& \frac{1}{\sqrt{-g}}\partial_{\rho}(\sqrt{-g}(\delta_{g} \Gamma_{\mu\nu}^{\rho}g^{\mu\nu}-g^{\rho\nu}\delta_{g} \Gamma_{\lambda\nu}^{\lambda})), \\&+2S\indices{_{\lambda\rho}^{\lambda}}(\delta_{g} \Gamma_{\mu\nu}^{\rho}g^{\mu\nu}-g^{\rho\nu}\delta_{g}\Gamma_{\lambda\nu}^{\lambda}).
\end{split}
\end{equation}
This divergence is neglected and taken out
%and explicitly mentioned by Hehl in the Appendix of 
(see the Appendix of \cite{Hehl}) so that the relevant parts of the $g^{\mu\nu}\delta_{g} R_{\mu\nu}$ term are now
\begin{equation}\label{MetricVariationR}
\begin{split}
g^{\mu\nu}\delta_{g} R_{\mu\nu}=&2S\indices{_{\lambda\rho}^{\lambda}}(\delta_{g} \Gamma^{\rho}_{\mu\nu}g^{\mu\nu}-g^{\rho\nu}\delta_{g} \Gamma_{\lambda\nu}^{\lambda})+2S\indices{_{\rho}^{\nu\lambda}}(\delta_{g} \Gamma_{\lambda\nu}^{\rho}),\\ = & 2(S\indices{_{\rho}^{\nu\lambda}}+\delta_{\rho}^{\lambda}S\indices{^{\nu\alpha}_{\alpha}}-g^{\lambda\nu}S\indices{_{\rho\alpha}^{\alpha}})(\delta_{g}\Gamma_{\lambda\nu}^{\rho}),\\ = & 2T\indices{_{\rho}^{\nu\lambda}}(\delta_{g} \Gamma^{\rho}_{\lambda\nu}),
\end{split}
\end{equation}
where we have used the modified torsion as defined earlier. For the variation of the connection with respect to the metric we use \cite{Ortin}
\begin{equation}
%\begin{split}
\delta_{g} \Gamma_{\lambda\nu}^{\rho}=  \frac{1}{2}g^{\rho\gamma}\left(\nabla_{\lambda}(\delta g_{\nu\gamma})+\nabla_{\nu}(\delta g_{\lambda\gamma})-\nabla_{\gamma}(\delta g_{\lambda\nu}) \right)
%\\ & -\left(-\delta S\indices{_{\lambda\nu}^{\rho}}+g^{\rho\mu}g_{\lambda\alpha}\delta S\indices{_{\nu\mu}^{\alpha}}-g^{\rho\mu}g_{\nu\alpha}\delta S\indices{_{\mu\lambda}^{\alpha}} \right)
%\end{split}
\end{equation}
and thus, we obtain
\begin{equation}
2T\indices{_{\rho}^{\nu\lambda}}\left(\delta \Gamma_{\lambda\nu}^{\rho} \right)=T^{\gamma\nu\lambda}(\nabla_{\lambda}(\delta g_{\nu\gamma})+\nabla_{\nu}(\delta g_{\lambda\gamma})-\nabla_{\gamma}(\delta g_{\lambda\nu})).
\end{equation}
We put this term into the integral after which, in order to integrate by parts, it is worth noting that for each of these three terms we get expressions of the form

\begin{equation}
\int d^{4}x \sqrt{-g}T^{\gamma\nu\lambda}(\nabla_{\lambda}(\delta g_{\nu\gamma}))=\int d^{4}x \sqrt{-g}\left[T^{\gamma\nu\lambda}\partial_{\lambda}(\delta g_{\nu\gamma})-\Gamma_{\lambda\nu}^{\alpha}\delta g_{\alpha\gamma}T^{\gamma\nu\lambda}-\delta g_{\nu\alpha}T^{\gamma\nu\lambda}\Gamma_{\lambda\gamma}^{\alpha}\right].
\end{equation}
Integrating by parts for the first term on the right hand side gives
\begin{equation}
\begin{split}
\int d^{4}x \sqrt{-g}T^{\gamma\nu\lambda}(\nabla_{\lambda}(\delta g_{\nu\gamma}))=& -\int d^{4}x\sqrt{-g}\left(\nabla_{\lambda}T^{\gamma\nu\lambda}-2S\indices{_{\lambda\alpha}^{\lambda}}T^{\gamma\nu\alpha} \right)\delta g_{\nu\gamma}\\ & +\int d^{4}x\partial_{\alpha}\left(\sqrt{-g}T^{\gamma\nu\alpha}\delta g_{\nu\gamma} \right),
\end{split}
\end{equation}
where the last term on the right is a full divergence and can be neglected. Finally, using the definition of the star derivative $\{ \accentset{\star}{\nabla}_{\lambda} \}$ and adding up the three terms we have hence
%
%\begin{equation}
%\int d^{4}x \sqrt{-g}2T\indices{_{\rho}^{\nu\lambda}}\left(\delta_{g}\Gamma_{\lambda\nu}^{\rho} \right)=-\int d^{4}x \sqrt{-g} \accentset{\star}{\nabla}_{\lambda}%\left(T^{\gamma\nu\lambda}+T^{\gamma\lambda\nu}-T^{\lambda\nu\gamma} \right) \delta g_{\nu\gamma}
%\end{equation}
%
\begin{equation}
\int d^{4}x \sqrt{-g} 2T\indices{_{\rho}^{\nu\lambda}}\left(\delta_{g}\Gamma_{\lambda\nu}^{\rho} \right)=\int d^{4}x \sqrt{-g}\accentset{\star}{\nabla}_{\lambda}\left(T\indices{_{\nu\mu}^{\lambda}}+T\indices{_{\nu}^{\lambda}_{\mu}}-T\indices{^{\lambda}_{\mu\nu}}\right)\delta g^{\mu\nu}.
\end{equation}
We note that
\begin{equation}
R_{\mu\nu}\delta_{g}\mathfrak{g}^{\mu\nu}=\sqrt{-g}\left(R_{\mu\nu}-\frac{1}{2}g_{\mu\nu} R \right)\delta g^{\mu\nu}=\sqrt{-g}G_{(\mu\nu)}\delta g^{\mu\nu}.
\end{equation}
We recall that the new Einstein tensor is $G_{\mu\nu}\equiv R_{\mu\nu}-\frac{1}{2}g_{\mu\nu}R$ in analogy to its GR equivalent. The symmetrization 
of this new Einstein tensor (EC tensor) 
is due to the fact that we are contracting it with a variation of the metric, making any antisymmetric part of the tensor $G_{\mu\nu}$ arbitrary and thus not relevant 
\cite{Variations}. The remaining terms of the variation
\begin{equation}
2T\indices{_{\rho}^{\nu\lambda}}\delta_{g}\Gamma^{\rho}_{\lambda\nu}=\sqrt{-g}\accentset{\star}{\nabla}_{\lambda}\left(-T\indices{_{\mu\nu}^{\lambda}}+T\indices{_{\nu}^{\lambda}_{\mu}}-T\indices{^{\lambda}_{\mu\nu}}\right)\delta g^{\mu\nu}
\end{equation}
should also be symmetrized. After this we arrive at
\begin{equation}
\frac{1}{\sqrt{-g}}\frac{\delta\mathfrak{R}}{\delta g^{\mu\nu}}=G_{(\mu\nu)}+\accentset{\star}{\nabla}_{\lambda}\left(T\indices{_{\nu}^{\lambda}_{\mu}}-T\indices{^{\lambda}_{\mu\nu}} \right),
\end{equation}
where we have left only the symmetric terms. Due to our previous result (\ref{AntisymmetricRicciG}) $G_{[\mu \nu]}= R_{[\mu\nu]}=\accentset{\star}{\nabla}_{\lambda}T\indices{_{\mu\nu}^{\lambda}}$, 
the variation with respect to the metric is simply \cite{HehlPRD} 
\begin{equation}
\frac{1}{\sqrt{-g}}\frac{\delta \mathfrak{R}}{\delta g^{\mu\nu}}=G_{\mu\nu}+\accentset{\star}{\nabla}_{\lambda}\left(-T\indices{_{\mu\nu}^{\lambda}}+T\indices{_{\nu}^{\lambda}_{\mu}}-T\indices{^{\lambda}_{\mu\nu}} \right).
\end{equation} 
On the other hand, for the variation with respect to contortion we have
\begin{equation}
\delta_{K}\mathfrak{R}=\sqrt{-g}\left[2T\indices{_{\rho}^{\nu\lambda}}(\delta_{K}\Gamma\indices{^{\rho}_{\lambda\nu}})\right],
\end{equation}
where we have used the results from the variation with respect to the metric shown in equation (\ref{MetricVariationR}). %(Sergio: which)
Furthermore, we find
\begin{equation}
\delta_{K}\Gamma_{\lambda\nu}^{\rho}=-\delta K\indices{_{\lambda\nu}^{\rho}}.
\end{equation}
With this result we are led to
\begin{equation}
\frac{1}{\sqrt{-g}}\frac{\delta \mathfrak{R}}{\delta K\indices{_{\mu\nu}^{\lambda}}}=-2T\indices{_{\lambda}^{\nu\mu}}.
\end{equation}
Concentrating on the matter sector of the Lagrangian $\mathcal{L}_{m}$, we  define 
the \emph{metric} energy--momentum tensor $\sigma_{\mu\nu}$ and the source $\tau\indices{_{\alpha}^{\nu\mu}}$ as
\begin{equation}\label{EMTensors}
\sigma_{\mu\nu}:=-\frac{2}{\sqrt{-g}}\frac{\delta \mathcal{L}_{m}}{\delta g^{\mu\nu}}\mbox{ }\mbox{ } \mbox{ , } \mbox{ } \mbox{ } \tau\indices{_{\alpha}^{\nu\mu}}:=\frac{1}{\sqrt{-g}}\frac{\delta \mathcal{L}_{m} }{\delta K\indices{_{\mu\nu}^{\alpha}}}.
\end{equation}
With all the results we can set up the field equations as
\begin{equation}\label{EinsteinHehl0}
G_{\mu\nu}+\accentset{\star}{\nabla}_{\lambda}\left(-T\indices{_{\mu\nu}^{\lambda}}+T\indices{_{\nu}^{\lambda}_{\mu}}-T\indices{^{\lambda}_{\mu\nu}} \right)=\kappa \sigma_{\mu\nu},
\end{equation}
\begin{equation}\label{TorsionHehl}
S\indices{_{\lambda\nu}^{\mu}}+\delta_{\lambda}^{\mu}S\indices{_{\nu\alpha}^{\alpha}}-\delta^{\mu}_{\nu}S\indices{_{\lambda\alpha}^{\alpha}}=\kappa \tau\indices{_{\lambda\nu}^{\mu}}.
\end{equation}
Note that the variation with respect to contortion is just an algebraic equation for torsion in terms of its source $\tau\indices{_{\mu\nu}^{\alpha}}$ which can be also
cast into a form using the modified torsion
\begin{equation}
T\indices{_{\mu\nu}^{\alpha}}=\kappa\tau\indices{_{\mu\nu}^{\alpha}}.
\end{equation}
Referring to $\Sigma_{\mu\nu}$ as the non-symmetric total energy-momentum tensor or 
as the canonical energy-momentum tensor defined by
\begin{equation}\label{CanonicalEMT}
\Sigma^{\mu\nu}:=\sigma^{\mu\nu}+\accentset{\star}{\nabla}_{\lambda}\left(\tau^{\mu\nu\lambda}-\tau^{\nu\lambda\mu}+\tau^{\lambda\mu\nu} \right),
\end{equation}
leads us to equations of motion of the Einstein form
\begin{equation}\label{EinsteinHehlFull}
G_{\mu\nu}=\kappa \Sigma_{\mu\nu}.
\end{equation}
%The left hand side of this equation is given by the Ricci tensor and scalar, which using the torsion equation (\ref{TorsionHehl}) can be written in terms of the spin angular momentum tensor using the following results:
From the algebraic torsion equation it follows that
\begin{equation}\label{TorsionSource1}
-2S\indices{_{\alpha\lambda}^{\lambda}}=\kappa \tau\indices{_{\alpha\lambda}^{\lambda}}, \mbox{ } \mbox{  } \mbox{ } \mbox{ } \mbox{ }
S\indices{_{\alpha\nu}^{\mu}}=\kappa \left(\tau\indices{_{\alpha\nu}^{\mu}}+\frac{1}{2}\delta_{\alpha}^{\mu}\tau\indices{_{\nu\lambda}^{\lambda}}-\frac{1}{2}\delta^{\mu}_{\nu}\tau\indices{_{\alpha\lambda}^{\lambda}} \right),
\end{equation}
\begin{equation}\label{TorsionSource2}
K\indices{_{\mu\nu}^{\alpha}}=\kappa\left(-\tau\indices{_{\mu\nu}^{\alpha}}+\tau\indices{_{\nu}^{\alpha}_{\mu}}-\tau\indices{^{\alpha}_{\mu\nu}}-\delta_{\mu}^{\alpha}\tau\indices{_{\nu\lambda}^{\lambda}}+g_{\mu\nu}\tau\indices{^{\alpha}_{\lambda}^{\lambda}} \right).
\end{equation}
It is, of course, mandatory to be able to express the torsion or contortion in terms of the source.
Splitting the symmetric part of the modified Einstein tensor into its torsionless part plus terms related to torsion , i.e. $G_{(\mu \nu)}=\accentset{\circ}{G}_{\mu \nu} +
H_{\mu \nu}$, using equation (\ref{AntisymmetricRicciG}) for the anti-symmetric part expressed through the star covariant derivative  
and finally expressing all torsion terms through the source we arrive at an equation of the form $G = \accentset{\circ}{G}+ H +\accentset{\star}{\nabla}$
which reads explicitly 
\begin{equation}
\label{EHF}
\begin{split}
G_{\mu\nu}=&\accentset{\circ}{R}_{\mu\nu}-\frac{1}{2}g_{\mu\nu}\accentset{\circ}{R}-\kappa\left\{\accentset{\star}{\nabla}_{\lambda}(-\tau\indices{_{\mu\nu}^{\lambda}}+\tau\indices{_{\nu}^{\lambda}_{\mu}}-\tau\indices{^{\lambda}_{\mu\nu}}) \right\}\\&+\kappa^{2}\left\{2 \tau\indices{_{\beta\nu}^{\beta}}\tau\indices{_{\mu\beta}^{\beta}}+\tau_{\alpha\lambda\mu}\tau\indices{^{\lambda\alpha}_{\nu}}+2\tau_{\mu\lambda\alpha}\tau\indices{_{\nu}^{\lambda\alpha}}+2\tau_{\mu\lambda\alpha}\tau\indices{_{\nu}^{\alpha\lambda}} \right\}\\&-\frac{1}{2}g_{\mu\nu}\kappa^{2}\left\{-2 \tau\indices{^{\beta\alpha}_{\alpha}}\tau\indices{_{\beta\lambda}^{\lambda}}+\tau^{\alpha\lambda\beta}\tau_{\alpha\lambda\beta}-2\tau_{\lambda\beta\alpha}\tau^{\alpha\lambda\beta} \right\}.
\end{split}
\end{equation}
This means that the dynamical equation of motion for the metric can be brought into the form
\begin{equation}\label{EinsteinHehl}
\accentset{\circ}{G}_{\mu\nu}\equiv \accentset{\circ}{R}_{\mu\nu}-\frac{1}{2}g_{\mu\nu}\accentset{\circ}{R}=\kappa \tilde{\sigma}_{\mu\nu}.
\end{equation}
Here, $\tilde{\sigma}_{\mu\nu}$ will include the torsion source terms 
\begin{equation}\label{HehlSigma}
\begin{split}
\tilde{\sigma}_{\mu\nu}:=\sigma_{\mu\nu}+\kappa & \left\{-4\tau\indices{_{\mu\lambda}^{[\alpha}}\tau\indices{_{\nu\alpha}^{\lambda]}}-2\tau_{\mu\lambda\alpha}\tau\indices{_{\nu}^{\lambda\alpha}}+\tau_{\alpha\lambda\mu}\tau\indices{^{\alpha\lambda}_{\nu}} \right. \\ & \left. +\frac{1}{2}g_{\mu\nu}\left(4\tau\indices{_{\lambda}^{\beta}_{[\alpha}}\tau\indices{^{\lambda\alpha}_{\beta]}}+\tau^{\alpha\lambda\beta}\tau_{\alpha\lambda\beta} \right) \right\}.
\end{split}
\end{equation}
%Sergio:
%and the $\accentset{\star}{\nabla}_{\lambda}$ terms are absorbed into $\sigma_{\mu\nu}$ by using the definition of $\Sigma_{\mu\nu}$.

All equivalent reformulations have only a physical meaning after we specifically state what is the source of the dynamical equation.
Indeed, looking at Eqs. (\ref{CanonicalEMT}), (\ref{EinsteinHehlFull}) and (\ref{EHF}) it is clear that the term proportional to $\accentset{\star}{\nabla}$ 
cancels from the Einstein tensor and the canonical energy-momentum tensor unless we identify/fix the canonical energy-momentum tensor $\Sigma$ as
the physical source or, equivalently we insist that the equations of motion must have the Einstein form $G=\kappa \Sigma$. 
In such a case the equations of motion are of the form $\accentset{\circ}{G}=\kappa \Sigma -H -\accentset{\star}{\nabla}$ with all torsion terms
expressed through it source. The metric energy-momentum tensor drops out from the picture here.
Though consistent and
logically possible, this is not stringent.  

\subsection{Alternative Variational Approach}
To demonstrate how the equations of motion come out naturally in a non-Einstein form let us apply an alternative method to the
variational principle splitting, right from the beginning, total divergences. 
The starting point of an alternative approach to the variation of the Lagrangian is precisely the same, namely we shall take the action 
%
%Let us start by taking $\mathcal{L}_{G} \sim \sqrt{-g}R[\Gamma]$ to be the gravitational part of the Lagrangian for the Einstein-Cartan theory, so that our full action is given as
%
%While Hehl does the variation of the contracted Riemann tensor in Einstein-Cartan theory, we will do it slightly differently by expressing $R$ in terms of the standard GR lagrangian $\accentset{\circ}{R}$ and contractions of the contorsion $K\indices{_{\mu\nu}^{\alpha}}$. Now we will describe the approach that we have considered to obtain the field equations. If we are to take essentially the same action as Hehl, namely:
%
\begin{equation}
S=\frac{1}{2\kappa}\int d^{4}x \sqrt{-g}R+\int d^{4}x \mathcal{L}_{m},
\end{equation}
where $\mathcal{L}_{m}$ is the matter Lagrangian and $\kappa=8\pi G$ (we have taken $c=1$).  
We write the Ricci scalar in Riemann--Cartan theory explicitly in terms of its torsionless portion and parts related to the torsion so that 
in the process a total divergence as in equation (\ref{RicciContorsion}) appears which we  discard. We define the divergence-less part of the Ricci scalar, $\tilde{R}$, as
 %
%An easier way to obtain the variation with respect to contortion is to use the Ricci scalar written in explicit terms of contorsion as in equation (\ref{RicciContorsion}) but neglecting the full divergence
\begin{equation} \label{extrax}
\tilde{R}\equiv\accentset{\circ}{R}-2K\indices{_{\lambda\alpha}^{\lambda}}K\indices{_{\beta}^{\alpha\beta}}+K\indices{_{\rho}^{\alpha\lambda}}K\indices{_{\lambda\alpha}^{\rho}}-K\indices{_{\alpha}^{\alpha\lambda}}K\indices{_{\rho\lambda}^{\rho}},
\end{equation}
which is just a divergence-less version of equation (\ref{RicciTorsion}).
We interpret it as an equivalent Lagrangian whose action we call %say $\mathcal{L}---$. 
\begin{equation}
S_{T}=\frac{1}{2\kappa}\int d^{4}x\sqrt{-g}\tilde{R}=\frac{1}{2\kappa}\int d^{4}x\sqrt{-g}\left(\accentset{\circ}{R}
+K\indices{_{\rho}^{\alpha\lambda}}K\indices{_{\lambda\alpha}^{\rho}}-K\indices{_{\alpha}^{\lambda\alpha}}K\indices{_{\rho\lambda}^{\rho}}\right).
\end{equation}
Varying this action with respect to contortion yields exactly the same algebraic equation as in the standard approach \cite{Hehl,HehlPRD,Variations,Hammond,PoplawskiInflation}%above
\begin{equation}
-K\indices{_{\alpha}^{\mu\nu}}-K\indices{^{\nu}_{\alpha}^{\mu}}+\delta_{\alpha}^{\mu}K\indices{_{\lambda}^{\lambda\nu}}+g^{\mu\nu}K\indices{_{\beta\alpha}^{\beta}}=-2\kappa \tau\indices{_{\alpha}^{\nu\mu}},
\end{equation}
%
%which after some simple manipulation can be written in the form
%
\begin{equation}
S\indices{_{\lambda\nu}^{\mu}}+\delta_{\lambda}^{\mu}S\indices{_{\nu\alpha}^{\alpha}}-\delta_{\nu}^{\mu}S\indices{_{\lambda\alpha}^{\alpha}}=\kappa \tau\indices{_{\lambda\nu}^{\mu}}, \quad \mbox{ where again } \quad \tau\indices{_{\alpha}^{\nu\mu}}:=\frac{1}{\sqrt{-g}}\frac{\delta \mathcal{L}_{m} }{\delta K\indices{_{\mu\nu}^{\alpha}}}.
\end{equation}
%The second term in the action is a full divergence and as such should not contribute to the variation of the action; we shall neglect it from now on and do the variation of $\tilde{R}$, namely
%\begin{equation}
%\tilde{R}=\accentset{\circ}{R}+g^{\alpha\beta}\left(K\indices{_{\rho\beta}^{\lambda}}K\indices{_{\lambda\alpha}^{\rho}}-K\indices{_{\lambda\beta}^{\lambda}}K\indices{_{\rho\alpha}^{\rho}} \right),
%\end{equation}
%where we have left the contorsion terms in the form they are defined, i.e. in the form $K\indices{_{\mu\nu}^{\alpha}}$, in order to do the variation. We show this variation with respect to the metric explicitly
Next we vary this action with respect to the metric, which simply yields
\begin{equation}
\delta_{g} \left(g^{\alpha\beta}(K\indices{_{\rho\beta}^{\lambda}}K\indices{_{\lambda\alpha}^{\rho}}-K\indices{_{\lambda\beta}^{\lambda}}K\indices{_{\rho\alpha}^{\rho}}) \right)=\left( K\indices{_{\rho\nu}^{\lambda}}K\indices{_{\lambda\mu}^{\rho}}-K\indices{_{\lambda\nu}^{\lambda}}K\indices{_{\rho\mu}^{\rho}} \right)\delta g^{\mu\nu}.
\end{equation}
along with the standard result,
\begin{equation}
\sqrt{-g}\delta \accentset{\circ}{R}=\sqrt{-g}\left(\accentset{\circ}{R}_{\mu\nu}\delta g^{\mu\nu}+g^{\mu\nu}\delta \accentset{\circ}{R}_{\mu\nu} \right).
\end{equation}
Examining the last term on the l.h.s. we note that it is a full divergence as in standard GR and will not contribute to the action. %can write 
Taking into account that
%\begin{equation}
%\begin{split}
%\sqrt{-g}g^{\mu\nu}\delta \accentset{\circ}{R}_{\mu\nu}&=\sqrt{-g}\left\{\accentset{\circ}{\nabla}_{\lambda}\left(g^{\mu\nu}\delta_{g}\accentset{\circ}{\Gamma}_{\mu\nu}^{\lambda} \right)-\accentset{\circ}{\nabla}_{\mu}\left(g^{\mu\nu}\delta \accentset{\circ}{\Gamma}_{\lambda\nu}^{\lambda}\right) \right\}\\
%& = \partial_{\lambda}\left(\sqrt{-g}g^{\mu\nu}\delta\accentset{\circ}{\Gamma}_{\mu\nu}^{\lambda}-\sqrt{-g}g^{\lambda\nu}\delta \accentset{\circ}{\Gamma}_{\alpha\nu}^{\alpha} \right),
%\end{split}
%\end{equation}
%
%which is clearly a full divergence which will not contribute to the action. On the other hand the terms coming from the variation of $\sqrt{-g}$ are
%
\begin{equation}
(\delta_{g}\sqrt{-g})\tilde{R}=-\frac{1}{2}\sqrt{-g}\left(\accentset{\circ}{R}+g^{\alpha\beta}\left(K\indices{_{\rho\beta}^{\lambda}}K\indices{_{\lambda\alpha}^{\rho}}-K\indices{_{\lambda\beta}^{\lambda}}K\indices{_{\rho\alpha}^{\rho}}\right) \right) g_{\mu\nu} \delta g^{\mu\nu}
\end{equation}
we obtain then the following field equations, where $\sigma_{\mu\nu}$ is the metric energy--momentum tensor we have already defined,
\begin{equation}
\accentset{\circ}{R}_{\mu\nu}-\frac{1}{2}g_{\mu\nu}\accentset{\circ}{R}=\kappa \sigma_{\mu\nu}-\left(K\indices{_{\rho\nu}^{\lambda}}K\indices{_{\lambda\mu}^{\rho}}-K\indices{_{\lambda\nu}^{\lambda}}K\indices{_{\rho\mu}^{\rho}} \right)+\frac{1}{2}g_{\mu\nu}\left(K\indices{_{\rho}^{\alpha\lambda}}K\indices{_{\lambda\alpha}^{\rho}}-K\indices{_{\lambda\beta}^{\lambda}}K\indices{_{\rho\alpha}^{\rho}} \right).
\end{equation}
We can abbreviate the term on the right hand side involving only contortion tensors  as the symmetric tensor $-H_{\mu \nu}$ (see below) in which case 
our equations of motion have the form  $\accentset{\circ}{G} +H=\kappa  \sigma$.
Given that the algebraic equation for torsion is the same as we obtained in the standard approach, we can make use of the results given in equations (\ref{TorsionSource1}) and (\ref{TorsionSource2}) to write
%One can take the algebraic equation for torsion in terms of the source and, taking the trace of the equation (\ref{}) we get $-S\indices{_{\alpha\lambda}^{\lambda}}=\kappa \tau\indices{_{\alpha\lambda}^{\lambda}}$ and may write torsion and contorsion in terms of its source
%
%CITE EQUATIONS
%\begin{eqnarray}
%S\indices{_{\alpha\nu}^{\mu}}&=&\kappa \left(\tau\indices{_{\alpha\nu}^{\mu}}+\frac{1}{2}\delta_{\alpha}^{\mu}\tau\indices{_{\nu\lambda}^{\lambda}}-\frac{1}{2}\delta_{\nu}^{\mu}\tau\indices{_{\alpha\lambda}^{\lambda}} \right)\\
%K\indices{_{\mu\nu}^{\alpha}}&=&\kappa \left(-\tau\indices{_{\mu\nu}^{\alpha}}+\tau\indices{_{\nu}^{\alpha}_{\mu}}-\tau\indices{^{\alpha}_{\mu\nu}}-\delta_{\mu}^{\alpha}\tau\indices{_{\nu\lambda}^{\lambda}}+g_{\mu\nu}\tau\indices{^{\alpha}_{\lambda}^{\lambda}} \right)
%\end{eqnarray}
%
%We may now write the torsion parts of the field equations explicitly in terms of the source of torsion $\tau\indices{_{\alpha\nu}^{\mu}}$, as follows
%
%Now, in order to compare it with the result given by Hehl, we will write this result in terms of the angular momentum tensor $\tau\indices{_{\mu\nu}^{\alpha}}$. The terms with $g_{\mu\nu}$ are somewhat straightforward:
%
\begin{equation}
\left[g^{\alpha\beta}\left(K\indices{_{\rho\beta}^{\lambda}}K\indices{_{\lambda\alpha}^{\rho}}-K\indices{_{\lambda\beta}^{\lambda}}K\indices{_{\rho\alpha}^{\rho}}\right)  \right]=\kappa^{2}\left( \tau^{\alpha\lambda\beta}\tau_{\alpha\lambda\beta}-2\tau_{\lambda\beta\alpha}\tau^{\alpha\lambda\beta}-2\tau\indices{_{\alpha\lambda}^{\lambda}}\tau\indices{^{\alpha\beta}_{\beta}} \right),
\end{equation}
\begin{eqnarray*}
K\indices{_{\rho\nu}^{\lambda}}K\indices{_{\lambda\mu}^{\rho}}&=&\kappa^{2}\left\{ \tau\indices{_{\beta\nu}^{\beta}}\tau\indices{_{\mu\beta}^{\beta}}+\tau_{\alpha\lambda\mu}\tau\indices{^{\lambda\alpha}_{\nu}}+2\tau_{\mu\lambda\alpha}\tau\indices{_{\nu}^{\lambda\alpha}}+2\tau_{\mu\lambda\alpha}\tau\indices{_{\nu}^{\alpha\lambda}} \right\},\\
-K\indices{_{\lambda\nu}^{\lambda}}K\indices{_{\rho\mu}^{\rho}}&=&\kappa^{2}\tau\indices{_{\lambda\nu}^{\lambda}}\tau\indices{_{\mu\alpha}^{\alpha}}.
\end{eqnarray*}
%For the right hand side of the field equations we have $\sigma_{\mu\nu}$ as given by equation (\ref{EMTensors}), then we can write the full field equations as follows:
These results mean that in terms of the tensor $\tau\indices{_{\mu\nu}^{\alpha}}$ our equations of motion are of the form
%
%\begin{equation}
%\begin{split}
%\accentset{\circ}{G}_{\mu\nu}=\kappa\sigma_{\mu\nu}-\kappa^{2}&\left\{2\tau\indices{_{\beta\nu}^{\beta}}\tau\indices{_{\mu\beta}^{\beta}}+\tau_{\alpha\lambda\mu}\tau\indices{^{\lambda\alpha}_{\nu}}+2\tau\indices{_{\mu\lambda\alpha}}\tau\indices{_{\nu}^{\lambda\alpha}}+2\tau_{\mu\lambda\alpha}\tau\indices{_{\nu}^{\alpha\lambda}}\right. \\
%& \left. -\frac{1}{2}g_{\mu\nu}\left(\tau^{\alpha\lambda\beta}\tau_{\alpha\lambda\beta}-2\tau_{\lambda\beta\alpha}\tau^{\alpha\lambda\beta}-2\tau\indices{_{\alpha\lambda}^{\lambda}}\tau\indices{^{\alpha\beta}_{\beta}} \right) \right\}\\
%\end{split}
%\end{equation}
\begin{equation}\label{FEU}
\begin{split}
\accentset{\circ}{G}_{\mu\nu}=\kappa\sigma_{\mu\nu}+\kappa^{2} & \left\{-4\tau\indices{_{\mu\lambda}^{[\alpha}}\tau\indices{_{\nu\alpha}^{\lambda]}}-2\tau_{\mu\lambda\alpha}\tau\indices{_{\nu}^{\lambda\alpha}}+\tau_{\alpha\lambda\mu}\tau\indices{^{\alpha\lambda}_{\nu}} \right. \\ & \left. +\frac{1}{2}g_{\mu\nu}\left(4\tau\indices{_{\lambda}^{\beta}_{[\alpha}}\tau\indices{^{\lambda\alpha}_{\beta]}}+\tau^{\alpha\lambda\beta}\tau_{\alpha\lambda\beta} \right) \right\}.
\end{split}
\end{equation}
Note that in this derivation no terms involving 
$\accentset{\star}{\nabla}_{\lambda}(\tau)$ are present and yet the equation above is formally equivalent to the one obtained in \cite{HehlPRD,HehlPart1,HehlPart2}. 
In order to account for this 
equivalence we point out that adding on both sides a term proportional $\accentset{\star}{\nabla}$ i.e.,
$
\Sigma^{\mu\nu}:=\sigma^{\mu\nu}+\accentset{\star}{\nabla}_{\lambda}\left(\tau^{\mu\nu\lambda}-\tau^{\nu\lambda\mu}+\tau^{\lambda\mu\nu} \right)
$
we obtain the Einstein form
%
%Note that Hehl  and others [REFS] write this equation in the following way
%
%\begin{equation}
\[
G_{\mu\nu}=R_{\mu\nu}-\frac{1}{2}g_{\mu\nu}R=\kappa \Sigma_{\mu\nu}.
\]
With the following tensor
\begin{equation}
\begin{split}\label{Hmunu}
H_{\mu\nu}\equiv\kappa^{2}&\left\{2\tau\indices{_{\beta\nu}^{\beta}}\tau\indices{_{\mu\beta}^{\beta}}+\tau_{\alpha\lambda\mu}\tau\indices{^{\lambda\alpha}_{\nu}}+2\tau\indices{_{\mu\lambda\alpha}}\tau\indices{_{\nu}^{\lambda\alpha}}+2\tau_{\mu\lambda\alpha}\tau\indices{_{\nu}^{\alpha\lambda}}\right. \\
& \left. -\frac{1}{2}g_{\mu\nu}\left(\tau^{\alpha\lambda\beta}\tau_{\alpha\lambda\beta}-2\tau_{\lambda\beta\alpha}\tau^{\alpha\lambda\beta}-2\tau\indices{_{\alpha\lambda}^{\lambda}}\tau\indices{^{\alpha\beta}_{\beta}} \right) \right\}
\end{split}
\end{equation}
there is  indeed a formal equivalence between the two approaches, namely,
\begin{equation}
\accentset{\circ}{G}_{\mu\nu}+H_{\mu\nu}=\kappa \sigma_{\mu\nu} \mbox{ } \mbox{ } \mbox{ } \Leftrightarrow \mbox{ } \mbox{ } \mbox{ } G_{\mu\nu}=\kappa \Sigma_{\mu\nu},
\end{equation}
where, to get the equation on the right we need to add the term $\accentset{\star}{\nabla}_{\lambda}\left(\tau^{\mu\nu\lambda}-\tau^{\nu\lambda\mu}+\tau^{\lambda\mu\nu} \right)$ on both sides of the equation. However, it is also clear that the equation of motion comes out in the form $\accentset{\circ}{G}+H=\kappa \sigma$ and only the insistence on the
analogous Einstein form leads to $G=\kappa  \Sigma$. Of course, formally these forms are equivalent. The two approaches will differ once we identify the physical source as $\Sigma$ or
$\sigma$. This will be outlined in the next sub-section. 

\subsection{Identification of the Energy--Momentum Tensor}
If the matter Lagrangian is uniquely given by choosing it to be the Dirac or Klein--Gordon Lagrangian
then, the answer to the question of which of the aforementioned approaches one should use is easily answerable.
The matter Lagrangian fixes the metric energy-momentum tensor $\sigma$ and therefore, the terms proportional to $\accentset{\star}{\nabla}$ cancel on both sides
of the equation of motion given in the form $G=\kappa \Sigma$ leaving us with $\accentset{\circ}{G} +H=\kappa \sigma$. One could argue that alone
for this reason the form   $\accentset{\circ}{G} +H=\kappa \sigma$ with $\sigma$ chosen to be the physical source is always preferable, over $G=\kappa \Sigma$ with $\Sigma$
the physical source, even if we handle a phenomenological energy-momentum tensor. However, even if we follow this path, there is still place
for new contributions to the phenomenological energy-momentum tensor coming from the source of the torsion. 
Especially, when turning to cosmology one is faced with the question of the choice of the energy-momentum tensor, 
given that in torsionless standard cosmology  one usually takes a perfect fluid
energy--momentum tensor. 

%Sergio:
We follow the approach taken by some authors of including spin into the energy--momentum
tensor, motivated by the fact that torsion is a manifestation of spin \cite{SabbataBook}.
%Some authors - driven by the notion that torsion could be a manifestation of spin \cite{SabbataBook}- have taken an energy momentum tensor with spin.
%;we will see that there are essentially different identifications for a spin fluid, being a Weyssenhoff fluid the starting phenomenological point.
Here, we are faced with two options which we will study independently, \textbf{(i)} we could take the construction given in \cite{HehlPRD,PoplawskiInflation,Variations}
following the full Einstein form as in equation (\ref{EinsteinHehlFull}) and work with the so called canonical energy--momentum tensor $\Sigma_{\mu\nu}$. 
Alternatively,  \textbf{(ii)} we could take the construction obtained from varying the Lagrangian after splitting off 
total divergences, namely equation (\ref{FEU}), and identify $\sigma$ as the physical source admitting room for contributions with spin. 
We will examine different choices of the energy--momentum tensor in
these two approaches below.

Our starting point is to determine the 
canonical energy--momentum tensor which we assume to be of the form \cite{Kopczynski1,Kopczynski2}
%
%As we have seen above, one can write the (explicitly symmetric) field equations as
%
%\begin{equation}
%\accentset{\circ}{R}_{\mu\nu}-\frac{1}{2}g_{\mu\nu}\accentset{\circ}{R}=\kappa \tilde{\sigma}_{\mu\nu},
%\end{equation}
%
%so in order to rewrite the them in this way we will first take the following energy-momentum tensor $\Sigma_{\mu\nu}$
%
\begin{equation}
\Sigma_{\mu\nu}=h_{\mu}u_{\nu}-p g_{\mu\nu},
\end{equation}
where $p$ is the pressure of the fluid, $u_{\nu}$ is the four-velocity and $h_{\mu}$ is the enthalpy density. 
An expression for the enthalpy density can be obtained from the identity (\ref{BianchiTorsion}) %to be \cite{Kopczynski1,Kopczynski2}
%
%\begin{equation}
%h_{\mu}=(\epsilon + P)u_{\mu}+2u^{\lambda}\accentset{\circ}{\nabla}_{\alpha}\left(u^{\alpha}s_{\mu\lambda} \right).
%\end{equation}
%
\begin{equation} \label{antisymmetricsigma}
\Sigma_{[\mu\nu]}=\accentset{\star}{\nabla}_{\lambda}\tau\indices{_{\mu\nu}^{\lambda}} \Rightarrow h_{[\mu}u_{\nu]}=\accentset{\star}{\nabla}_{\lambda}\tau\indices{_{\mu\nu}^{\lambda}},
\end{equation}
and the requirement (coined upon a similar result in the standard hydrodynamical energy-momentum tensor)
\begin{equation} \label{hydro}
\Sigma_{\mu\nu}u^{\mu}u^{\nu}=\rho.
\end{equation}
As a result, choosing the source of torsion as spin tensor multiplied by velocity
\begin{equation} \label{sourcespin}
\tau\indices{_{\mu\nu}^{\alpha}}=s_{\mu\nu}u^{\alpha},
\end{equation}
one obtains a spin fluid which was studied by Weyssenhoff and Raabe back 
in the 1940s \cite{Weyssenhoff} and further studied by Ray and Smalley in a series of papers \cite{RaySmalley1,RaySmalley2,RaySmalley3,RaySmalley4,RaySmalley5,RaySmalley6,SmalleyKrisch,Smalley,RaySmalleyKrisch}. To this end, we take into account the Frenkel conditions \cite{Frenkel} 
\begin{equation}\label{Frenkel}
s_{\mu\nu}u^{\nu}=0,
\end{equation} 
which by anti-symmetry of the spin tensor implies $s_{0i}v^i=0$.
%
%\begin{equation}
%
%\end{equation}
%
By virtue of (\ref{antisymmetricsigma}) we find an interesting relation, namely
\begin{equation}
h_{\mu}u_{\nu}-h_{\nu}u_{\mu}=2\nabla_{\lambda}(s_{\mu\nu}u^{\lambda})+4S\indices{_{\lambda\alpha}^{\alpha}}s_{\mu\nu}u^{\lambda},
\end{equation}
in which last term is zero given that $S\indices{_{\lambda\alpha}^{\alpha}}\propto  \tau\indices{_{\lambda\alpha}^{\alpha}}=s_{\lambda\alpha}u^{\alpha}=0$. The simplified version reads now
\begin{equation} \label{enthalpy}
\begin{split}
h_{\mu}u_{\nu}-h_{\nu}u_{\mu}&=2\accentset{\circ}{\nabla}_{\lambda}(s_{\mu\nu}u^{\lambda})+K\indices{_{\lambda\mu}^{\alpha}}s_{\alpha\nu}u^{\lambda}+K\indices{_{\lambda\nu}^{\alpha}}s_{\mu\alpha}u^{\lambda}-K\indices{_{\lambda\alpha}^{\lambda}}s_{\mu\nu}u^{\alpha}
\\
&=2\accentset{\circ}{\nabla}_{\lambda}(s_{\mu\nu}u^{\lambda}).
\end{split}
\end{equation}
Again, the last two terms with contortion and the spin tensor can be shown to yield zero by using equation (\ref{TorsionSource2}) along with the Frenkel conditions. 
Invoking as before $\Sigma_{\mu\nu}u^{\mu}u^{\nu}=\rho$, we find that
\begin{equation}
h_{\mu}u^{\mu}=(\rho+p).
\end{equation}
Finally, contracting equation (\ref{enthalpy}) with $u^{\mu}$, we obtain 
%
%\begin{equation}
%h_{\mu}u_{\nu}u^{\mu}-h_{\nu}u_{\mu}u^{\mu}=2u^{\mu}\accentset{\circ}{\nabla}_{\lambda}(s_{\mu\nu}u^{\lambda})
%\end{equation}
%
%and finally
%
%\begin{equation}
%(\epsilon + P)u_{\nu}-h_{\nu}=2u^{\mu}\accentset{\circ}{\nabla}_{\lambda}(u^{\lambda}s\indices{_{\mu\nu}})
%\end{equation}
\begin{equation}
h_{\nu}=(\rho + p)u_{\nu}+2u^{\mu}\accentset{\circ}{\nabla}_{\lambda}(u^{\lambda}s_{\nu\mu}),
\end{equation}
which fixes the canonical energy--momentum tensor
\begin{equation}
\Sigma_{\mu\nu}=(\rho+p)u_{\mu}u_{\nu}-pg_{\mu\nu}+2u^{\lambda}u_{\nu}\accentset{\circ}{\nabla}_{\alpha}(u^{\alpha}s_{\mu\lambda}).
\end{equation}
%
%We shall use this together with the so-called Frenkel conditions \cite{Frenkel}, namely
%
%\begin{equation}\label{Frenkel}
%\tau\indices{_{\mu\nu}^{\lambda}}=s_{\mu\nu}u^{\lambda} \mbox{ } \mbox{ } \mbox{ , } \mbox{ } \mbox{ } s_{\mu\nu}u^{\nu}=0
%\end{equation}
%
%and the square of the spin:
%
%\begin{equation}
%s_{\mu\nu}s^{\mu\nu}=\frac{1}{2}s^{2}
%\end{equation}
By identifying the canonical energy--momentum tensor as the physical source, the metric energy-momentum tensor (see equation (\ref{CanonicalEMT})) has dropped out from the picture, but it is useful
to define formally a similar expression
\begin{equation} \label{xxx}
\begin{split}
\bar{\sigma}^{\mu\nu}&\equiv\Sigma^{\mu\nu}-\accentset{\star}{\nabla}_{\lambda}\left(\tau^{\mu\nu\lambda}-\tau^{\nu\lambda\mu}+\tau^{\lambda\mu\nu} \right),
\\ &= (\rho + p)u^{\mu}u^{\nu}-p g^{\mu\nu}+2u^{\lambda}u^{(\nu}\accentset{\circ}{\nabla}_{\alpha}(u^{\alpha}s\indices{^{\mu)}_{\lambda}})-\accentset{\star}{\nabla}_{\lambda}\left(-\tau^{\nu\lambda\mu}+\tau^{\lambda\mu\nu} \right),
\end{split}
\end{equation}
where we have kept the terms symmetric in $(\mu\nu)$. For the sake of later references to this quantity, we will refer to the term proportional to
$\accentset{\circ}{\nabla}$ as the Weyssenhoff term and to the term proportional to $\accentset{\star}{\nabla}$ simply as the  $\accentset{\star}{\nabla}$-term. 
In (\ref{xxx}) we can replace the covariant derivative with respect to the Christoffel connection by the full
covariant derivative with respect to the full connection.

A few comments on the spin tensor $s_{\mu \nu}$ are in order. First note that $s_{\mu \nu}$ is related to the spin vector
by $s_{\alpha}=g^{-1/2}\epsilon_{\alpha\beta \gamma \delta}s^{\beta \gamma}u^{\delta}$. Interpreting the spin vector as an expectation
value of quantum mechanical spin operators, $s_{\alpha}$ will be proportional to $\hbar$ reminiscent of its quantum mechanical origin.
Relating the source of the torsion to quantum mechanics might help us to understand why the Einstein--Cartan cosmologies can avoid the
initial singularity.  

Secondly, the microscopical quantities within this theory will be fluctuating, thus it is common to take average values of these quantities. If indeed spin orientation 
is random, then the average of the spin will vanish, but not the terms that are quadratic in spin \cite{Hehl}. Given this non-vanishing quantity, the average of the square of the 
spin is taken to be\footnote{A different notation in \cite{PoplawskiInflation,PoplawskiSingular,PoplawskiBH}.}
\begin{equation}
\langle s_{\mu\nu}s^{\mu\nu} \rangle=\frac{1}{2}s^{2},
\end{equation}
where we follow the notation in \cite{HehlPRD} as far as the quantities with spin are concerned. 
Taking $\tau\indices{_{\mu \nu}^{\alpha}}=s_{\mu \nu} u^{\alpha}$ and respecting the Frenkel conditions 
we can write
\begin{equation} \label{xxxx}
\begin{split}
\bar{\sigma}^{\mu\nu}&=(p+\rho)u^{\mu}u^{\nu}-pg^{\mu\nu}-2u_{\lambda}u^{\alpha}\accentset{\circ}{\nabla}_{\alpha}\left(\left\langle s^{\lambda(\mu}\right\rangle u^{\nu)}\right)+\left[-\kappa s^{2}u^{\mu}u^{\nu}+2\accentset{\circ}{\nabla}_{\lambda}\left(\left\langle s^{\lambda(\mu} \right\rangle u^{\nu)} \right)+2\kappa \langle s^{\mu\lambda}s\indices{^{\nu}_{\lambda}}\rangle\right]
\\
&=(\rho+p-\kappa s^{2})u^{\mu}u^{\nu}-pg^{\mu\nu}-2\left(u_{\lambda}u^{\alpha}
+\delta^{\alpha}_{\lambda} \right)\accentset{\circ}{\nabla}_{\alpha}\left(\left\langle s^{\lambda(\mu}\right\rangle u^{\nu)}\right)+2\kappa \langle s^{\mu\lambda}s\indices{^{\nu}_{\lambda}}\rangle,
\end{split}
\end{equation}
where the $\accentset{\star}{\nabla}$-terms have contributed to the 
terms in the square brackets.
It is also worth noting that in cosmology (which we consider in the subsequent section) 
the comoving reference frame $u^{\mu}=(1,0,0,0)$ together with the Frenkel conditions yields a `new' condition, i.e. $s^{i0}=0$ \cite{ECWormhole}. Adding to $\tilde{\sigma}_{\mu\nu}$ the components of
$H_{\mu \nu}$ (see equation (\ref{Hmunu})) we arrive at
\begin{equation}
\begin{split}
\tilde{\sigma}_{\mu\nu}=\bar{\sigma}_{\mu\nu}+\kappa & \left\{-4\tau\indices{_{\mu\lambda}^{[\alpha}}\tau\indices{_{\nu\alpha}^{\lambda]}}-2\tau_{\mu\lambda\alpha}\tau\indices{_{\nu}^{\lambda\alpha}}+\tau_{\alpha\lambda\mu}\tau\indices{^{\alpha\lambda}_{\nu}} \right. \\ & \left. +\frac{1}{2}g_{\mu\nu}\left(4\tau\indices{_{\lambda}^{\beta}_{[\alpha}}\tau\indices{^{\lambda\alpha}_{\beta]}}+\tau^{\alpha\lambda\beta}\tau_{\alpha\lambda\beta} \right) \right\}
\end{split}
\end{equation}
The contributions of $H$ that yield non-zero expressions in the co-moving frame are
%
%\begin{equation}
%-2\tau\indices{^{\mu}_{\lambda}^{\alpha}}\tau\indices{^{\nu}_{\alpha}^{\lambda}}+2\tau\indices{^{\mu}_{\lambda}^{\lambda}}\tau\indices{^{\nu}_{\alpha}^{\alpha}}=-2s\indices{^{\mu}_{\lambda}}u^{\alpha}s\indices{^{\nu}_{\alpha}}u^{\lambda}+2s\indices{^{\mu}_{\lambda}}u^{\lambda}s\indices{^{\nu}_{\alpha}}u^{\alpha}=0
%\end{equation}
%
\begin{eqnarray} \label{freeaverage}
-2\tau\indices{^{\mu}_{\lambda\alpha}}\tau^{\nu\lambda\alpha}=&-2\langle s\indices{^{\mu}_{\lambda}}s^{\nu\lambda}\rangle u_{\alpha}u^{\alpha}&=-2\langle s\indices{^{\mu}_{\lambda}}s^{\nu\lambda}\rangle,\\
%\end{equation}
%
%\begin{equation}
\tau\indices{_{\alpha\lambda}^{\mu}}\tau^{\alpha\lambda\nu}=&\langle s_{\alpha\lambda}s^{\alpha\lambda}\rangle u^{\mu}u^{\nu}&=\frac{1}{2}s^{2}u^{\mu}u^{\nu},\\
%\end{equation}
%
%\begin{equation}
%2\tau\indices{_{\lambda}^{\beta}_{\alpha}}\tau\indices{^{\lambda\alpha}_{\beta}}-2\tau\indices{_{\lambda}^{\beta}_{\beta}}\tau\indices{^{\lambda\alpha}_{\alpha}}=2s\indices{_{\lambda}^{\beta}}u_{\alpha}s^{\lambda\alpha}u_{\beta}-2s\indices{_{\lambda}^{\beta}}u_{\beta}s^{\lambda\alpha}u_{\alpha}=0
%\end{equation}
%
%\begin{equation}
\tau^{\alpha\lambda\beta}\tau_{\alpha\lambda\beta}=&\langle s^{\alpha\lambda}s_{\alpha\lambda}\rangle u^{\beta}u_{\beta}&=\frac{1}{2}s^{2},
\end{eqnarray}
The term (\ref{freeaverage}) cancels with the last term in (\ref{xxxx}). The modified energy--momentum tensor $\tilde{\sigma}_{\mu\nu}$ will finally be 
\begin{equation}
\tilde{\sigma}_{\mu\nu}=\left(\rho + p -\frac{1}{2}s^{2}\right)u_{\mu}u_{\nu}-\left(p-\frac{1}{4}\kappa s^{2}\right)g_{\mu\nu}-2(u_{\lambda}u^{\alpha}
+\delta^{\alpha}_{\lambda})\accentset{\circ}{\nabla}_{\alpha}\left(\left \langle s\indices{^{\lambda}_{(\mu}}\right \rangle u_{\nu)} \right).
\end{equation}
The equations of motion here are of the form $\accentset{\circ}{G}=\kappa \tilde{\sigma}$. Next, we turn our attention to the second possibility of identifying the metric energy--momentum tensor as the physical source.
As outlined above in sub-section B the variation of the Lagrangian with respect to the metric leads to the equation
of motion $\accentset{\circ}{G} + H =\kappa \sigma$ . There is no doubt that this is the correct result and only formally equivalent to $G=\kappa \Sigma$.
If we view $\sigma$ as the physical source, we can repeat some steps from above. i.e., we write as an ansatz
\begin{equation} \label{sig1}
\sigma_{\mu \nu}=\frac{1}{2}(h_{\mu}u_{\nu} +h_{\nu} u_{\mu}) -pg_{\mu \nu}
\end{equation}
and impose $\sigma_{\mu \nu}u^{\mu}u^{\nu}=\rho$ as before. This gives us the equation $h_{\mu}u^{\mu}= \rho +p$. With the first choice being $h_{\mu}=u_{\mu}(\rho +p) + \tilde{h}_{\mu}$,
the other possible terms denoted here by $\tilde{h}$ have to give zero when contracted with the four-velocity $u^{\mu}$. Hence, 
by virtue of the Frenkel condition we can have 
$\tilde{h}_{\mu} \in \{s_{\mu \alpha} u^{\alpha}, 2u^{\lambda}\accentset{\circ}{\nabla}^{\beta}(u_{\beta} s_{\mu \lambda}), ....\}$.
We recognize in the second term the Weyssenhoff spin contribution mentioned before. But, in general, we can choose any combination of allowed $\tilde{h}_{\mu}$ contribution.
Motivated by the result found in \cite{ObukhovSpin,Obukhov} which claims that the spin contribution to the standard hydrodynamical
energy-momentum tensor is of the Weyssenhoff form, we opt here for 
\begin{equation} \label{xxx}
%\begin{split}
\sigma^{\mu\nu}%\equiv\Sigma^{\mu\nu}-\accentset{\star}{\nabla}_{\lambda}\left(\tau^{\mu\nu\lambda}-\tau^{\nu\lambda\mu}+\tau^{\lambda\mu\nu} \right)
= (\rho + p)u^{\mu}u^{\nu}-p g^{\mu\nu}+2u^{\lambda}u^{(\nu}\accentset{\circ}{\nabla}_{\alpha}(u^{\alpha}s\indices{^{\mu)}_{\lambda}}).
%\end{split}
\end{equation} 
The equations of motion are of the form now $\accentset{\circ}{G}=\kappa\sigma-H$.%=\kappa(\tilde{\sigma} +\accentset{\star}{\nabla}-terms)$. 
\section{A Note on the Connection between Torsion and Quantum Fields}
We focus throughout this paper on geometric aspects of torsion applied to cosmology.
It is, however, certainly worthwhile to mention briefly the connection of
torsion to (fermionic) quantum fields. The first such a link can be found
by noticing that the source of torsion, which in the present paper we treated
in a more global manner, can be also found in the spin part of the angular
momentum tensor which arises in quantum field theory by the invariance of the
matter Lagrangian with respect to proper Lorentz boosts \cite{Bjorken}. 
For example, for the Dirac field we would have $\tau_{\mu \nu \rho} \propto
\epsilon_{\mu \nu \rho \alpha}\bar{\psi}\gamma_5 \gamma^{\alpha}\psi$
where $\psi$ stands for the Dirac field \cite{Dirac1,Dirac2}.
Of course, by making this choice the full system to solve would consist of 
Einstein equations with torsion and the Dirac equation coupled to gravity.
Secondly, in the coupled system torsion gives rise to the so-called
Hehl--Datta term \cite{Kibble, Hehldatta} which is a `bare' four fermion 
interaction term of two axial currents. This has led to new lines of research
\cite{Fansworth, Weller, Poplawsky2001, Poplawsky2011, Poplawsky2011b,
Boos2017} speculating on new propagating massive degrees of freedom, 
new contributions to the matter-anti matter asymmetry and the cosmological
constant (or Dark energy in general) as well as the demonstration of
the possible formation of Cooper pairs due to the Hehl--Datta term.
Finally we point out the connection of torsion to the photon as discussed 
e.g., in \cite{SabbataBook}. 

\section{Standard Einstein--Cartan Cosmology} 

In this section we investigate the consequences of the standard Einstein--Cartan cosmology based on the identification of the canonical energy--momentum tensor $\Sigma$ as the physical source \cite{Atazadeh,Ritis}.
As discussed above our equations are $\accentset{\circ}{G}=\kappa \tilde{\sigma}$. 

In order to study this cosmological model we must obtain the Friedmann equations from the field equations (\ref{EinsteinHehl}). Let us take the Friedman--Lema\^itre--Robertson--Walker metric in its flat version, i.e. $k=0$, and in the sign convention $(+,-,-,-)$ as we have used before. This metric is given by the line element
\begin{equation}
ds^{2}=g_{\mu\nu}x^{\mu}x^{\nu}=dt^{2}-R^2(t)\left(dr^{2}+r^{2}d\theta+r^{2}\sin^{2}\theta d\phi^{2} \right)
\end{equation}
with $R(t)$ the scale factor \cite{Carroll}. 
The Friedmann equations are obtained from the \emph{time--time} and \emph{space--space} components of the field equations. Let us first write explicitly the field equations we will be working with, namely
\begin{equation}
\begin{split}
\accentset{\circ}{R}_{\mu\nu}-\frac{1}{2}\accentset{\circ}{R}= \kappa & \left[\left( \rho+p-\frac{1}{2}\kappa s^{2} \right)u_{\mu}u_{\nu}-\left(p-\frac{1}{4}\kappa s^{2} \right)g_{\mu\nu} \right. \\ & \left. -2\left(u_{\lambda}u^{\alpha}+\delta_{\lambda}^{\alpha} \right)\accentset{\circ}{\nabla}_{\alpha}(\langle s\indices{^{\lambda}_{(\mu}}\rangle u_{\nu)}) \right].
\end{split}
\end{equation}
%where we will also take the four velocity $u_{\mu}$ in the rest frame, namely $u^{0}=1$, $u^{i}=0$. 
We can now take the $00$ component of the field equations to obtain the 
first Friedmann equation. %, for the right hand side we have:
%
%\begin{equation}
%\accentset{\circ}{R}_{00}=-3\dot{H}-3H^{2}
%\end{equation}
%
%where $H\equiv\frac{\dot{a}(t)}{a(t)}$. We also have on the left hand side
%
%\begin{equation}
%\accentset{\circ}{R}=-6\dot{H}-12H^{2}.
%\end{equation}
%
%As for the left hand side, most of it is simple enough, we can check that the derivative term is zero
Let us concentrate on the r.h.s. given the Frenkel and the rest frame conditions (recall that $s^{i0}=0$), we note that the term with explicit spin tensor dependence is zero, i.e.
\begin{equation}
\accentset{\circ}{\nabla}_{\alpha}\left(s\indices{^{\alpha}_{0}}u_{0} \right)=0.
\end{equation}
This makes it quite straightforward to derive the first Friedmann equation %for this case
\begin{equation}\label{FP1}
H^{2}=\frac{\kappa}{3}\left(\rho -\frac{1}{4}\kappa s^{2} \right),
\end{equation}
where $H=\dot{R}/R$. %Sergio addition
This is one of the crucial forms of the Friedmann equations which avoid the singularity and leads to a bounce. It defines
a critical density when the right hand side of (\ref{FP1}) becomes zero which implies that at a certain time
the Hubble parameter is also zero. This suggests that the Hubble parameter can have a local minimum, a sign 
of a bouncing universe. Such forms appear in theories such as Loop Quantum Cosmology models \cite{LQCStatus}, Generalized Uncertainty Principle approaches \cite{MinimalLength} or quantum corrections in Newtonian Cosmology \cite{BBN1}.%Sergio%

For the second Friedman equation we take the space components of the field equations where again it is
worth noting that the last term (the Weyssenhoff term) on the right of the Einstein equations is zero because of the rest frame conditions. 
The second Friedmann equation is simply%Ricci tensor for the left hand side, i.e.
%
%\begin{eqnarray}
%\accentset{\circ}{R}_{11}&=&a\ddot{a}+2\dot{a}^{2}\\
%\accentset{\circ}{R}_{22}&=& (a\ddot{a}+2\dot{a}^{2})r^{2}\\
%\accentset{\circ}{R}_{33}&=& (a\ddot{a}+2\dot{a}^{2})r^{2}\sin^{2}\theta
%\end{eqnarray}
%
%which means on the left hand side we get:
%
%\begin{equation}
%\accentset{\circ}{R}_{ii}-\frac{1}{2}g_{ii}\accentset{\circ}{R}=\left(a\ddot{a}+2\dot{a}^{2}\right)\tilde{g}_{ii}-3\left(a\ddot{a}+\dot{a}^{2} \right)\tilde{g}_{ii},
%\end{equation}
%
%where
%
%\begin{equation}
% \tilde{g}_{ij}=\mbox{diag}(1,r^{2},r^{2}\sin \theta).
%\end{equation}
%
%On the right hand side, the term which includes the covariant Christoffel derivative, $\accentset{\circ}{\nabla}_{\alpha}\left(s^{\lambda i}u^{i}\right)$ is zero given the rest frame conditions. The non zero terms together with the above yield the Friedman equation
%
%\begin{equation}
%\left(-2a\ddot{a}-\dot{a}^{2}\right)\tilde{g}_{ii}=\kappa \left(P-\frac{1}{4}\kappa s^{2} \right)a^{2}\tilde{g}_{ii}
%\end{equation}
\begin{equation}\label{FP2}
\frac{\dot{a}^{2}}{a^{2}}+2\frac{\ddot{a}}{a}=3H^{2}+2\dot{H}=-\kappa \left(p-\frac{1}{4}\kappa s^{2} \right),
\end{equation}
where we defined $a=R(t)/R(t_0)=R/R_0$.
This Friedmann equation can be written using equation (\ref{FP1}) in the following form
%\begin{equation}%\label{P2}
%3H^{2}+2\dot{H}=-\kappa \left(P-\frac{1}{4}\kappa s^{2} \right),
%\end{equation}
\begin{equation}
\frac{\ddot{a}}{a}=H^{2}+\dot{H}=-\frac{\kappa}{6}\left(\rho+3p-\kappa s^{2}\right).
\end{equation}
Equations (\ref{FP1}) and (\ref{FP2}) match the ones given in \cite{PoplawskiInflation} with the choice $k=0$. Combining both equations one can also find that 
a conservation equation emerges 
\begin{equation}\label{ContinuityHehl1}
\frac{d}{dt}\left(\rho -\frac{\kappa s^{2}}{4} \right)=-3H\left(\rho + p -\frac{\kappa s^{2}}{2} \right).
\end{equation}

\begin{figure}[htbp] %  figure placement: here, top, bottom, or page
   \centering
   \includegraphics[width=5in]{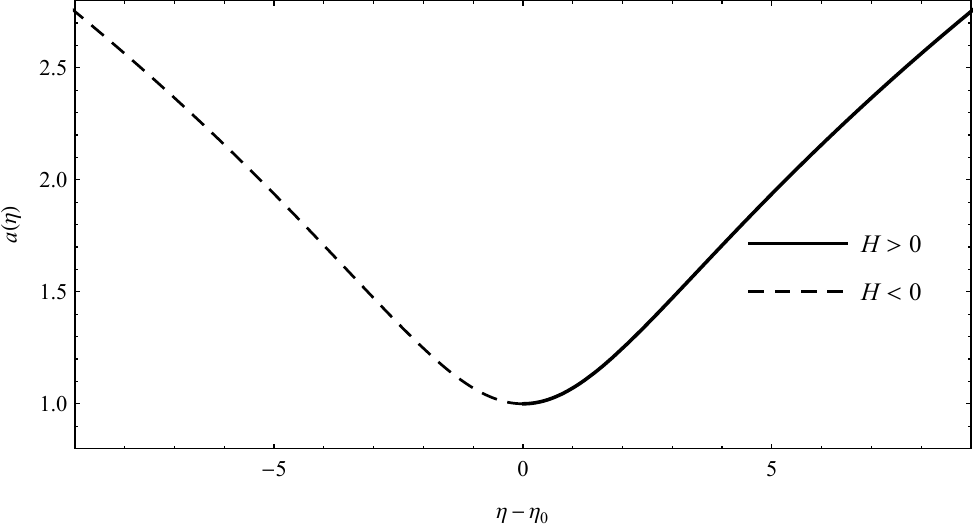} 
   \caption{Plot for the two signs of equation (\ref{PoplawskiFinal}) which correspond to the contracting ($H<0$) and expanding ($H>0$) solutions where
   the bounce at $\eta-\eta_{0}=0$ is indeed differentiable.}
   \label{fig:PoplawskiBounce}
\end{figure}

The relation between the particle number density $n$ and the energy density of the fluid can be obtained as \cite{PoplawskiInflation,Nurgaliev,Gasperini}
%Sergio:from where and how? (SERGIO EXTRA)% 
\begin{equation}
\frac{dn}{n}=\frac{d\rho}{(\rho + p)}.
\end{equation}
From this relation and by using an equation of state of the form $p= w \rho$, one obtains via integration
\begin{equation}
n=\frac{n_{0}}{\rho_{0}^{\frac{1}{1+w}}}\rho^{\frac{1}{1+w}}.
\end{equation}
We will also have the following relation for the spin density of a fluid of fermions with no spin polarization
\begin{equation}\label{spinnumber}
s^{2}=\frac{1}{8}(\hbar c n)^{2},
\end{equation}
which means that we can relate spin density to energy density according to
\begin{equation}\label{SpinDensity}
s^{2}=\frac{1}{8}(\hbar c)^{2} \frac{n_{0}^{2}}{\rho_{0}^{\frac{2}{1+w}}}\rho^{\frac{2}{1+w}}=B_{w}\rho^{\frac{2}{1+w}},
\end{equation}
%\begin{equation}
%s^{2}\propto \epsilon^{\frac{2}{1+w}}
%\end{equation}
where $B_{w}$ is a coefficient with dimension which depends on the value of $w$. Using this proportionality in the continuity equation, we may solve for $\rho$ in terms of $a$.
Let us start by taking equation (\ref{ContinuityHehl1}) in the form
\begin{equation}\label{ContTorsion1}
\frac{d}{dt}\left(\rho - \frac{\kappa B_{w}}{4}\rho^{\frac{2}{1+w}} \right)=-3H\left(\rho+w\rho-\frac{\kappa B_{w}}{2}\rho^{\frac{2}{1+w}} \right).
\end{equation}
Then by using $H=\dot{a}/a$ we can write (\ref{ContTorsion1}) as %can be put in the following form
%
%\begin{equation}
%d\rho \left(1-\frac{\kappa B_{w}}{4}\left(\frac{2}{1+w} \right) \rho^{\frac{1-w}{1+w}} \right)=-3\frac{da}{a}\left((1+w)\rho-\frac{\kappa B_{w}}{2}\rho^{\frac{2}{1+w}} \right).
%\end{equation}
%
\begin{equation}
d\rho \left(1-\frac{\kappa B_{w}}{2(1+w)} \rho^{\frac{1-w}{1+w}} \right)=-3(1+w)\rho \frac{da}{a}\left(1-\frac{\kappa B_{w}}{2(1+w)}\rho^{\frac{1-w}{1+w}} \right)
\end{equation}
which, as long as $\left(1-\frac{\kappa B_{w}}{2(1+w)}\rho^{\frac{1-w}{1+2}}\right)\neq 0$ (or rather $\rho^{\frac{1-w}{1+w}}\neq \frac{2(1+w)}{\kappa B_{w}}$) gives the standard
General Relativity result
\begin{equation}
\int_{\rho_{0}}^{\rho}\frac{d\rho'}{-3(1+w)\rho'}=\int_{a_{0}}^{a}\frac{da'}{a'}, \mbox{ } \mbox{ } \mbox{ } \Rightarrow \mbox{ } \mbox{ } \mbox{ }
%\end{equation}
%\begin{equation}
%\frac{1}{1+w}\log \left(\frac{\epsilon}{\epsilon_{0}} \right)=-3\log \left(\frac{a}{a_{0}}\right)
%\end{equation}
%\begin{equation}
\rho=\rho_{0}\left(\frac{a}{a_{0}} \right)^{-3(1+w)},
\end{equation}
%\begin{equation}
%\epsilon \propto a^{-3(1+w)}
%\end{equation}
%
%as is usually obtained in standard GR. 
From this point on, we will choose $a_{0}=1$. It is now easy to see that the proportionality of $s$ and $a$ is given by
\begin{equation}
s^{2}= B_{w}\rho_{0}^{\frac{2}{1+w}}a^{-6},
\end{equation}
where this proportionality is independent of $w$ given the relation found in equation (\ref{SpinDensity}). As we have stated before, the spin effects will be significant during high density 
eras, that means that we will be interested in the radiation dominated era, and so, from here on, we will take $w=\frac{1}{3}$. In this regime the first Friedmann
equation will be 
\begin{equation}
H^{2}=\frac{\kappa}{3}\left(  \rho_{0} a^{-4}-\frac{1}{4}\kappa B_{\frac{1}{3}}\rho_{0}^{\frac{3}{2}}a^{-6}\right).
\end{equation}
%Sergio:
Note that given the factor $\kappa^{2}$ of the spin part on the r.h.s.
of the equation above, these terms are expected to
be dominant at high densities of matter.
%Sergio-end
We can integrate the above equation to yield
%
%\begin{equation}
%\frac{1}{a}\frac{da}{dt}=\pm \sqrt{\frac{\kappa}{3}}\sqrt{\rho_{0} a^{-4}-\frac{1}{4}\kappa %B_{\frac{1}{3}}\rho_{0}^{\frac{3}{2}}a^{-6}}
%\end{equation}
%
%\begin{equation}
%\int_{a(T_{0})}^{a} d\bar{a} \frac{\bar{a}^{2}}{\sqrt{\rho_{0}\bar{a}^{2}-\frac{1}{4}\kappa %B_{\frac{1}{3}}\rho_{0}^{\frac{3}{2}}}}=\pm \int_{T_{0}}^{t} \sqrt{\frac{\kappa}{3}} dt
%\end{equation}
%to yield 
\begin{equation}
\begin{split}
\frac{a}{2}\sqrt{a^{2}-\frac{1}{4}\kappa B_{\frac{1}{3}}\sqrt{\rho_{0}}}-\frac{a(T_{0})}{2}\sqrt{a(T_{0})^{2}-\frac{1}{4}\kappa B_{\frac{1}{3}}\sqrt{\rho_{0}}}&\\
+\frac{1}{4}\kappa B_{\frac{1}{3}}\sqrt{\rho_{0}}\log \left[\frac{a+\sqrt{a^{2}-\frac{1}{4}\kappa B_{\frac{1}{3}}\sqrt{\rho_{0}}}}{a(T_{0})+\sqrt{a^{2}(T_{0})-\frac{1}{4}\kappa B_{\frac{1}{3}}\sqrt{\rho_{0}}}} \right]&=\pm \sqrt{\frac{\kappa \rho_{0}}{3}}(t-T_{0}),
\end{split}
\end{equation}
which according to \cite{PoplawskiSingular}, gives a non-singular universe.  
We confirm that indeed this is the case by choosing $T_{0}=t_{0}$ and thus $a(T_{0})=a_{0}=1$ which simplifies the above expression, more precisely
\begin{equation}\label{PopBounce}
\begin{split}
\frac{a}{2}\sqrt{a^{2}-\frac{1}{4}\kappa B_{\frac{1}{3}}\sqrt{\rho_{0}}}-\frac{1}{2}\sqrt{1-\frac{1}{4}\kappa B_{\frac{1}{3}}\sqrt{\rho_{0}}}&\\
+\frac{1}{4}\kappa B_{\frac{1}{3}}\sqrt{\rho_{0}}\log \left[\frac{a+\sqrt{a^{2}-\frac{1}{4}\kappa B_{\frac{1}{3}}\sqrt{\rho_{0}}}}{1+\sqrt{1-\frac{1}{4}\kappa B_{\frac{1}{3}}\sqrt{\rho_{0}}}}\right]&=\pm \sqrt{\frac{\kappa \rho_{0}}{3}}(t-t_{0}).
\end{split}
\end{equation}
As already mentioned we have a critical density at which $H^{2}$ becomes zero. For $w=\frac{1}{3}$ it is given by
\begin{equation}
\rho_{crit}=\frac{16}{\kappa^{2}B_{\frac{1}{3}}^{2}}.
\end{equation}
We also see a restriction upon the value of $a$ in the solution above, namely
\begin{equation}
a_{\min}=\sqrt{\frac{\frac{1}{4}\kappa B_{\frac{1}{3}}\rho_{0}^{\frac{3}{2}}}{\rho_{0}}}=\sqrt{\frac{1}{4}\kappa B_{\frac{1}{3}}\rho_{0}^{\frac{1}{2}}}=\left(\frac{\rho_{0}}{\rho_{crit}}\right)^{1/4}
\le 1,
\end{equation}
by virtue of $\rho \le  \rho_{crit}$. 
We note that by making the obvious choice $\rho_0=\rho_{crit}$ we have $a(T_0)=1$ and this also implies 
\begin{equation}
a_{\min}=1
\end{equation}
The expression for $a$ given in equation (\ref{PopBounce}) simplifies to
\begin{equation}\label{PoplawskiFinal}
\frac{a}{2}\sqrt{a^{2}-1}+\log (a+\sqrt{a^{2}-1})=\pm\frac{1}{\sqrt{3}}(\eta-\eta_{0})
\end{equation}
where $\eta\equiv \sqrt{\kappa \rho_{0}}t$. We plot this expression
for the two possible signs, which refer to an expanding universe ($H>0$) 
or a contracting universe ($H<0$), and we note that the branches can be 
taken as one solution given that they are differentiable at the place where they match (see Figure \ref{fig:PoplawskiBounce}).
This represents the aforementioned bouncing universe.

%SEgio: is a plot in the same figure with rho_0 < rho_crit possible?
Within this theory the possibility of inflation has also been considered in \cite{Gasperini,SpinInflation}, and more recently in \cite{PoplawskiInflation}. 
We briefly outline the basics here.
One of the possible
inflation stages occurs when $\ddot{a}>0$ and $\dot{H}>0$, referred to as `super-inflation' \cite{SuperInflation,Lucchin}. Since we have
\begin{equation}
\ddot{a}=-\frac{\kappa}{6}\left(\rho + 3p - \kappa s^{2} \right)a,
\end{equation}
taking the radiation case and the relation between $s^{2}$ and $\rho$ given above, we obtain that this is possible as long as
\begin{equation}
\rho > \frac{4}{\kappa^{2}B^{2}}\equiv \rho_{f},
\end{equation}
where we must also note that the second requirement comes from
\begin{equation}
\dot{H}=-\frac{\kappa}{2}(\rho+p-\frac{1}{2}\kappa s^{2}),
\end{equation}
which implies $\dot{H}>0$ whenever
\begin{equation}
\rho > \frac{64}{9}\frac{1}{\kappa^{2}B_{\frac{1}{3}}^{2}}\equiv \rho_{t}.
\end{equation}
%Sergio: should not the above waution have  a B with an index as before? (Done)
The bound is a strong requirement but still within the possibilities of the model. The transitional stage of this process will be determined by an exponential stage of inflation
which has been suggested by Guth \cite{InflationGuth}. Here, it corresponds to the particular moment at which $\dot{H}=0$, or equivalently when the density is $\rho_{t}$. 
Thus, we have the following scenario
\begin{itemize}
\item{when $\rho_{crit}>\rho>\rho_{t}$, we have the superinflation stage,}
\item{when $\rho=\rho_{t}$, we have the exponential inflation,}
\item{when $\rho_{t}>\rho>\rho_{f}$ we have a power law inflation \cite{Liddle}.}
\end{itemize}
The model expressed above is the standard approach to Einstein--Cartan cosmology 
\cite{Gasperini, PoplawskiBounce}. In the following we will explore how a different choice of the physical source may affect these results.
\\
Indeed, it is almost mandatory to pin down the term which is responsible for the nice feature (especially the avoidance of the initial
singularity) of the theory. To this end we simply drop by hand all the $\accentset{\star}{\nabla}$-terms in (\ref{xxx}).
%which in our notation is equivalent to set
%$\tilde{\sigma}_{\mu \nu}=\sigma_{\mu \nu}$ 
%In this identification we ask if $\sigma_{\mu\nu}$ as our metric energy momentum tensor could receive some corrections coming from spin. Let us first examine what happens 
%when $\sigma_{\mu\nu}$ is simply taken to be the perfect hydrodynamical fluid with spin corrections. A fluid with spin contributions was first studied by Weyssenhoff and Raabe back in the 1940s \cite{Weyssenhoff} when they established a phenomenological basis for the
%study of such a fluid. As we have seen earlier, spin and torsion are related quantities and because of this spin fluids have been of interest in general relativity theories
%with the inclusion of torsion. The construction of an energy momentum tensor for a spin fluid was adressed in \cite{Obukhov} by Obukhov and Korotky (O.K.), who developed the variational theory of an ideal spinning fluid (Weyssenhoff fluid) in Einstein-Cartan theory, obtaining the following 
%expression for the Energy Momentum of spinning fluid without charge:
Hence, we will work simply with
\begin{equation}
\begin{split}
\tilde{\sigma}_{\mu\nu}&=(\rho + p)u_{\mu}u_{\nu}-pg_{\mu\nu}+2u^{\alpha}u_{(\mu}\nabla_{\beta}\tau\indices{_{\alpha|\nu)}^{\beta}}\\
&= (\rho + p)u_{\mu}u_{\nu}-pg_{\mu\nu}+2u^{\alpha}u_{(\mu}\nabla_{\beta}(s_{\alpha|\nu)}u^{\beta}),
\end{split}
\end{equation}
where we have already used one of the Frenkel conditions to write the source of torsion in terms of the spin tensor. 
It is clear to see that this result is basically the perfect fluid part plus
an additional term % added to the energy momentum 
%tensor is 
due to Weyssenhoff (this term is symmetric as would be expected) \cite{Ziale,Oliveira}. % It is plain to see that this is quite similar what was obtained by Hehl and others to be
% the canonical energy momentum tensor (safe for a sign difference -- CHECK!--).
Once we implement it into cosmology the result is the same as if we would have a perfect fluid identification. This is due to the
Frenkel conditions and the comoving frame condition. %This would be equivalent to identifying $\sigma_{\mu\nu}$ with the symmetric part of what Hehl chooses as $\Sigma_{\mu\nu}$.
%Given that
%
%\begin{equation}
%\sigma_{00}=\epsilon \mbox{ } \mbox{ } \mbox{ , } \mbox{ } \mbox{ } \sigma_{ij} = -P\tilde{g}_{ij}
%\end{equation}
%

\begin{figure}
\centering
\begin{minipage}{.45\textwidth}
\centering
   \includegraphics[width=3.3in]{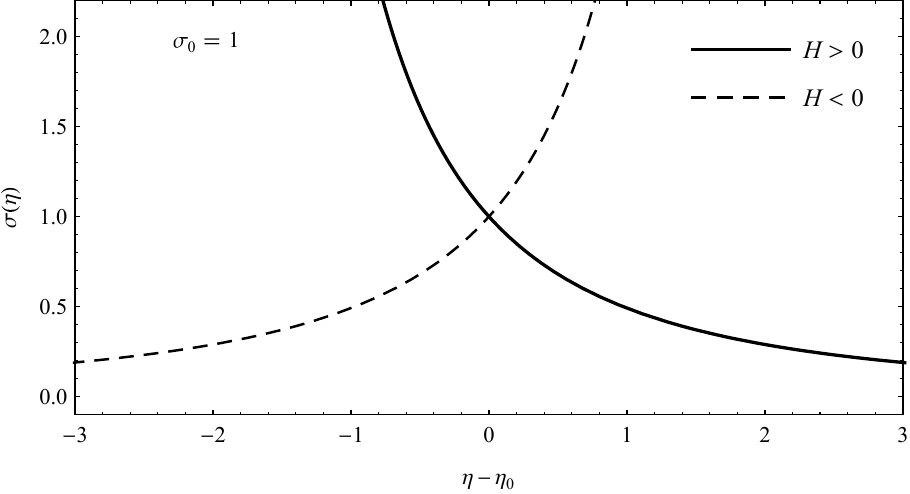} 
   \caption{Plot of $\sigma(\eta)$ as given by equation (\ref{sigmaeta1}) with both sign choices. Note that the dashed line corresponds to a contracting universe.}
   \label{fig:1SigmaEta}
\end{minipage}\hfill
\begin{minipage}{.45\textwidth}
\centering
   \includegraphics[width=3.3in]{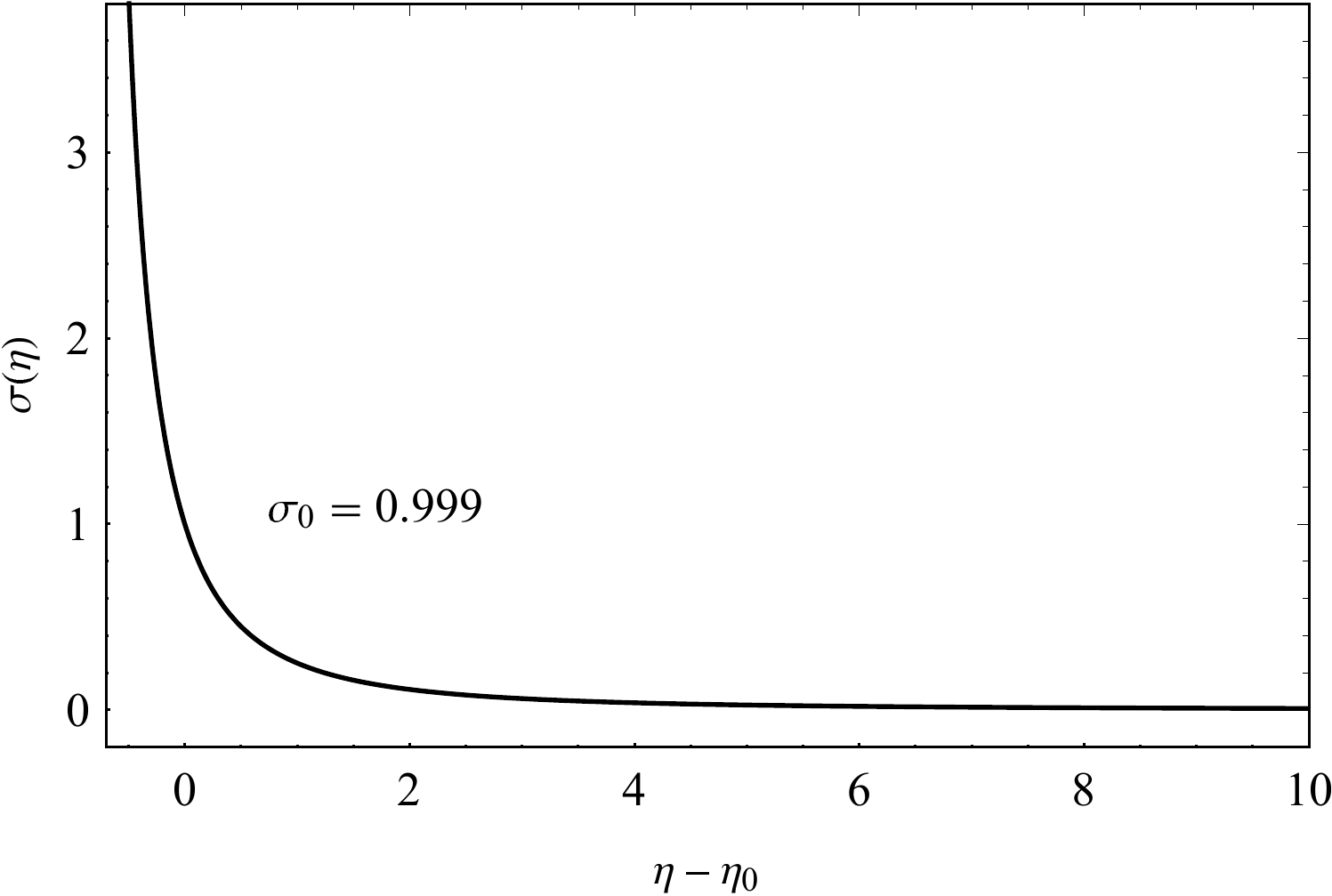} 
   \caption{Plot of $\sigma(\eta)$ as the numerical solution of equation (\ref{DESigma}) with initial value $\sigma(\eta)=0.999$.}
   \label{fig:1SigmaEtaNum}
\end{minipage}\hfill
\end{figure}

\begin{figure}[htbp] %  figure placement: here, top, bottom, or page
   \centering
	   \includegraphics[width=5in]{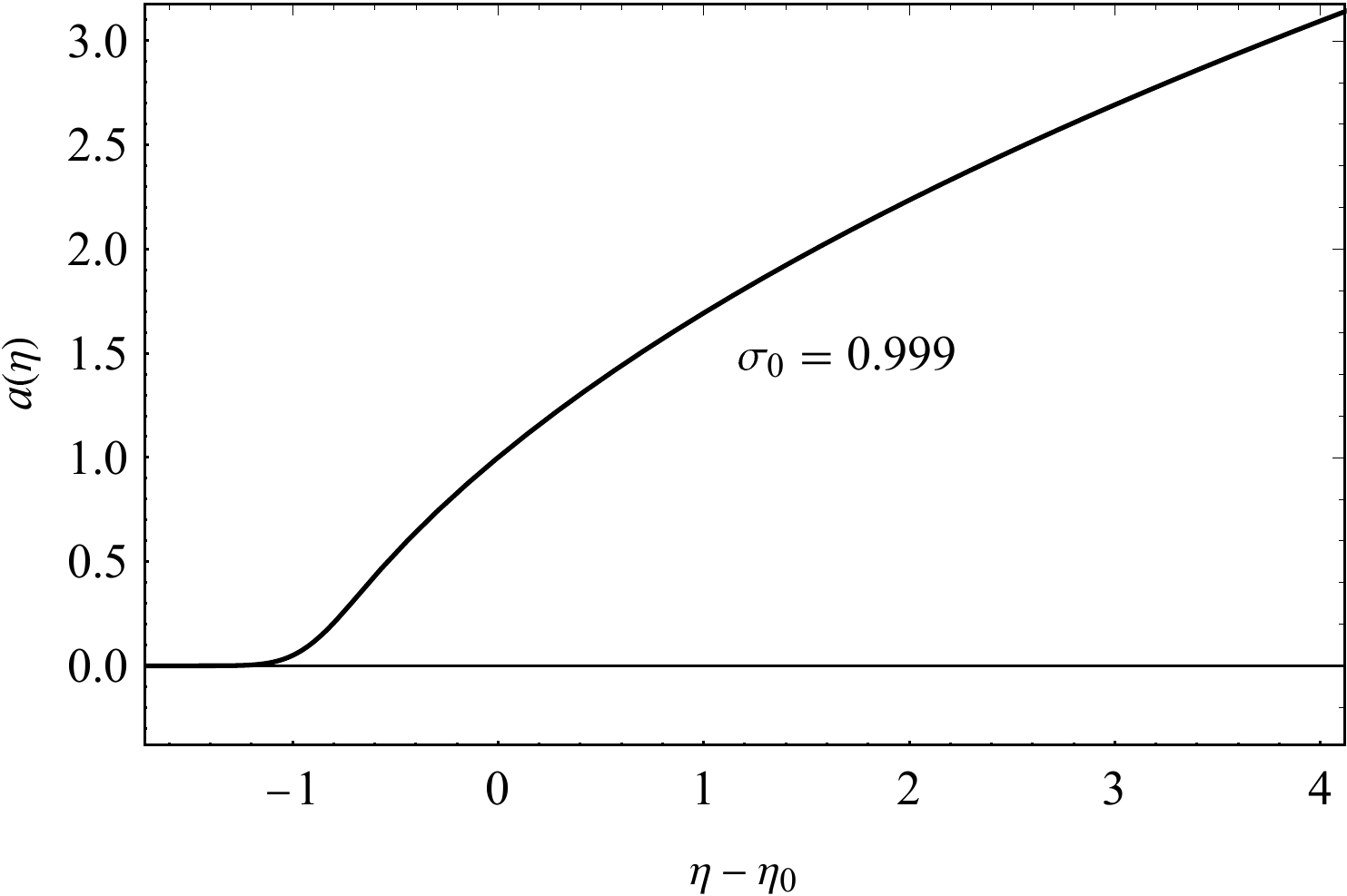} 
   \caption{Plot of the numerical solution of $a(\eta)$ from the positive sign of equation (\ref{aeta}) for an initial value $\sigma(0)=0.999$.}
   \label{fig:1AEta}
\end{figure}

Its effect upon the field equations gives an effective energy--momentum tensor of the form
\begin{equation}\label{EffectiveEM1}
\tilde{\sigma}_{\mu\nu}=\left( \rho+p+\frac{1}{2}\kappa s^{2} \right)u_{\mu}u_{\nu}-\left(p-\frac{1}{4}\kappa s^{2} \right)g_{\mu\nu}-2\kappa \langle s_{\mu\lambda}s\indices{_{\nu}^{\lambda}}\rangle,
\end{equation}
where we have used the Frenkel conditions \cite{Frenkel}, i.e. equations (\ref{Frenkel}).
%
%where $s_{\mu\nu}$ is the antisymmetric spin tensor. 
Now apart from this we have also assumed that the average over the contraction of two spin tensors is not negligible and actually takes the form we established before  $\langle s^{\mu\nu}s_{\mu\nu}\rangle = \frac{1}{2}s^{2}$. Furthermore, we will follow Gasperini's result and take
\begin{equation}
\langle s\indices{^{\lambda}_{\mu}}s\indices{_{\lambda}^{\nu}} \rangle=0,
\end{equation}
which is a result given implicitly in \cite{Gasperini} \footnote{For the interested reader we suggest checking equations (4) through (9) in \cite{Gasperini} and noting that his result indeed suggests
the value for this average.}. With these conditions the Friedmann equations are given by
\begin{equation}
H^{2}=\frac{1}{3}\kappa\left(\rho +\frac{3}{4}\kappa s^{2} \right), \mbox{ } \mbox{ } \mbox{  } \mbox{ } \mbox{ } 3H^{2}+2\dot{H}=-\kappa \left(p-\frac{1}{4}\kappa s^{2} \right).
\end{equation}
Using both of these equations we can write
\begin{equation}
\ddot{a}=-\frac{\kappa}{6}(\rho + 3p)a.
\end{equation}
%Sergio: why fermions (I will include a reference)
Under the assumption that we are dealing with fermions and that there is no polarization of spins in an ideal fluid  \cite{Nurgaliev}, we can use the relation between spin density and 
particle number density obtained in equations (\ref{spinnumber}) and (\ref{SpinDensity}). %for...). 
Now we can construct the continuity equation 
\begin{equation}
\frac{d}{dt}\left(\rho+\frac{3}{4}\kappa s^{2}\right)=-3H\left(\rho+p+\frac{1}{2}\kappa s^{2} \right),
\end{equation}
 which is obtained from the 
Friedmann equations we found above. %If we then use the proportionality of $s^{2}$ with $\rho$, we can obtain a differential equation for $\rho$, %and $a$,
%where we have chosen the equation of state for radiation $w=1/3$, since we are interested in the early universe behaviour 
Equivalently, we can write it in the following form
\begin{equation}
\frac{d\rho}{dt}\left(1+\frac{9}{8}\kappa B_{\frac{1}{3}}\rho^{\frac{1}{2}}\right)=-4H\rho\left(1+\frac{3}{8}\kappa B_{\frac{1}{3}}\rho^{\frac{1}{2}} \right).
\end{equation}
%\begin{equation}
%d\rho+ \frac{3}{2}\kappa B_{w}\frac{1}{1+w} \rho^{\frac{1-w}{1+w}}d\rho=-3\frac{da}{a}\left(\rho + P -\frac{1}{2}\kappa B_{w} \rho ^{\frac{2}{1+w}} \right)
%\end{equation}
%. From this choice we may obtain an expression for $a$ in terms of $\rho$ (in the case in which $\rho\neq\left(\frac{8}{3}\right)^{2}\frac{1}{\kappa^{2} B^{2}_{\frac{1}{3}}}$)
%Note that the right hand side of this equation goes to zero when $\rho\rightarrow\rho(t=t_{c})=\left(\frac{8}{3}\right)^{2}\frac{1}{\kappa^{2} B^{2}_{1/3}}\equiv \rho_{c}$, this suggests that $\dot{\rho}
%(t=t_{c})=0$ and so $\rho$ will have a critical value at $t_{c}$ which should be reflected in the solution. Unfortunately the standard way in which we have attempted to solve 
%this equation does not seem to reproduce this result, and we have identified the issue with our attempt. Our first approach was to take $H=\frac{1}{a}\frac{da}{dt}$ and have the 
%differential equation be written in the following way
The solution of the differential equation is
%\begin{equation}
%\frac{\left(1+\frac{9}{8}\kappa B_{\frac{1}{3}}\rho^{\frac{1}{2}}\right)}{\rho\left(1+\frac{3}{8}\kappa B_{\frac{1}{3}}\rho^{\frac{1}{2}} \right)}d\rho=-4\frac{da}{a}
%\end{equation}
%%is solved by integration, i.e, 
%\begin{equation}\label{arho}
% 2\ln\left(\sqrt{\frac{\rho}{\rho_{0}}}\right)+4\ln \left(\frac{8+3\kappa B_{\frac{1}{3}}\sqrt{\rho}}{8+3\kappa B_{\frac{1}{3}}\sqrt{\rho_{0}}} \right)=-4\ln \left(a \right)
%\end{equation}
\begin{equation}\label{PerfectRho}
a=\left(\frac{8+3\kappa B_{\frac{1}{3}}\sqrt{\rho_{0}}}{8+3\kappa B_{\frac{1}{3}}\sqrt{\rho}} \right)\left(\frac{\rho_{0}}{\rho} \right)^{\frac{1}{4}}.
\end{equation}
We can see that, while it is not exactly the standard GR case (see Appendix B), equation (\ref{PerfectRho}) shows that when the density goes to high
values $\rho\rightarrow \infty$, the scale parameter will approach zero $a\rightarrow 0$ as in GR. There is no restriction upon $\rho$ and thus we may 
interpret that we are not avoiding the singular behaviour. In order to study this system further, we will make the change of variables 
\begin{equation}
\rho_{1}^{1/2}\equiv\frac{4}{3}\frac{1}{\kappa B_{\frac{1}{3}}} ,
\quad \sigma \equiv \frac{\rho}{\rho_{1}} \mbox{ }\mbox{ } \mbox{ }\mbox{ and } \mbox{ } \mbox{ } \mbox{ } \eta\equiv \sqrt{\kappa \rho_{1}}t.
\end{equation}
With these new variables the first Friedmann equation turns out to be
\begin{equation}\label{firstfriedmanneta}
\frac{1}{a}\frac{da}{d\eta}=\pm \sqrt{\frac{1}{3}}\sqrt{\sigma(1+\sigma^{\frac{1}{2}})}.
\end{equation}
We replace this in the differential equation for $\rho$ and rewrite as
\begin{equation}\label{DESigma}
\frac{d\sigma}{d\eta}=\mp \frac{4}{\sqrt{3}}\frac{\sigma^{\frac{3}{2}}\sqrt{(1+\sigma^{\frac{1}{2}})}(1+\frac{1}{2}\sigma^{\frac{1}{2}})}{1+\frac{3}{2}\sigma^{\frac{1}{2}}}\equiv \mp f(\sigma)
\end{equation}
This equation can be solved analytically giving us 
\begin{equation}\label{sigmaeta1}
-\frac{2\sqrt{1+\sigma^{1/2}}}{\sqrt{\sigma}}-4\tan^{-1}[\sqrt{1+\sqrt{\sigma}}]-2\coth^{-1}\left[\sqrt{1+\sqrt{\sigma}}\right]+K=\mp \frac{4}{\sqrt{3}}\left(\eta-\eta_{0} \right),
\end{equation}
where $K$ is a constant given by the integration limit $\sigma_0$
\begin{equation}
K=\frac{2\sqrt{1+\sigma_{0}^{1/2}}}{\sqrt{\sigma_{0}}}+4\tan^{-1}[\sqrt{1+\sqrt{\sigma_{0}}}]+2\coth^{-1}\left[\sqrt{1+\sqrt{\sigma_{0}}}\right].
\end{equation}
We choose the sign that reproduces an expanding universe. With this in mind and looking at the plot of $\sigma(\eta)$ for this 
expanding solution (Figure \ref{fig:1SigmaEta}),
it is clear that $\sigma$ grows arbitrarily as $\eta$ approaches negative values, which reinforces the idea of the density approaching an infinite value at early times.
Equation (\ref{sigmaeta1}) is as far as we may go analytically, but we can get further insight confirming our conclusions so far by solving the equations numerically.
%\begin{figure}[htbp] %  figure placement: here, top, bottom, or page
%   \centering
%   \includegraphics[width=5in]{1SigmaEta.eps} 
%   \caption{Plot of $\sigma(\eta)$ as given by equation (\ref{sigmaeta1}) with both sign choices. Note that the dashed line corresponds to a contracting Universe.}
%   \label{fig:1SigmaEta}
%\end{figure}
With the help of MATHEMATICA we first obtain a numerical solution for the solution for $\sigma(\eta)$ that corresponds to an expanding universe, 
which we have plotted in Figure \ref{fig:1SigmaEtaNum}. 
%\begin{figure}[htbp] %  figure placement: here, top, bottom, or page
%   \centering
%   \includegraphics[width=5in]{1SigmaEtaNum.eps} 
%   \caption{Plot of $\sigma(\eta)$ as the numerical solution of equation (\ref{DESigma}) with initial value $\sigma(\eta)=0.999$.}
%   \label{fig:1SigmaEtaNum}
%\end{figure}
The behaviour of $a$ in terms of $\eta$ is also obtained numerically by taking equation (\ref{firstfriedmanneta}) and integrating
%this has been plotted for different initial values of $\sigma$, as well as for the signs present in the differential equation, and put into figures \ref{fig:sigmanegative} and \ref{fig:sigmapositive}. 
%\begin{figure}[htbp] %  figure placement: here, top, bottom, or page
%   \centering
%   \includegraphics[width=3.5in]{NegativeSigmaEta.eps} 
%   \caption{Plot of the numerical solution of $\sigma$ from the negative sign of equation (\ref{DESigma}) for different initial values $\sigma(0)$, with the distinction of the critical value $\sigma=1$. }
%   \label{fig:sigmanegative}
%\end{figure}
%\begin{figure}[htbp] %  figure placement: here, top, bottom, or page
%   \centering
%   \includegraphics[width=3.5in]{PositiveSigmaEta.eps} 
%   \caption{Plot of the numerical solution of $\sigma$ from the positive sign of equation (\ref{DESigma}) for different initial values $\sigma(0)$, with the distinction of the critical value $\sigma=1$. }
%   \label{fig:sigmapositive}
%\end{figure}
%
%which we would then use to solve for $a(t)$ since 
%we have that the first Friedmann equation is of the form
%\begin{equation}
%\frac{da}{dt}=g(\rho(t))a,
%\end{equation}
%
%Given this solution we are able to obtain the behaviour of $a$ in terms of $\eta$ by writing the first Friedmann equation as 
%\begin{equation}
%\frac{1}{a}\frac{da}{d\eta}=\pm \sqrt{\frac{1}{3}}\sqrt{\sigma(1+2\sigma^{\frac{1}{2}})},
%\end{equation}
%and integrating it numerically in the following way
\begin{equation}\label{aeta}
a=\exp\left[\pm \int_{\eta_{0}}^{\eta}d\eta' \sqrt{\frac{1}{3}}\sqrt{\sigma(1+\sigma^{\frac{1}{2}})} \right].
\end{equation}
We plot the numerical solution of $a(\eta)$ in Figure \ref{fig:1AEta}.
%\begin{figure}[htbp] %  figure placement: here, top, bottom, or page
%   \centering
%   \includegraphics[width=3.5in]{NegativeAEta2.eps} 
%   \caption{Plot of the numerical solution of $a(\eta)$ from the positive sign of equation (\ref{aeta}) for different initial values $\sigma(0)>1$, where it is clear that
%    the solution gets truncated after certain values of $\eta$. }
%   \label{fig:negativea2}
%\end{figure}
%\begin{figure}[htbp] %  figure placement: here, top, bottom, or page
%   \centering
%   \includegraphics[width=3.5in]{PositiveAEta1.eps} 
%   \caption{Plot of the numerical solution of $a(\eta)$ from the negative sign of equation (\ref{aeta}) for different initial values $\sigma(0)<1$. }
%   \label{fig:positivea1}
%\end{figure}
%\begin{figure}[htbp] %  figure placement: here, top, bottom, or page
%   \centering
%   \includegraphics[width=3.5in]{PositiveAEta2.eps} 
%   \caption{Plot of the numerical solution of $a(\eta)$ from the negative sign of equation (\ref{aeta}) for different initial values $\sigma(0)>1$, where it is clear that
%    the solution gets truncated after certain values of $\eta$. }
%   \label{fig:positivea2}
%\end{figure}
%
Note that for the expanding universe solution in this model, i.e. the one with $H>0$, as one goes to the past ($\eta<0$) $a$ reaches
zero unavoidably, suggesting it does not avoid the initial singularity. In conclusion, it is the term proportional to $\accentset{\star}{\nabla}$ which is responsible for the
nice features of the model. In the next section we will present models which do not include this term, yet display an absence of the singularity, and eventually also an accelerated behaviour
of the expansion in the early universe. To this end we will make use of different Lagrangians.

\section{Alternative Lagrangians} 
Following our aim of studying alternative treatments of space time with torsion, an important observation is that the choice of the Lagrangian in ECT is not stringent to start with the identification $\mathcal{L}\sim R[\Gamma]$, 
although it is perhaps aesthetic. Indeed in space-time with torsion we have the following independent diffeomorphic  invariants
%Let us take the following invariants of space-time with torsion
%
\begin{equation}
\mathcal{A}_{1}=\left\{\accentset{\circ}{R},S_{\lambda}S^{\lambda},S_{\alpha\beta\gamma}S^{\gamma\beta\alpha} ,S_{\alpha\beta\gamma}S^{\alpha\beta\gamma}\right\}.
\end{equation}
Therefore, a more general Lagrangian could be constructed out of a linear combination of these terms for an alternative approach to Einstein--Cartan Theory. Our final goal is to
examine the cosmology based on the following 
Lagrangian
% Using them we can construct the following first order lagrangian with arbitrary coefficients, which would account
%for an alternative treatment of Einstein-Cartan,
\begin{equation}\label{torsioninv}
\mathcal{L}_{AT}=\sqrt{-g}\left(\accentset{\circ}{R}+b_{1}S_{\lambda}S^{\lambda}+b_{2}S_{\alpha\beta\gamma}S^{\gamma\beta\alpha}+b_{3}S_{\alpha\beta\gamma}S^{\alpha\beta\gamma} \right),
\end{equation}
\begin{equation}\label{contorsioninv}
\begin{split}
\mathcal{L}_{AT}=\sqrt{-g}&\left( \accentset{\circ}{R}+\frac{b_{1}}{4}K\indices{_{\alpha\lambda}^{\alpha}}K\indices{_{\rho}^{\lambda\rho}}+\left(\frac{3b_{2}}{4}-\frac{b_{3}}{2} \right)K_{\alpha\beta\gamma}K^{\gamma\beta\alpha} \right. \\
& \left. +\left(\frac{b_{3}}{2}-\frac{b_{2}}{4} \right)K_{\alpha\beta\gamma}K^{\alpha\beta\gamma}\right),
\end{split}
\end{equation}
where the $b_{i}$ are dimensionless coefficients. Note also that we have explicitly neglected the total divergence term present in standard ECKS. The particular 
choices of  $b_{1}=-4$, $b_{2}=2$ and $b_{3}=1$ would account for the case 
explored in the preceding section. Indeed, a linear combination of invariants is similar to the starting point of Teleparallel Theory 
and particularly, the Teleparallel Equivalent of General Relativity (TEGR) in which
the effects attributed to curvature are now replaced by effects of torsion \cite{TeleparallelBook,TEGR}. This theory has been quite studied in recent years after the Hayashi construction \cite{Hayashi1979} 
of scalar fields behaving differently in TEGR which gives rise to alternative views of dark matter behaviour \cite{Otalora}. We mention also the possibilities
of $f(T)$ theories with $T$ the scalar constructed from torsion invariants  \cite{FTTheories,TomiTeleparallel} which matches the behaviour of $\accentset{\circ}{R}$ in GR.

Inspecting the form of equation (\ref{contorsioninv}) we see that a simpler version is also possible.
Before embarking on the general case, it is instructive to use first this simplified version of the above invariants using only contortion, namely
\begin{equation}
\mathcal{A}'_{1}=\left\{\accentset{\circ}{R},K\indices{_{\rho}^{\alpha\lambda}}K\indices{_{\lambda\alpha}^{\rho}},K\indices{_{\alpha}^{\lambda\alpha}}K\indices{_{\rho\lambda}^{\rho}} \right\},
\end{equation}
where it is relevant to note that this approach, while including essentially the same invariants, is a simplification of the more general case with a certain relation between the
coefficients. From these invariants $\mathcal{A'}_{1}$ we can construct the following Lagrangian with arbitrary coefficients $a_{1}$ and $a_{2}$ (this case corresponds to the 
particular choice $b_{2}=2b_{3}$ in equation (\ref{contorsioninv}) and taking $b_{1}\rightarrow - 4 a_{2}$ and $b_{3}\rightarrow a_{1}$)
%Sergio: which choice of b_1 etc leads to the simpler case?(the a_1 cases) (DONE)
%Let us take the following form of Lagrangian in Einstein-Cartan, where we have given arbitrary coefficients to the invariants constructed from contracting contorsions given
%that we shall eventually make the standard spin fluid identification,
%
\begin{equation}
\mathcal{L}_{AT2}=\sqrt{-g}\left(\accentset{\circ}{R}+a_{1}K\indices{_{\rho}^{\alpha\lambda}}K\indices{_{\lambda\alpha}^{\rho}}-a_{2}K\indices{_{\alpha}^{\lambda\alpha}}K\indices{_{\rho\lambda}^{\rho}}\right).
\end{equation}
The Lagrangian  is now $\mathcal{L}_{AT2}=\sqrt{-g}L_{AT2}$ with the action
\begin{equation}
S_{AT}=\frac{1}{2\kappa}\int d^{4}x \sqrt{-g} L_{AT}+\int d^{4}x \mathcal{L}_{m},
\end{equation}
where $\mathcal{L}_{m}$ is the matter Lagrangian. We first vary the action with respect to the metric to obtain
the field equations which may be written as
\begin{equation}
\begin{split}
\accentset{\circ}{R}_{\mu\nu}-\frac{1}{2}g_{\mu\nu}\accentset{\circ}{R} = & \kappa \sigma_{\mu\nu} +\frac{1}{2}g_{\mu\nu}\left[a_{1}g^{\alpha\beta}K\indices{_{\rho\beta}^{\lambda}}K\indices{_{\lambda\alpha}^{\rho}}-a_{2}g^{\alpha\beta}K\indices{_{\lambda\beta}^{\lambda}}K\indices{_{\rho\alpha}^{\rho}} \right] \\ & -a_{1}K\indices{_{\lambda\mu}^{\rho}}K\indices{_{\rho\nu}^{\lambda}}+a_{2}K\indices{_{\alpha\nu}^{\alpha}}K\indices{_{\rho\mu}^{\rho}}.
\end{split}
\end{equation}
The algebraic relation which emerges upon variation of the contortion
can be put in the following form
\begin{equation}
a_{1}\left(K\indices{^{\nu}_{\alpha}^{\mu}}+K\indices{_{\alpha}^{\mu\nu}}\right)-a_{2}\left(\delta^{\mu}_{\alpha}K\indices{_{\lambda}^{\lambda\nu}}+g^{\mu\nu}K\indices{_{\rho\alpha}^{\rho}}\right)=2\kappa \tau\indices{_{\alpha}^{\nu\mu}},
\end{equation}
\begin{equation}
a_{1}S\indices{_{\alpha\nu}^{\mu}}+a_{2}\delta_{\alpha}^{\mu}S\indices{_{\nu\lambda}^{\lambda}}-a_{2}\delta_{\nu}^{\mu}S\indices{_{\alpha\rho}^{\rho}}=\kappa \tau\indices{_{\alpha\nu}^{\mu}},
\end{equation}
We recover the ECKS standard case when taking the proper values for the coefficients $a_{i}$, namely $a_{1}=a_{2}=1$. By taking the trace of the spin energy-momentum tensor in the 
equation above we get
\begin{equation}
\kappa \tau\indices{_{\alpha\lambda}^{\lambda}}=(a_{1}-3a_{2})S\indices{_{\alpha\lambda}^{\lambda}},
\end{equation}
which means that we can write the torsion tensor in terms of its source $\tau\indices{_{\mu\nu}^{\alpha}}$ as
\begin{equation}
S\indices{_{\alpha\nu}^{\mu}}=\frac{\kappa}{a_{1}}\left\{ \tau\indices{_{\alpha\nu}^{\mu}}-\left(\frac{a_{2}}{a_{1}-3a_{2}} \right)\delta_{\alpha}^{\mu}\tau\indices{_{\nu\lambda}^{\lambda}}+\left(\frac{a_{2}}{a_{1}-3a_{2}} \right)\delta_{\nu}^{\mu}\tau\indices{_{\alpha\rho}^{\rho}} \right\},
\end{equation}
and similarly, the contortion in terms of $\tau\indices{_{\mu\nu}^{\alpha}}$
\begin{equation}
K\indices{_{\mu\nu}^{\alpha}}=\frac{\kappa}{a_{1}}\left\{-\tau\indices{_{\mu\nu}^{\alpha}}+\tau\indices{_{\nu}^{\alpha}_{\mu}}-\tau\indices{^{\alpha}_{\mu\nu}}+2\left(\frac{a_{2}}{a_{1}-3a_{2}} \right)\delta_{\mu}^{\alpha}\tau\indices{_{\nu\lambda}^{\lambda}}-2\left( \frac{a_{2}}{a_{1}-3a_{2}}\right)g_{\mu\nu}\tau\indices{^{\alpha\rho}_{\rho}} \right\}.
\end{equation}
The last equations show that we can solve the algebraic equation in terms of the source.
With this at hand we can write the full field equations coming from the variation with respect to the metric in terms of the spin energy-momentum tensor as
\begin{equation}
\begin{split}
\accentset{\circ}{R}_{\mu\nu}-\frac{1}{2}g_{\mu\nu}\accentset{\circ}{R}= & \kappa \sigma_{\mu\nu}+\frac{1}{2}g_{\mu\nu}\left[  \frac{\kappa^{2}}{a_{1}} \left\{ 4\left(\frac{a_{2}}{a_{1}-3a_{2}} \right)\tau\indices{^{\alpha\beta}_{\beta}}\tau\indices{_{\alpha\sigma}^{\sigma}}\left(3\left(\frac{a_{2}}{a_{1}-3a_{2}} \right)+2 \right)\right. \right.
\\ &\left. \left. +\tau_{\alpha\beta\gamma}\tau^{\alpha\beta\gamma}-2\tau^{\alpha\beta\gamma}\tau_{\gamma\alpha\beta} \right\}
-\frac{\kappa^{2}}{a_{1}^{2}}a_{2}\left\{2+6\left(\frac{a_{2}}{a_{1}-3a_{2}} \right)^{2} \right\}\tau\indices{^{\alpha\lambda}_{\lambda}}\tau\indices{_{\alpha\rho}^{\rho}}\right]\\
& - \frac{\kappa^{2}}{a_{1}}\left\{ 2\tau_{\mu\lambda\rho}\tau\indices{_{\nu}^{\rho\lambda}}+2\tau_{\mu\lambda\rho}\tau\indices{_{\nu}^{\lambda\rho}}-\tau\indices{^{\rho\lambda}_{\mu}}\tau_{\rho\lambda\nu}\right. \\
& \left.+ 2\left( \frac{a_{1}}{a_{1}-3a_{2}}\right)\left[4\tau\indices{_{\mu\lambda}^{\lambda}}\tau\indices{_{\nu\beta}^{\beta}}+6\tau\indices{_{\mu\alpha}^{\alpha}}\tau\indices{_{\nu\beta}^{\beta}}\left(\frac{a_{2}}{a_{1}-3a_{2}} \right) \right] \right\}\\
&+\frac{\kappa^{2}}{a_{1}^{2}}a_{2}\left\{ 4\tau\indices{_{\alpha\nu}^{\alpha}}\tau\indices{_{\beta\mu}^{\beta}}+24\left(\frac{a_{2}}{a_{1}-3a_{2}} \right)\tau\indices{_{\alpha\nu}^{\alpha}}\tau\indices{_{\beta\mu}^{\beta}}+36\left(\frac{a_{2}}{a_{1}-3a_{2}} \right)^{2}\tau\indices{_{\alpha\nu}^{\alpha}}\tau\indices{_{\beta\mu}^{\beta}}  \right\}.
\end{split}
\end{equation}
The Frenkel conditions (\ref{Frenkel}) lead to a major simplification of these equations and one obtains
\begin{equation}
\accentset{\circ}{G}_{\mu\nu}=\kappa \sigma_{\mu\nu}+\frac{\kappa^{2}}{4a_{1}}s^{2}g_{\mu\nu}+\frac{\kappa^{2}}{2a_{1}}s^{2}u_{\mu}u_{\nu}.
\end{equation}

\subsection{Cosmology in the Alternative Lagrangian}
\begin{figure}[htbp] %  figure placement: here, top, bottom, or page
   \centering
   \includegraphics[width=3.5in]{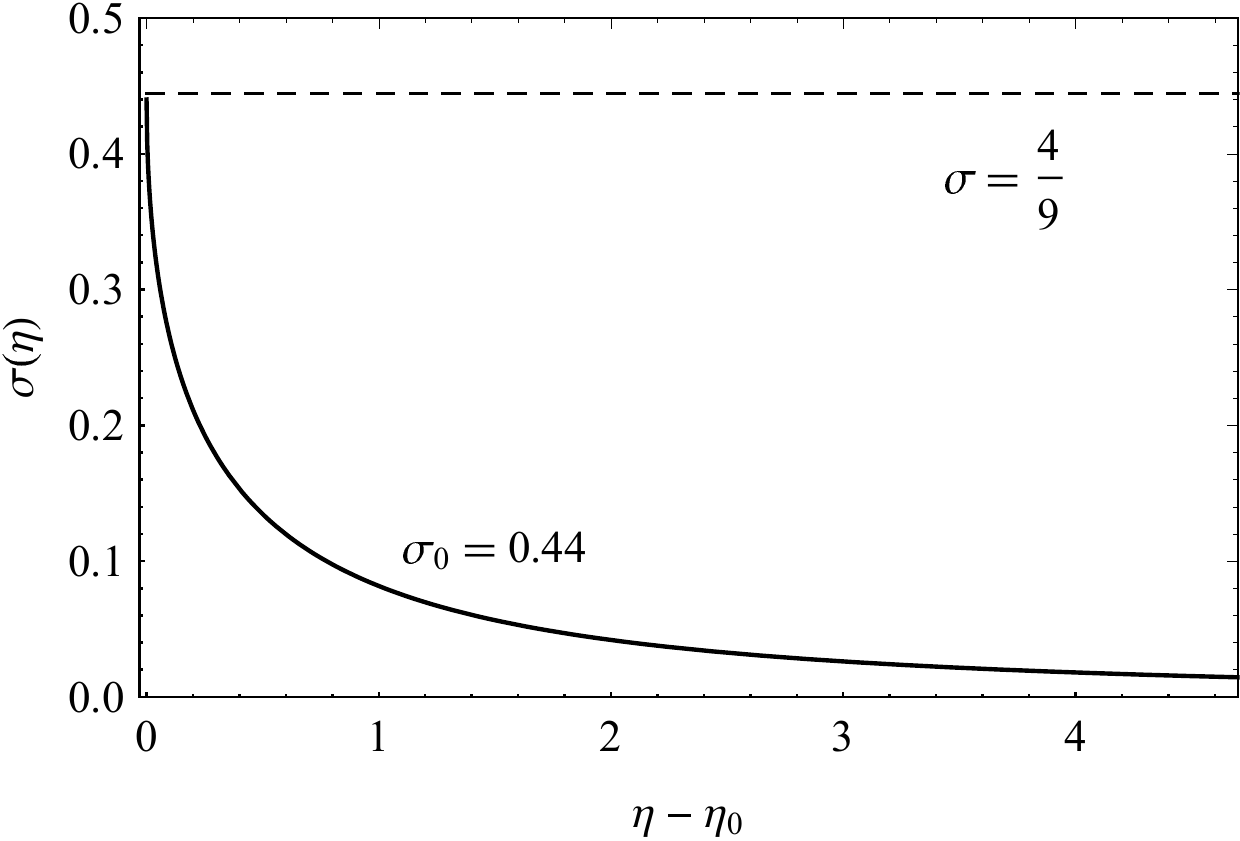} 
   \caption{Numerical solution of equation (\ref{DEqSigma}) with the negative sign and initial value $\tilde{\sigma}(\eta=0)=0.44$. We have drawn the maximum allowed density $\tilde{\sigma}=\frac{4}{9}$ as well.}
   \label{fig:sigma1}
\end{figure}

\begin{figure}
\centering
\begin{minipage}{.45\textwidth}
\centering
   \includegraphics[width=3.2in]{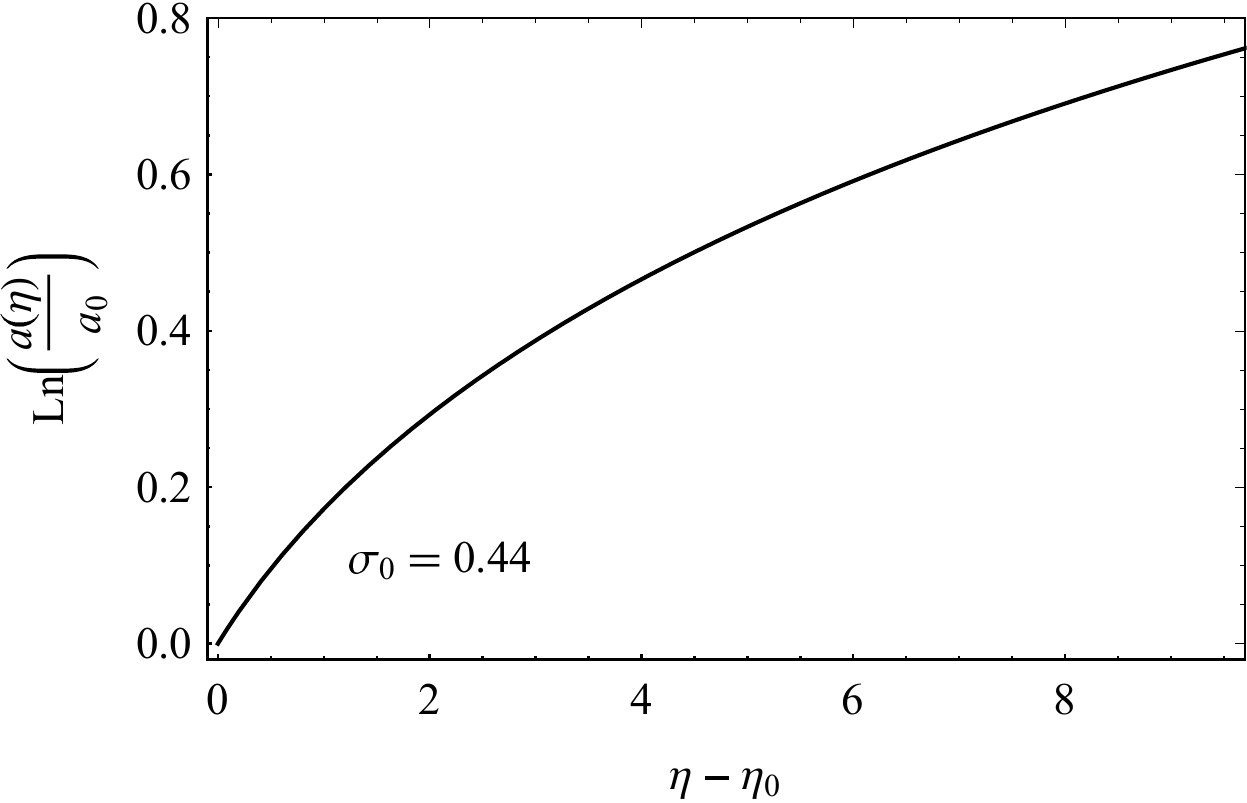} 
   \caption{Plot of the numerical solution to equation (\ref{aSigma}) with initial value $\tilde{\sigma}(\tilde{\eta}=0)=0.44$. We show both solutions, with the growing solution the one corresponding to the positive sign.}
   \label{fig:sigma2}
\end{minipage}\hfill
\begin{minipage}{.45\textwidth}
\centering
   \includegraphics[width=3.2in]{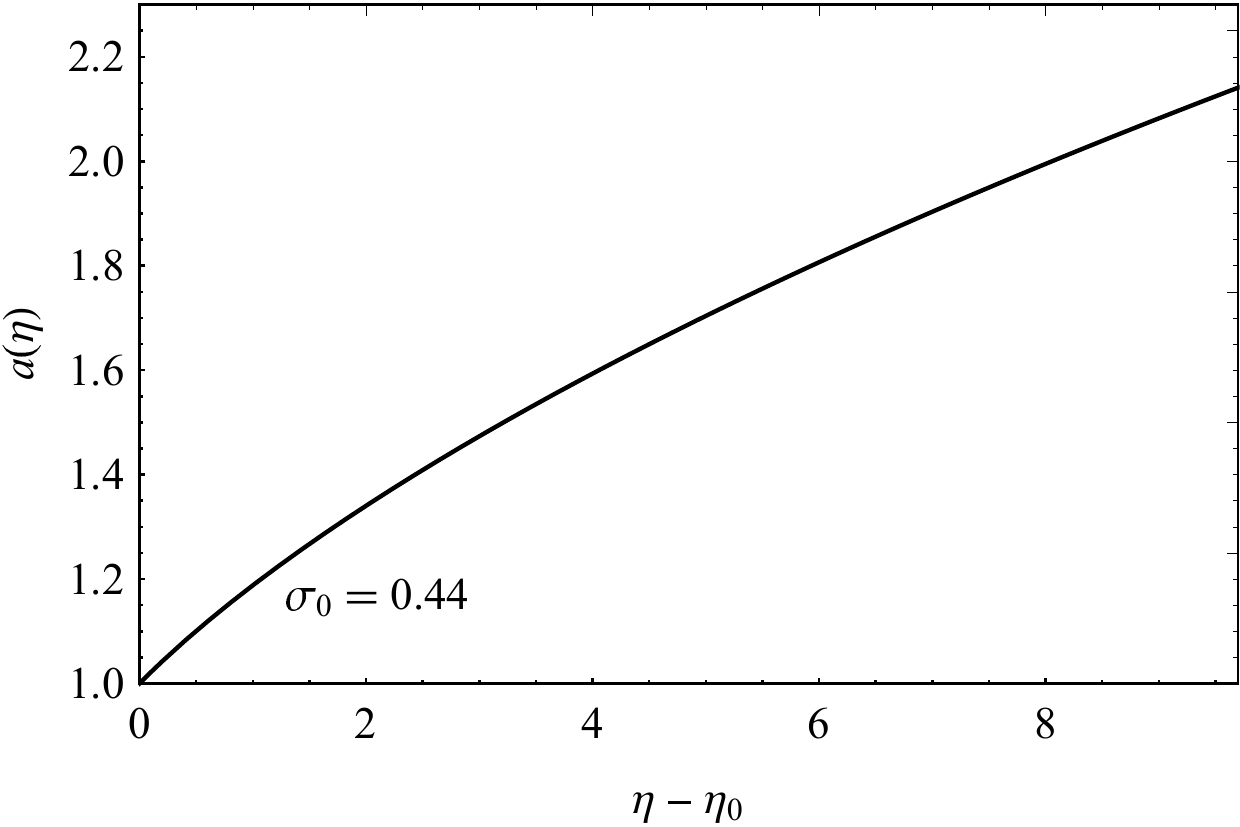} 
   \caption{Plot of the numerical solution to equation (\ref{aSigma}) with initial value $\sigma(\eta=0)=0.44$.}
   \label{fig:sigma3}
\end{minipage}\hfill
\end{figure}

%As we have seen before the choice of the Energy Momentum tensor might influence the results, so we will study three different choices of energy momentum tensor, starting
%from the standard perfect fluid to the possible fluids with spin.
The standard choice of General Relativity is taking the standard hydrodynamic perfect fluid energy--momentum tensor to which a spin-dependent Weyssenhoff
term is added.
We must note that this further term will, however, 
not contribute to the cosmological equations because of the Frenkel conditions. For a perfect fluid with spin we have the following energy--momentum tensor
\begin{equation}
\sigma_{\mu\nu}=\left(\rho+p \right)u_{\mu}u_{\nu}-pg_{\mu\nu}+2u^{\alpha}u_{(\mu}\nabla_{\beta}(s_{\alpha|\nu)}u^{\beta}).
\end{equation}
Setting this into the field equation and using both Frenkel and comoving frame conditions yields
\begin{equation}
\accentset{\circ}{G}_{\mu\nu}=\kappa\left[ \left(\rho+p+\frac{\kappa}{2a_{1}}s^{2} \right)u_{\mu}u_{\nu}-\left(p-\frac{\kappa}{4a_{1}}s^{2} \right)g_{\mu\nu} \right].
\end{equation}
As usual the Friedmann equations will be obtained by taking the diagonal components of the field equations. From the time--time components we get
\begin{equation}
H^{2}=\frac{\kappa}{3}\left(\rho+\frac{3}{4}\frac{\kappa}{a_{1}}s^{2} \right),
\end{equation}
while the space--space diagonal terms give
\begin{equation}
3H^{2}+2\dot{H}=-\kappa \left(p-\frac{\kappa}{4a_{1}}s^{2} \right) \rightarrow H^{2}+\dot{H}=-\frac{\kappa}{6}\left(\rho+3p \right)=\frac{\ddot{a}}{a}.
\end{equation}
From the last equation it is clear that we cannot have an accelerated stage in this model.
Both Friedmann equations result in the continuity law of the form
\begin{equation}
\left(p-\frac{\kappa}{4a_{1}}s^{2} \right)\frac{d}{dt}(a^{3})+\frac{d}{dt}\left[\left(\rho+\frac{3}{4}\frac{\kappa}{a_{1}}s^{2} \right)a^{3} \right]=0.
\end{equation}
Using the relation $s^{2}\propto \rho^{\frac{2}{1+w}}$ for the particular case of radiation given that we are interested in the early universe behaviour 
we take $w=1/3$ and rewrite the continuity equation as
\begin{equation}
\frac{d}{dt}\left(\rho + \frac{3}{4a_{1}}\kappa B_{\frac{1}{3}}\rho^{\frac{3}{2}} \right)=-3H\left(\rho+p+\frac{1}{2a_{1}} \kappa B_{\frac{1}{3}}\rho^{\frac{3}{2}} \right).
\end{equation}
A quick examination of the Friedmann equations and the continuity equation suggests that a positive value for $a_{1}$ will just be a slight generalization upon
the solutions obtained at the end of section IV 
with a singularity present at the beginning of the universe.
But, if we consider the case in which $a_{1}<0$, we see that this choice will have an essential impact on the Friedmann equations. Moreover, it will indeed define
a particular maximum density given the form of the first Friedmann equation. We already mentioned that the possibility of the Hubble parameter becoming
zero at some time can eventually lead to a model of a bouncing universe. With a negative $a_1$ we would have $H^2 \propto (\rho -\rho_{max})$,
but the model exhibits a more complicated hierarchy of three different critical densities which prevent 
the situation in which $\rho=\rho_{max}$. To see that, we start by expressing the scale parameter through the density by integrating the continuity equation
\begin{equation}
\int_{\rho_{0}}^{\rho}\left(\frac{1+\frac{9}{8 a_{1}}\kappa B \rho^{\frac{1}{2}}}{\frac{4}{3}\rho+\frac{1}{2a_{1}}\kappa B \rho^{\frac{3}{2}}} \right)=-3\int_{a_{0}=1}^{a}\frac{da'}{a'}
\end{equation}
leading to
\begin{equation}
\log\left[\frac{8a_{1}+3\kappa B \sqrt{\rho}}{8a_{1}+3\kappa B \sqrt{\rho_{0}}} \right]+\frac{1}{4}\log \left[\frac{\rho}{\rho_{0}} \right]=-\log \left(a \right).
\end{equation}
As a result 
$a$ can be rewritten in terms of $\rho$ as
\begin{equation} \label{yyy}
a=\left(\frac{8a_{1}+3\kappa B\sqrt{\rho_{0}}}{8a_{1}\rho^{\frac{1}{4}}+3\kappa B \rho^{\frac{3}{4}}} \right)\rho_{0}^{\frac{1}{4}}.
\end{equation}
To obtain $\rho$ in terms of $a$ would imply solving a fourth order equation which will not give any useful insight into the Friedman equations.
We will therefore not pursue this path to solve the Friedmann equations. 
Instead, we will use the continuity equation of the form
\begin{equation} \label{ccc}
\frac{\dot{a}}{a}=-\left(\frac{2a_{1}\rho^{-\frac{3}{4}}+\frac{9}{4}\kappa B \rho^{-\frac{1}{4}}}{8a_{1}\rho^{\frac{1}{4}}+3\kappa B \rho^{\frac{3}{4}}} \right)\frac{d\rho}{dt}.
\end{equation}
Replacing therein, the first Friedman equation
\begin{equation}
H=\pm \sqrt{\frac{\kappa}{3}}\sqrt{\rho+\frac{3}{4a_{1}}\kappa B \rho^{\frac{3}{2}}}
\end{equation}
it is possible to arrive at a relation between $\rho$ and $t$ from
\begin{equation}
-\left[\frac{a_{1}(8a_{1}+9\kappa B \rho^{\frac{5}{4}})}{(4a_{1}+3\kappa B \sqrt{\rho})(8a_{1}+3\kappa B\sqrt{\rho})\rho^{2}} \right]d\rho =\pm \sqrt{\frac{\kappa}{3}}dt
\end{equation}
The scale factor can be obtained either by integration of (\ref{ccc}) and by replacing $\rho$ into (\ref{yyy}). Before embarking on this programme it is mandatory to probe into critical densities of this model.
We recall that under the condition 
$a_{1}<0$ the first Friedmann equation acquires a form reminiscent of bouncing cosmologies, i.e.
\begin{equation}
H^{2}=\frac{\kappa}{3}\left(\rho-\frac{3}{4|a_{1}|}\kappa B \rho^{\frac{3}{2}} \right),
\end{equation}
and obviously suggests a possible critical density. Given that $\rho>0$ we have
\begin{equation}
\rho^{\frac{1}{2}}<\frac{4|a_{1}|}{3\kappa B}\equiv \rho_{max}^{\frac{1}{2}}.
\end{equation}
%which may be written in terms of the Planck density $\rho_{pl}$ in the following way
%\begin{equation}
%\rho_{c}=\frac{16}{9\pi^{2}}a_{1}\frac{\zeta^{3}}{\eta^{4}}\rho_{pl}
%\end{equation}
%with $\zeta$ and $\eta$ dimensionless parameters given by
%\begin{eqnarray}
%\zeta&\equiv& \frac{\epsilon_{0}}{\rho_{pl}} \\
%\eta&\equiv& n_{0}(l_{pl})^{3}
%\end{eqnarray}
%\begin{equation}
%\rho_{c}=\left(\frac{16a_{1}^{2}}{9\pi^{2}m_{pl}^{3}}\frac{\epsilon_{0}^{3}}{n_{0}^{4}}\right)\rho_{pl},
%\end{equation}
%where $m_{pl}$ is the Planck mass.
It is interesting to see the value of the scale factor $a$ corresponding to this $\rho_{max}$ 
\begin{equation}
a=\frac{\rho_{0}^{1/4}\sqrt{\kappa B}\sqrt{3}}{24}\left(\frac{8a_{1}+3\kappa B \sqrt{\rho_{0}}}{a_{1}^{\frac{3}{2}}} \right),
\end{equation}
where the appearance of complex value of $a_1^{3/2}$ shows that this is not possible. 
Indeed, the model has three different critical densities out of which one is the $\rho_{max}$ defined above.
For a negative $a_1$ the continuity equation reads
\begin{equation}
\dot{\rho}\left(1-\frac{9}{8|a_{1}|}\kappa B_{\frac{1}{3}}\rho^{\frac{1}{2}} \right)=-4H\rho\left(1-\frac{3}{8}\frac{\kappa B_{\frac{1}{3}}}{|a_{1}|}\rho^{\frac{1}{2}} \right)
\end{equation}
or equivalently,
\begin{equation}\label{CEqa1}
\dot{\rho}\left( 1-\frac{\rho^{1/2}}{\rho_{c_{1}}^{1/2}}\right)=-4H\rho\left(1-\frac{\rho^{1/2}}{\rho_{c_{2}}^{1/2}} \right),
\end{equation}
where
\begin{equation}
\rho_{c_{1}}^{1/2}\equiv \frac{8|a_{1}|}{9}\frac{1}{\kappa B_{\frac{1}{3}}},\quad
\rho_{c_{2}}^{1/2}\equiv \frac{8|a_{1}|}{3}\frac{1}{\kappa B_{\frac{1}{3}}}.
\end{equation}
The hierarchy of the three critical densities is
\begin{equation}
\rho_{c_{2}}>\rho_{max}>\rho_{c_{1}}.
\end{equation}
We see that only $\rho(t_{c_{1}})=\rho_{c1}$ is physically allowed since the inequality $\rho < \rho_{max}$ must be satisfied to avoid complex values of the
Hubble parameter. When this happens, i.e., $\rho$ hits $\rho_{c1}$, the r.h.s. of the continuity equation vanishes, and this
suggests that either $\rho=0$ or $H=0$ or $\rho=\rho_{c_{2}}$.
Indeed, $\rho=\rho_{c2}$ and $\rho=0$ are not possible whereas $H=0$ would imply $\rho=\rho_{max}$ which leads also
to a contradiction.   
Mathematically, we can have $\rho < \rho_{c1} < \rho_{max}$ or $\rho_{max} > \rho > \rho_{c1}$. We will see that a physically viable
expanding universe must obey the first chain of inequalities thus making a bouncing universe impossible. This, however, does not exclude a scenario where the universe is born at a non-zero scale factor.  
With this in mind we write the continuity equation (\ref{CEqa1}) in a dimensional-less form by introducing the following variables 
\begin{equation}
\tilde{\sigma} \equiv \frac{\rho}{\rho_{max}}=\frac{9\kappa^{2}B_{\frac{1}{3}}^{2}}{16a_{1}^{2}}\rho,\quad
\tilde{\eta} \equiv \sqrt{\kappa \rho_{max}}t=\frac{4|a_{1}|}{3B_{\frac{1}{3}}}\sqrt{\kappa} t
\end{equation}
The continuity equation takes form
\begin{equation}\label{sigmaeta-a1}
\left(1-\frac{3}{2}\tilde{\sigma}^{\frac{1}{2}} \right)\frac{d\tilde{\sigma}}{d\tilde{\eta}}=-4\frac{1}{a}\tilde{\sigma}\left(1-\frac{1}{2}\tilde{\sigma}^{\frac{1}{2}} \right)\frac{da}{d\eta}
\end{equation}
where the strongest bound on $\tilde{\sigma}$ is $\tilde{\sigma}<1$. The physical solution will reveal a stronger bound, namely
\begin{equation}
\tilde{\sigma}<\frac{4}{9}.
\end{equation}
The unphysical bound involving $\rho_{c_{2}}$ corresponds to $\tilde{\sigma}=4$. Next, we will make use of the first Friedmann equation in this approach
$
H^{2}=\left(\frac{da}{dt}\frac{1}{a}\right)^{2}=\frac{\kappa}{3}\left(\rho-\frac{3}{4|a_{1}|}\kappa B_{\frac{1}{3}}\rho^{\frac{3}{2}} \right),
$
and rewrite it in the form
\begin{equation}
\frac{da}{d\tilde{\eta}}\frac{1}{a}=\pm \sqrt{\frac{1}{3}}\sqrt{\tilde{\sigma}(1-\tilde{\sigma}^{\frac{1}{2}})}.
\end{equation}
Replacing this in the continuity equation we obtain a differential equation for $\sigma(\eta)$, namely,
\begin{equation}\label{DEqSigma}
\frac{d\tilde{\sigma}}{d\tilde{\eta}}\left(1-\frac{3}{2}\tilde{\sigma}^{1/2} \right)=\mp 4\sqrt{\frac{1}{3}}\tilde{\sigma}^{3/2}\left(1-\tilde{\sigma}^{1/2} \right)^{1/2}\left(1-\frac{1}{2}\tilde{\sigma}^{1/2} \right),
\end{equation}
where all three critical densities are manifest.
We have solved this equation numerically and plotted the results for a particular 
initial value of $\tilde{\sigma}$ in Figure \ref{fig:sigma1}.%....... [INCLUDE PLOTS]
The solution has an initial value slightly smaller than $\rho_{c1}$ corresponding to $\tilde{\sigma}=4/9$. It is evident that $\tilde{\sigma}$ does not exceed the critical value $4/9$.
We will come to this point below. 
 In these plots we have also included the critical values mentioned earlier, where we note that $\tilde{\sigma}=\frac{4}{9}$ is definitely a forbidden value, and we have that either
$\tilde{\sigma}<\frac{4}{9}$ or $\frac{4}{9}<\tilde{\sigma}<1$, which evidently truncates the solutions obtained.
 We may also take this solution for $\tilde{\sigma}(\tilde{\eta})$ and use it in the first Friedmann equation.
We find a solution for $a(\tilde{\eta})$ given by
%\begin{equation}\label{LogaSigma}
%\int_{a_{0}=1}^{a(t)} \frac{da'}{a'}=\pm \int_{\tilde{\eta}_{0}}^{\tilde{\eta}}d\eta'\sqrt{\frac{1}{3}}\sqrt{\tilde{\sigma}(\eta')(1-\tilde{\sigma}^{\frac{1}{2}}(\eta'))}
%\end{equation}
\begin{equation}\label{aSigma}
a(\eta)=\exp\left\{\pm \int_{\tilde{\eta}_{0}}^{\tilde{\eta}}d\eta' \sqrt{\frac{\tilde{\sigma}(\eta')}{3}\left(1-\tilde{\sigma}^{1/2}(\eta') \right)} \right\}
\end{equation}
which we solve numerically. We plot these numerical solutions in Figures \ref{fig:sigma2} and \ref{fig:sigma3}.
%\begin{figure}[htbp] %  figure placement: here, top, bottom, or page
%   \centering
%   \includegraphics[width=3.5in]{LogaPlot1.eps} 
%   \caption{Plot of the numerical solution to equation (\ref{LogaSigma}) with initial value $\tilde{\sigma}(\tilde{\eta}=0)=0.44$. We show both solutions, with the growing solution the one corresponding to the positive sign.}
%   \label{fig:sigma2}
%\end{figure}
%\begin{figure}[htbp] %  figure placement: here, top, bottom, or page
%   \centering
%   \includegraphics[width=3.5in]{aPlot1.eps} 
%   \caption{Plot of the numerical solution to equation (\ref{aSigma}) with initial value $\sigma(\eta=0)=0.44$ and $a_{0}=1$.}
%   \label{fig:sigma3}
%\end{figure}
Alternatively, if we keep in mind that $\tilde{\sigma}\leq 1$, we obtain using equation (\ref{sigmaeta-a1})  
\begin{equation}\label{asigmatilde}
a=\left(\frac{\tilde{\sigma}_{0}}{\tilde{\sigma}} \right)^{\frac{1}{4}}\left(\frac{\sqrt{\tilde{\sigma}_{0}}-2}{\sqrt{\tilde{\sigma}}-2} \right)
\end{equation}
Computing $\dot{a}/a$ from equation (\ref{asigmatilde}) and
%\begin{equation}
%\frac{\dot{a}}{a}=-\frac{1}{2}\left(\frac{\frac{3}{2}\tilde{\sigma}^{\frac{1}{2}}-1}{\tilde{\sigma}^{\frac{3}{2}}-2\tilde{\sigma}} \right)\dot{\tilde{\sigma}}
%\end{equation}
changing time in terms of $\eta$, it is possible to get an implicit expression for  $\tilde{\rho}$
which when integrated gives
\begin{equation}\label{SigmaEta}
\mp \sqrt{\frac{1}{3}}(\eta-\eta_{0})=\frac{1}{2}\left[-\frac{\sqrt{1-(\sigma')^{1/2}}}{\sqrt{\sigma'}} +2\tan^{-1}(\sqrt{1-(\sigma')^{1/2}})+\tanh^{-1}(\sqrt{1-(\sigma')^{1/2}})\right] \Bigg|_{\tilde{\sigma}_{0}}^{\tilde{\sigma}}.
\end{equation}

\begin{figure}
\centering
\begin{minipage}{.45\textwidth}
\centering
   \centering
   \includegraphics[width=3.5in]{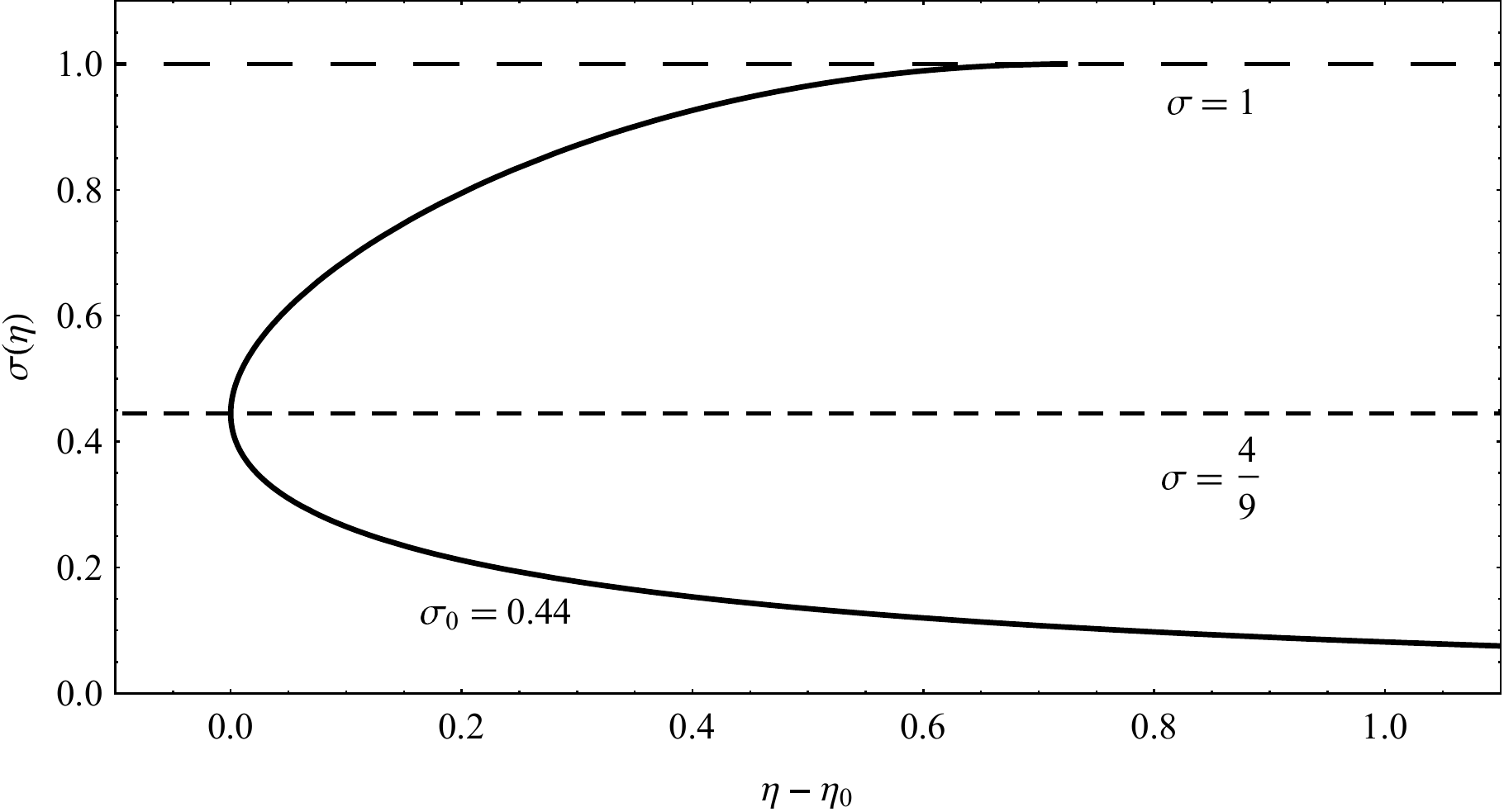} 
   \caption{Plot of the upper sign of equation (\ref{SigmaEta}) which corresponds to an expanding Universe and suggests the behaviour of $\tilde{\sigma}$ in terms of $\eta$. We have plotted the critical values $\tilde{\sigma}=1$ and $\tilde{\sigma}=\frac{4}{9} > 0.44$. Consistently we have to take the lower branch of the curve.}
   \label{fig:sigmaeta1}
\end{minipage}\hfill
\begin{minipage}{.45\textwidth}
   \centering
   \includegraphics[width=3.5in]{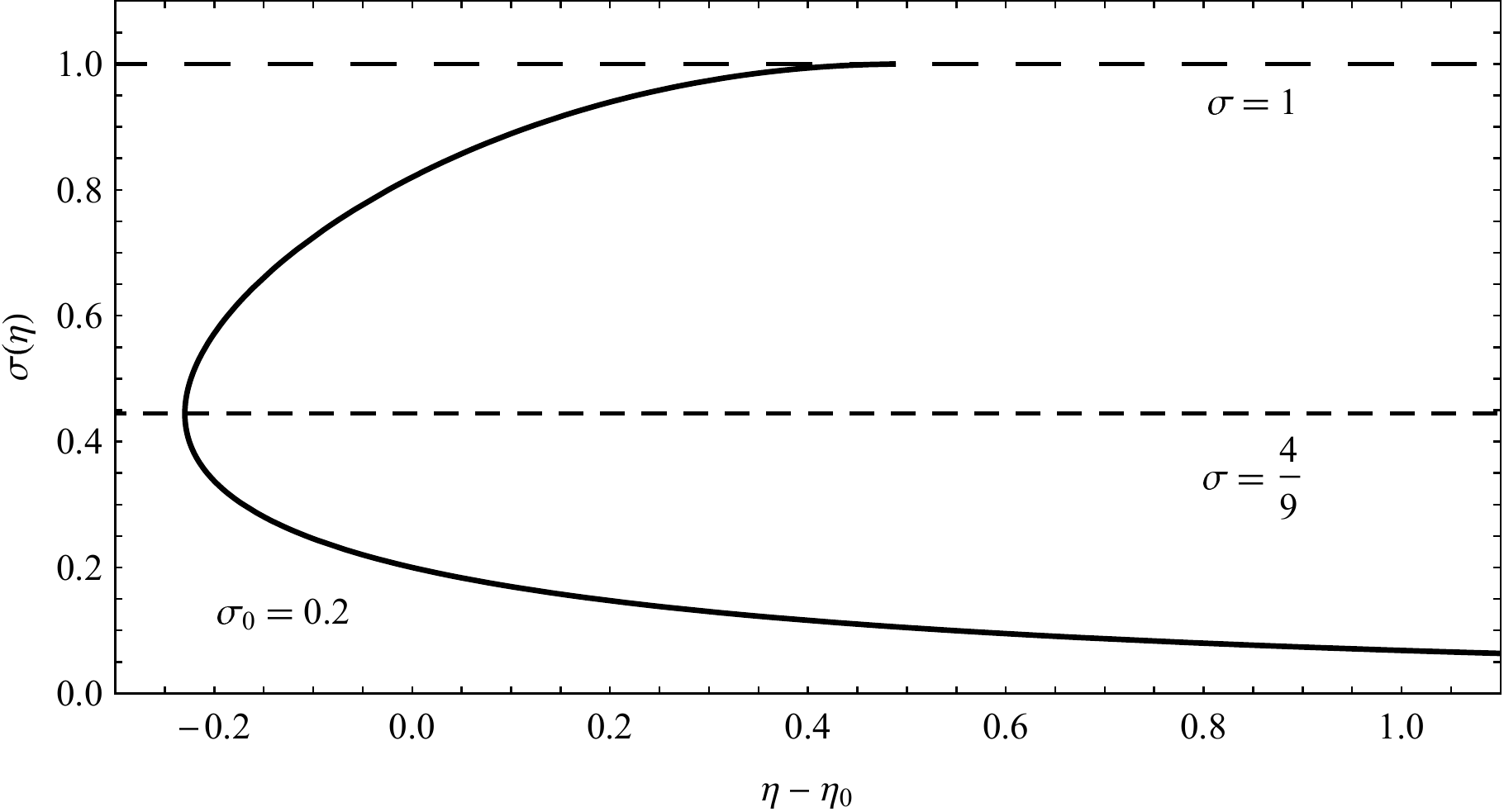} 
   \caption{Same as Figure \ref{fig:sigmaeta1} with a different initial value $\tilde{\sigma}_{0}<4/9$.}
   \label{fig:sigmaeta2}
\end{minipage}\hfill
\end{figure}

\begin{figure}[htbp] %  figure placement: here, top, bottom, or page
   \centering
   \includegraphics[width=3.5in]{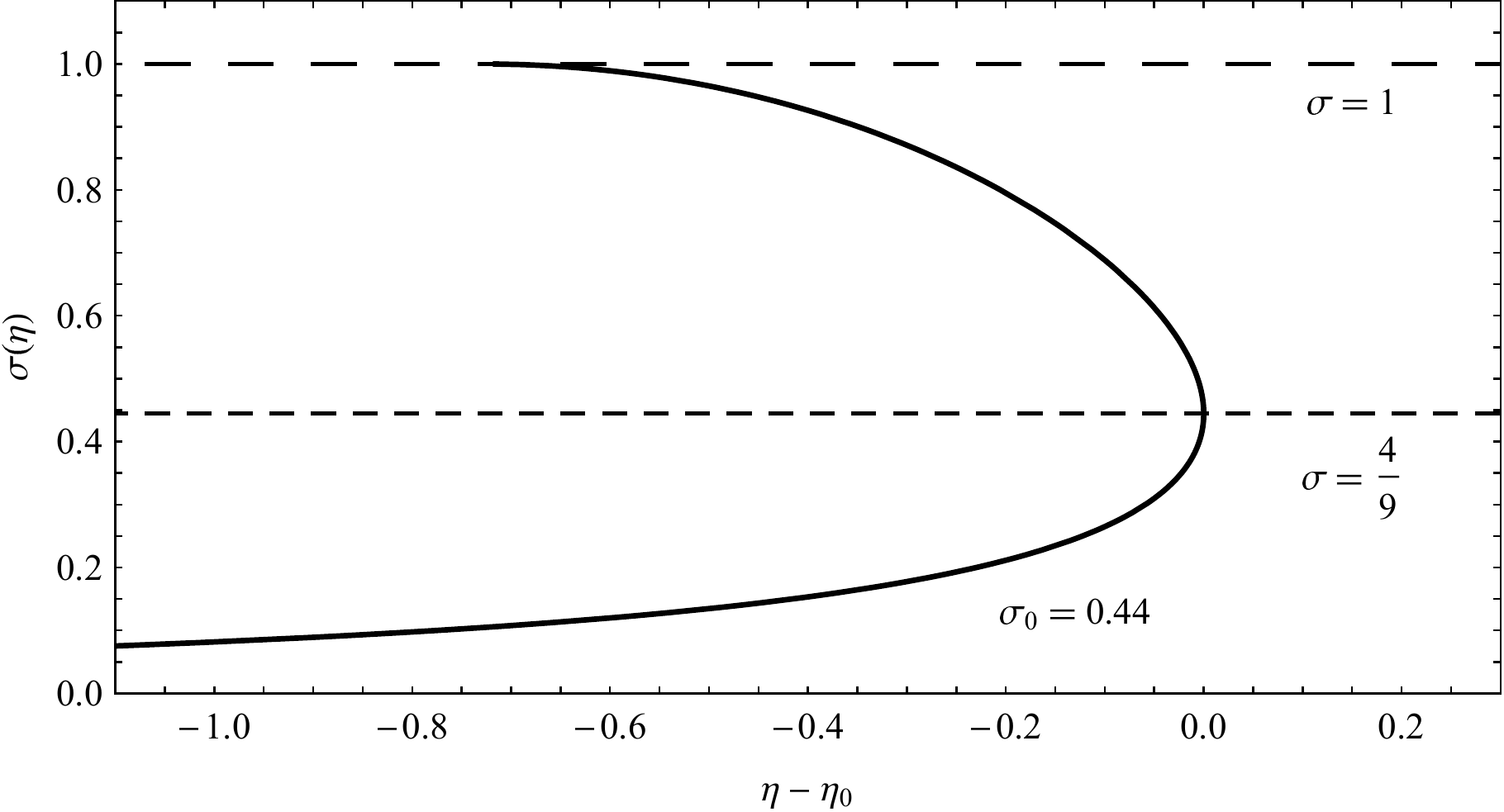} 
   \caption{Plot of the lower sign of equation (\ref{SigmaEta}) which corresponds to a
   contracting Universe. In this case we also choose the lower branch.}
   \label{fig:sigmaeta3}
\end{figure}
Note that in the plot of equation (\ref{SigmaEta}) (see Figures \ref{fig:sigmaeta1} and \ref{fig:sigmaeta2}, where the sign has been chosen for an expanding universe), 
a non-invertible behaviour is evident from the point when it reaches 
the first critical density $\tilde{\sigma}=\frac{4}{9}$. We also note that 
the maximum density defined from the first Friedmann equation, given by $\tilde{\sigma}=1$, will indeed be a value beyond which $\tilde{\sigma}$ will never go.
Mathematically, $1> \tilde{\sigma} > 4/9$ is possible, but can be excluded on physical grounds for an expanding universe. We have also plotted the solution for a contracting Universe in Figure \ref{fig:sigmaeta3}.
%Sergio: 1plot for \tilde{\sigma}_00-03. 0.2 possible for expanding universe? 2. plot for conracting universe) 

%\subsection{Perfect Fluid with Spin term}

%As we have seen before, we can have a Perfect Fluid with spin in the following way [Ref \`a-la Weyssenhoff (Obukhov)?]
%
%\begin{equation}
%\begin{split}
%\sigma_{\mu\nu}&=(\epsilon + P)u_{\mu}u_{\nu}-Pg_{\mu\nu}+2u^{\alpha}u_{(\mu}\nabla_{\beta}\tau\indices{_{\alpha|\nu)}^{\beta}}\\
%&= (\epsilon + P)u_{\mu}u_{\nu}-Pg_{\mu\nu}+2u^{\alpha}u_{(\mu}\nabla_{\beta}(s_{\alpha|\nu)}u^{\beta})
%\end{split}
%\end{equation}

%DO THE SAME WITH TORSION INVARIANTS, THREE COEFFICIENTS.

%\newpage
\subsection{Cosmology in the Alternative Lagrangians}
\begin{figure}[h] %  figure placement: here, top, bottom, or page
   \centering
   \includegraphics[width=4in]{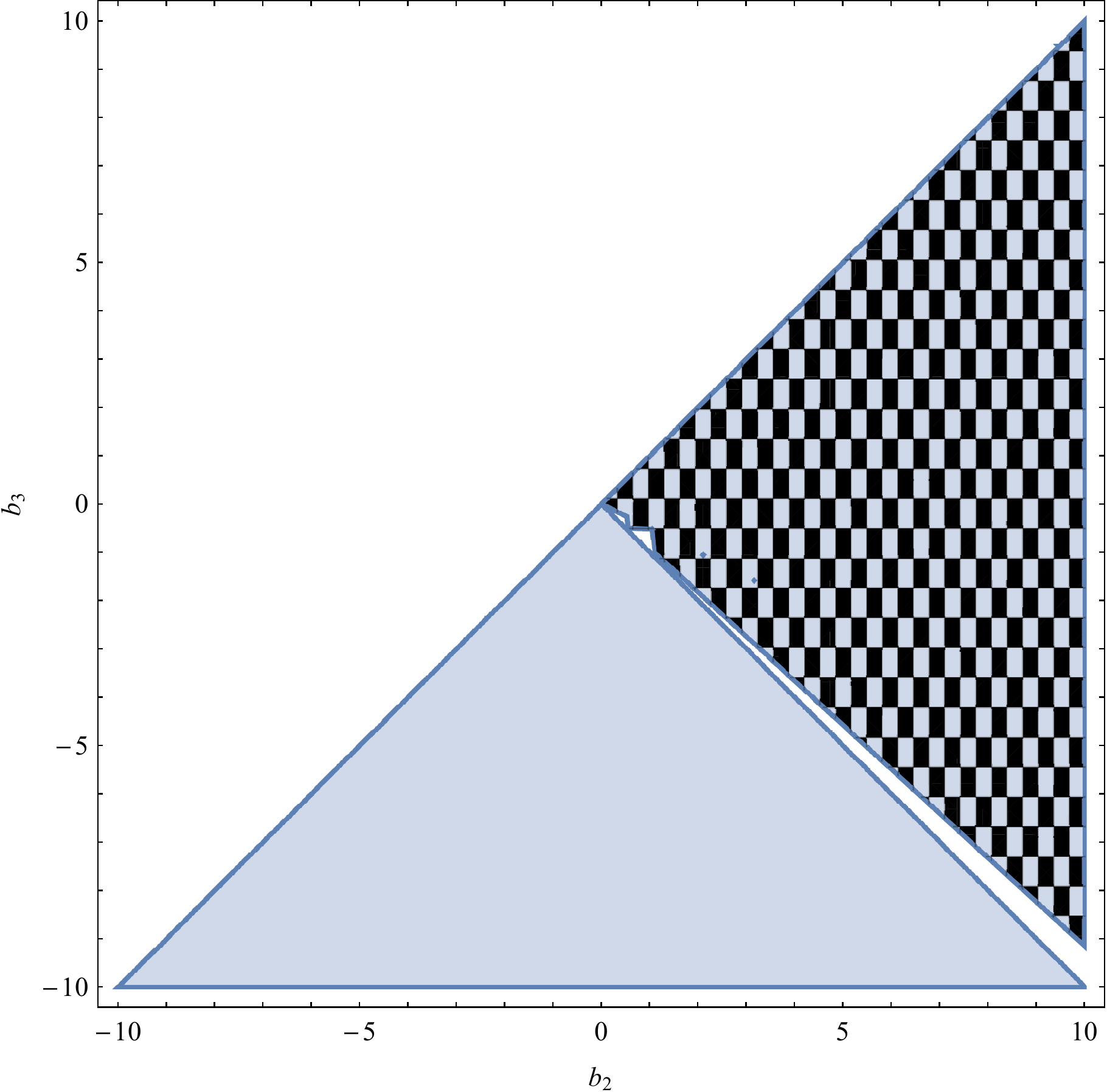} 
   \caption{Region of parameters $b_{2}$ and $b_{3}$ which avoid the third critical value for the density. The region with the squared pattern corresponds to $|C_{1}|/|C_{2}|>3/4$ and the plain region to $C_{2}<0$.}
   \label{fig:Regionc2}
\end{figure}

Let us go back and consider the Lagrangian $\mathcal{L}_{AT}$ in equation (\ref{torsioninv}).
In order to obtain the equations of motion we vary the Lagrangian with respect to the metric tensor and 
the 
contortion tensor.  The variation of $\mathcal{L}_{AT}=\sqrt{-g}L_{AT}$ with respect to 
the inverse metric will yield the following field equations
%\begin{equation}
%\begin{split}
%\delta_{g}\mathcal{L}_{AT}=\delta_{g}\left(\sqrt{-g} \right)L_{AT}+\sqrt{-g}&\left( \delta_{g}\accentset{\circ}{R}+b_{1}\delta_{g} \left(S_{\lambda}S^{\lambda} \right)+b_{2}\delta_{g}\left(S_{\alpha\beta\gamma}S^{\gamma\beta\alpha} \right)\right.\\ & \left. +b_{3}\delta_{g}\left(S_{\alpha\beta\gamma}S^{\alpha\beta\gamma} \right) \right)
%\end{split}
%\end{equation}
%
%The following terms will be relevant
%
%\begin{eqnarray}
%\delta_{g}\left( S_{\lambda}S^{\lambda}\right)&=&\left( S\indices{_{\alpha\mu}^{\alpha}}S\indices{_{\rho\nu}^{\rho}}\right)\delta g^{\mu\nu},\\
%\delta_{g}\left(S_{\alpha\beta\gamma}S^{\gamma\beta\alpha} \right)&=&\left(S\indices{_{\alpha\nu}^{\gamma}}S\indices{_{\gamma\mu}^{\alpha}} \right)\delta g^{\mu\nu},\\
%\delta_{g}\left(S_{\alpha\beta\gamma}S^{\alpha\beta\gamma} \right)&=& \left(2S\indices{_{\mu\beta}^{\gamma}}S\indices{_{\nu}^{\beta}_{\gamma}}-S_{\alpha\beta\mu}S\indices{^{\alpha\beta}_{\nu}} \right)\delta g^{\mu\nu}.
%\end{eqnarray}
%
%And so the full field equations can be written as
%
\begin{equation}\label{FEqs}
\begin{split}
\accentset{\circ}{R}_{\mu\nu}-\frac{1}{2}g_{\mu\nu}\accentset{\circ}{R}-\frac{1}{2}g_{\mu\nu}\left[ b_{1}S_{\lambda}S^{\lambda}+b_{2}S_{\alpha\beta\gamma}S^{\gamma\beta\alpha}+b_{3}S_{\alpha\beta\gamma}S^{\alpha\beta\gamma} \right] & \\
+b_{1}S\indices{_{\alpha\mu}^{\alpha}}S\indices{_{\rho\nu}^{\rho}}+b_{2}S\indices{_{\alpha\nu}^{\gamma}}S\indices{_{\gamma\mu}^{\alpha}}+b_{3}\left(2S\indices{_{\mu\beta}^{\gamma}}S\indices{_{\nu}^{\beta}_{\gamma}}-S_{\alpha\beta\mu}S\indices{^{\alpha\beta}_{\nu}} \right)&=\kappa \sigma_{\mu\nu},
\end{split}
\end{equation}
where $\sigma_{\mu\nu}$ is the metric energy--momentum tensor defined in (\ref{EMTensors}).
%The lagrangian $\mathcal{L}_{AT}$ can also be written in terms of contorsion is as follows
%\begin{equation}
%\begin{split}
%\mathcal{L}_{AT}=\sqrt{-g}&\left( \accentset{\circ}{R}+\frac{b_{1}}{4}K\indices{_{\alpha\lambda}^{\alpha}}K\indices{_{\rho}^{\lambda\rho}}+\left(\frac{3b_{2}}{4}-\frac{b_{3}}{2} \right)K_{\alpha\beta\gamma}K^{\gamma\beta\alpha} \right. \\
%& \left. +\left(\frac{b_{3}}{2}-\frac{b_{2}}{4} \right)K_{\alpha\beta\gamma}K^{\alpha\beta\gamma}\right),
%\end{split}
%\end{equation}
The variation of the Lagrangian with respect to contortion results in the following algebraic equation in terms of contortion and its source
\begin{equation}\label{ALcontorsion}
\begin{split}
\frac{b_{1}}{8}\left(\delta_{\nu}^{\mu}K\indices{_{\rho\alpha}^{\rho}}+\delta_{\alpha}^{\mu}K\indices{_{\gamma}^{\gamma}_{\nu}} \right)+\left(\frac{b_{3}}{2}-\frac{b_{2}}{4}  \right)K\indices{^{\mu}_{\alpha\nu}}&\\
+\left(\frac{3b_{2}}{8}-\frac{b_{3}}{4} \right)\left(K\indices{_{\alpha}^{\mu}_{\nu}}+K\indices{_{\nu\alpha}^{\mu}} \right)&=\kappa \tau\indices{_{\alpha\nu}^{\mu}},
\end{split}
\end{equation}
Alternatively, in terms of torsion (\ref{ALcontorsion}) reads
\begin{equation}\label{ALTorsion}
\begin{split}
\left(\frac{b_{2}}{2}\right)S\indices{_{\alpha\nu}^{\mu}}+\frac{b_{1}}{4}\left(-\delta_{\alpha}^{\mu}S\indices{_{\nu\gamma}^{\gamma}}+\delta_{\nu}^{\mu}S\indices{_{\alpha\rho}^{\rho}}\right)&\\
-\left(\frac{b_{3}}{2}-\frac{b_{2}}{4} \right)\left(S\indices{^{\mu}_{\alpha\nu}}+S\indices{_{\nu}^{\mu}_{\alpha}} \right)&=\kappa\tau\indices{_{\alpha\nu}^{\mu}}.
\end{split}
\end{equation}
By taking the trace of (\ref{ALTorsion}) we get
\begin{equation}
\left(\frac{3b_{1}}{4}+\frac{b_{2}}{4}+\frac{b_{3}}{2} \right)S\indices{_{\alpha\lambda}^{\lambda}}=\kappa \tau\indices{_{\alpha\lambda}^{\lambda}}.
\end{equation}
We can now relate the source of torsion in the following way
\begin{equation}\label{FinalAlTorsion}
\begin{split}
\left(\frac{b_{2}}{2} \right)S\indices{_{\alpha\nu}^{\mu}}-\left(\frac{b_{3}}{2}-\frac{b_{2}}{4} \right)\left(S\indices{^{\mu}_{\alpha\nu}}+S\indices{_{\nu}^{\mu}_{\alpha}}\right) =&\kappa \tau\indices{_{\alpha\nu}^{\mu}}+\left( \frac{\kappa b_{1}}{3b_{1}+b_{2}+2b_{3}}\right)\delta_{\alpha}^{\mu}\tau\indices{_{\nu\gamma}^{\gamma}}\\&-\left(\frac{\kappa b_{1}}{3b_{1}+b_{2}+2b_{3}} \right)\delta_{\nu}^{\mu}\tau\indices{_{\alpha\rho}^{\rho}}.
\end{split}
\end{equation}
Using (\ref{FinalAlTorsion}) it is possible to write the torsion in terms of its source only. We have
%\begin{equation}
%S\indices{_{\nu}^{\mu}_{\alpha}}=\frac{\kappa}{b_{3}-b_{2}}\left[ -\left(\frac{2b_{3}-b_{2}}{2b_{3}+b_{2}}\right)W\indices{_{\alpha\nu}^{\mu}}-\left(\frac{2b_{3}-b_{2}}{2b_{3}+b_{2}}\right) W\indices{^{\mu}_{\alpha\nu}} + \left(\frac{2b_{3}-3b_{2}}{2b_{3}+b_{2}} \right)W\indices{_{\nu}^{\mu}_{\alpha}} \right]
%\end{equation}
\begin{equation}
S\indices{_{\alpha\nu}^{\mu}}=\frac{\kappa}{b_{3}-b_{2}}\left[\left(\frac{2b_{3}-3b_{2}}{2b_{3}+b_{2}} \right)W\indices{_{\alpha\nu}^{\mu}}-\left(\frac{2b_{3}-b_{2}}{2b_{3}+b_{2}} \right)\left\{ W\indices{^{\mu}_{\alpha\nu}}+W\indices{_{\nu}^{\mu}_{\alpha}}\right\} \right],
\end{equation}
where we have defined a tensor $W\indices{_{\alpha\nu}^{\mu}}$ as
\begin{equation}
W\indices{_{\alpha\nu}^{\mu}}\equiv\tau\indices{_{\alpha\nu}^{\mu}}+\left(\frac{b_{1}}{3b_{1}+b_{2}+2b_{3}}\right)\left[\delta_{\alpha}^{\mu}\tau\indices{_{\nu\gamma}^{\gamma}}-\delta_{\nu}^{\mu}\tau\indices{_{\alpha\rho}^{\rho}} \right]=-W\indices{_{\nu\alpha}^{\mu}}.
\end{equation}
The contortion will be written in terms of $W\indices{_{\alpha\nu}^{\mu}}$ as
\begin{equation}
K\indices{_{\mu\nu}^{\alpha}}=\frac{\kappa}{b_{3}-b_{2}}\left\{\left(\frac{2b_{3}-3b_{2}}{2b_{3}+b_{2}} \right)\left[-W\indices{_{\mu\nu}^{\alpha}}+W\indices{_{\nu}^{\alpha}_{\mu}}-W\indices{^{\alpha}_{\mu\nu}} \right]+\left(\frac{2b_{3}-b_{2}}{2b_{3}+b_{2}} \right)\left[2W\indices{_{\nu}^{\alpha}_{\mu}} \right] \right\}.
\end{equation}
This result implies that we can write the metric field equations in terms of the source $\tau\indices{_{\mu\nu}^{\alpha}}$. 
Indeed, after some manipulations and by defining the following coefficients
%\begin{equation}
%S_{\lambda}S^{\lambda}=\frac{16\kappa^{2}}{(3b_{1}+b_{2}+2b_{3})^{2}}\tau\indices{_{\lambda\alpha}^{\alpha}}\tau\indices{^{\lambda\rho}_{\rho}}
%\end{equation}
%
%\begin{equation}
%S_{\alpha\beta\gamma}S^{\alpha\beta\gamma}=\frac{\kappa^{2}}{(b_{3}-b_{2})^{2}}\left[\left\{\left(\frac{2b_{3}-3b_{2}}{2b_{3}+b_{2}} \right)^{2}+2\left(\frac{2b_{3}-b_{2}}%{2b_{3}+b_{2}} \right)^{2} \right\}W_{\alpha\beta\gamma}W^{\alpha\beta\gamma}+4\left(\frac{2b_{3}-3b_{2}}{2b_{3}+b_{2}} \right)\left(\frac{2b_{3}-b_{2}}{2b_{3}+b_{2}} \right) %W_{\alpha\beta\gamma}W^{\alpha\gamma\beta} \right]
%\end{equation}
%
%\begin{equation}
%\begin{split}
%S_{\alpha\beta\gamma}S^{\gamma\beta\alpha}=\frac{\kappa^{2}}{(b_{3}-b_{2})^{2}} & \left[\left\{2\left(\frac{2b_{3}-b_{2}}{2b_{3}+b_{2}} \right)\left(\frac{2b_{3}-3b_{2}}{2b_{3}+b_{2}} \right)+\left(\frac{2b_{3}-b_{2}}{2b_{3}+b_{2}} \right)^{2} \right\}W_{\alpha\beta\gamma}W^{\alpha\beta\gamma}\right. \\ & \left. +\left\{4\left(\frac{2b_{3}-b_{2}}{2b_{3}+b_{2}} \right)^{2}-2\left(\frac{2b_{3}-b_{2}}{2b_{3}+b_{2}} \right)\left(\frac{2b_{3}-3b_{2}}{2b_{3}+b_{2}} \right) \right\}W_{\alpha\beta\gamma}W^{\alpha\gamma\beta} \right]
%\end{split}
%\end{equation}
%
%\begin{equation}
%S\indices{_{\alpha\mu}^{\alpha}}S\indices{_{\rho\nu}^{\rho}}=\frac{16\kappa^{2}}{(3b_{1}+b_{2}+2b_{3})^{2}}\tau\indices{_{\mu\lambda}^{\lambda}}\tau\indices{_{\nu\alpha}%^{\alpha}}
%\end{equation}
%
%we define
%
%\begin{equation}
\[
A\equiv \left(\frac{2b_{3}-3b_{2}}{2b_{3}+b_{2}} \right), \mbox{ } \mbox{ } \mbox{ } \mbox{ } \mbox{ } \mbox{ } B\equiv \left(\frac{2b_{3}-b_{2}}{2b_{3}+b_{2}} \right)
\]
we get
\begin{equation}\label{FullFEqs}
\begin{split}
\accentset{\circ}{R}_{\mu\nu}-\frac{1}{2}g_{\mu\nu}\accentset{\circ}{R}=& \kappa \sigma_{\mu\nu}+\frac{1}{2}g_{\mu\nu}\left[b_{1}\frac{16\kappa^{2}}{(3b_{1}+b_{2}+2b_{3})^{2}}\tau\indices{_{\lambda\alpha}^{\alpha}}\tau\indices{^{\lambda\rho}_{\rho}}\right. \\ & \left.+ b_{2}\frac{\kappa^{2}}{(b_{3}-b_{2})^{2}}  \left[\left\{2BA+B^{2} \right\}W_{\alpha\beta\gamma}W^{\alpha\beta\gamma} +\left\{4B^{2}-2BA \right\}W_{\alpha\beta\gamma}W^{\alpha\gamma\beta} \right]\right. \\ & \left. +b_{3} \frac{\kappa^{2}}{(b_{3}-b_{2})^{2}}\left[\left\{A^{2}+2B^{2} \right\}W_{\alpha\beta\gamma}W^{\alpha\beta\gamma}+4AB W_{\alpha\beta\gamma}W^{\alpha\gamma\beta} \right] \right]\\
& -b_{1}\frac{16\kappa^{2}}{(3b_{1}+b_{2}+2b_{3})^{2}}\tau\indices{_{\mu\lambda}^{\lambda}}\tau\indices{_{\nu\alpha}^{\alpha}}\\
&-b_{2}\frac{\kappa^{2}}{(b_{3}-b_{2})^{2}} \left\{ (2AB)W_{\mu\alpha\gamma}W\indices{_{\nu}^{\alpha\gamma}}+\left[A^{2}+B^{2} \right]W_{\mu\alpha\gamma}W\indices{_{\nu}^{\gamma\alpha}}\right. \\ & \left. +(AB-B^{2})\left[W\indices{^{\alpha\gamma}_{\mu}}W_{\nu\alpha\gamma}+W_{\mu\alpha\gamma}W\indices{^{\alpha\gamma}_{\nu}} \right]-(B^{2})W_{\alpha\gamma\mu}W\indices{^{\alpha\gamma}_{\nu}} \right\}\\
& -2b_{3}\frac{\kappa^{2}}{(b_{3}-b_{2})^{2}}\left\{ \left[A^{2}+B^{2} \right]W\indices{_{\mu\beta}^{\gamma}}W\indices{_{\nu}^{\beta}_{\gamma}}+(2AB)W\indices{_{\mu\beta}^{\gamma}}W\indices{_{\nu\gamma}^{\beta}} \right. \\ & \left. \left(B^{2}-AB \right)\left[W\indices{_{\mu\beta}^{\gamma}}W\indices{^{\beta}_{\gamma\nu}}+W\indices{_{\beta}^{\gamma}_{\mu}}W\indices{_{\nu}^{\beta}_{\gamma}} \right]+(B)^{2}W_{\beta\gamma\mu}W\indices{^{\beta\gamma}_{\nu}} \right\}
\\ & +b_{3}\frac{\kappa^{2}}{(b_{3}-b_{2})^{2}}  \left\{2(B)^{2}W_{\mu\alpha\beta}W\indices{_{\nu}^{\alpha\beta}}-2(B)^{2}W_{\mu\alpha\beta}W\indices{_{\nu}^{\beta\alpha}}\right. \\ &\left. -2AB\left[ W_{\mu\alpha\beta}W\indices{^{\alpha\beta}_{\nu}}+W_{\alpha\beta\mu}W\indices{_{\nu}^{\alpha\beta}} \right] + (A)^{2}W_{\alpha\beta\mu}W\indices{^{\alpha\beta}_{\nu}}\right\}.
\end{split}
\end{equation}
which, upon invoking the conditions for the source represented by (\ref{Frenkel}) and performing an averaging over the spin terms,
%\begin{equation}           
%\tau\indices{_{\mu\nu}^{\alpha}}=s_{\mu\nu}u^{\alpha}\mbox{ , } s_{\mu\nu}u^{\nu}=0
%\end{equation}
can be considered satisfied. For instance, already the $W$ tensor takes a simpler form
%, which under Frenkel conditions will be simply given by
\begin{equation}
W\indices{_{\mu\nu}^{\alpha}}=\tau\indices{_{\mu\nu}^{\alpha}}.
\end{equation}
%furthermore this implies upon the expression found above for the field equations
%\begin{equation}
%S_{\lambda}S^{\lambda}=0
%\end{equation}
%\begin{equation}
%S_{\alpha\beta\gamma}S^{\alpha\beta\gamma}=\frac{\kappa^{2}}{(b_{3}-b_{2})^{2}}\left\{\left[(A')^{2}+2(B')^{2} \right]s_{\alpha\beta}u_{\gamma}s^{\alpha\beta}u^{\gamma} \right\}
%\end{equation}
%\begin{equation}
%S_{\alpha\beta\gamma}S^{\gamma\beta\alpha}=\frac{\kappa^{2}}{(b_{3}-b_{2})^{2}}\left\{\left[2 A'B'+(B')^{2}\right]s_{\alpha\beta}u_{\gamma}s^{\alpha\beta}u^{\gamma} \right\}
%\end{equation}
%\begin{equation}
%S\indices{_{\alpha\mu}^{\alpha}}S\indices{_{\rho\nu}^{\rho}}=0
%\end{equation}
%\begin{equation}
%S\indices{_{\alpha\nu}^{\gamma}}S\indices{_{\gamma\mu}^{\alpha}}=\frac{\kappa^{2}}{(b_{3}-b_{2})^{2}}\left\{2A'B' s_{\mu\alpha}u_{\gamma}s\indices{_{\nu}^{\alpha}}u^{\gamma}-(B')^{2}s_{\alpha\gamma}u_{\mu}s^{\alpha\gamma}u_{\nu} \right\}
%\end{equation}
%\begin{equation}
%S\indices{_{\mu\beta}^{\gamma}}S\indices{_{\nu}^{\beta}_{\gamma}}=\frac{\kappa^{2}}{(b_{3}-b_{2})^{2}}\left\{[(A')^{2}+(B')^{2}]s_{\mu\beta}u^{\gamma}s\indices{_{\nu}^{\beta}}u_{\gamma}+(B')^{2}s_{\beta\gamma}u_{\mu}s^{\beta\gamma}u_{\nu} \right\}
%\end{equation}
%\begin{equation}
%S_{\alpha\beta\mu}S\indices{^{\alpha\beta}_{\nu}}=\frac{\kappa^{2}}{(b_{3}-b_{2})^{2}} \left\{2(B')^{2}s_{\mu\alpha}u_{\beta}s\indices{_{\nu}^{\alpha}}u^{\beta}+(A')^{2}s_{\alpha\beta}u_{\mu}s^{\alpha\beta}u_{\nu} \right\}
%\end{equation}
Following \cite{Gasperini}, expressions of the form $\langle s_{\mu\alpha}s\indices{_{\nu}^{\alpha}}\rangle $, i.e., spin expression with an open index, 
average to zero while $\langle s_{\mu\nu}s^{\mu\nu}\rangle$ averages to $\frac{1}{2}s^{2}$. The final result is
\begin{equation}
\begin{split}
\accentset{\circ}{R}_{\mu\nu}-\frac{1}{2}g_{\mu\nu}\accentset{\circ}{R}=  \kappa \sigma_{\mu\nu}+& \frac{1}{4}g_{\mu\nu}\frac{\kappa^{2}}{(b_{3}-b_{2})^{2}}\left[ b_{2}\left(2AB+B^{2} \right)+b_{3}\left(A^{2}+2B^{2} \right) \right] s^{2}\\
& +\frac{\kappa^{2}}{2(b_{3}-b_{2})^{2}}\left\{b_{2}B^{2}s^{2}+b_{3}(A^{2}-2B^{2})s^{2} \right\}u_{\mu}u_{\nu}.
\end{split}
\end{equation}
To proceed further we need to specify the energy-momentum tensor  
which we take to be that of a perfect fluid plus a spin contribution of the Weyssenhoff form.
Hence, we get
\begin{equation}
\begin{split}
\accentset{\circ}{R}_{\mu\nu}-\frac{1}{2}g_{\mu\nu}\accentset{\circ}{R}=&\kappa\left[ \left(\rho+p+\frac{\kappa}{2(b_{3}-b_{2})^{2}}\left[(b_{2}-2b_{3})B^{2}+b_{3}A^{2} \right]s^{2} \right)u_{\mu}u_{\nu}\right. \\ & \left.  - \left(p - \frac{\kappa}{4(b_{3}-b_{2})^{2}}\left[b_{3}A^{2}+2b_{2}AB+(b_{2}+2b_{3})B^{2} \right]s^{2} \right)g_{\mu\nu} \right].
\end{split}
\end{equation}
The Friedmann equations in this case can be obtained formally as before. Therefore, we just quote the result
\begin{equation}\label{FFriedmannAL}
H^{2}=\frac{\kappa}{3}\left(\rho +\frac{\kappa}{4(b_{3}-b_{2})^{2}}\left[3b_{3}(A)^{2}+2b_{2}AB+3b_{2}(B)^{2}-2b_{3}B^{2} \right] s^{2} \right),
\end{equation}
\begin{equation}\label{SFriedmannAL}
3H^{2}+2\dot{H}=-\kappa \left(p - \frac{\kappa s^{2}}{4(b_{3}-b_{2})^{2}}\left[b_{3}(A)^{2}+2b_{2}AB+(b_{2}+2b_{3})(B)^{2} \right] \right).
\end{equation}
A small manipulation gives
\begin{equation}\label{SFriedmannAL1}
\frac{\ddot{a}}{a}=H^{2}+\dot{H}=-\frac{\kappa}{6}\left(\rho+3p-\frac{\kappa s^{2}}{4(b_{3}-b_{2})^{2}}\left[4b_{2}AB+8b_{3}(B)^{2} \right] \right),
\end{equation}
\begin{equation}
\dot{H}=-\frac{\kappa}{2}\left( \rho+p+\frac{\kappa s^{2}}{4(b_{3}-b_{2})^{2}}\left[2b_{3}A^{2}+(2b_{2}-4b_{3})B^{2} \right] \right).
\end{equation}
In order to rewrite the most relevant expressions obtained from the Friedmann equations, let us define the following constants which all depend upon the parameters $b_{2}$ and
$b_{3}$
\begin{equation}\label{c1def}
C_{1}\equiv \frac{1}{4(b_{3}-b_{2})^{2}}\left\{3b_{3}A^{2}+2b_{2}AB+(3b_{2}-2b_{3})B^{2} \right\},
\end{equation}
\begin{equation}
C_{2}\equiv \frac{1}{4(b_{3}-b_{2})^{2}}\left\{2b_{3}A^{2}+(2b_{2}-4b_{3})B^{2} \right\},
\end{equation}
\begin{equation}
C_{3}\equiv \frac{1}{4(b_{3}-b_{2})^{2}}(4b_{2}AB+8b_{3}B^{2}).
\end{equation}
The cosmological equations (\ref{FFriedmannAL}), (\ref{SFriedmannAL}) and (\ref{SFriedmannAL1}) will be
\begin{equation}\label{FirstFriedmannALA}
H^{2}=\frac{\kappa}{3}(\rho+C_{1}\kappa s^{2}),
\end{equation}
\begin{equation}\label{SecondFriedmannALA}
\ddot{a}=-\frac{\kappa}{6}(\rho+3p-C_{3}\kappa s^{2})a, \mbox{ } \mbox{ } \mbox{ }  \mbox{ } \mbox{ } \mbox{ } \dot{H}=-\frac{\kappa}{2}(\rho+p+C_{2}\kappa s^{2}).
\end{equation}
The continuity equation (\ref{ContinuityEqAL1}) is derived in the case of radiation, $w=\frac{1}{3}$, by using the relation (\ref{SpinDensity})
\begin{equation}\label{ContinuityEqAL1}
\frac{d}{dt}\left(\rho+C_{1}\kappa B_{\frac{1}{3}}\rho^{\frac{3}{2}} \right)=-3H\left(\frac{4}{3}\rho+C_{2}\kappa B_{\frac{1}{3}}\rho^{\frac{3}{2}} \right).
\end{equation}
This can be written as a differential equation for $\rho$ of the form
\begin{equation}\label{ContALB}
\frac{d\rho}{dt}\left(1+\frac{3}{2}C_{1}\kappa B_{\frac{1}{3}}\rho^{\frac{1}{2}} \right)=-4H\rho\left(1+\frac{3}{4}C_{2}\kappa B_{\frac{1}{3}}\rho^{\frac{1}{2}} \right),
\end{equation}
We draw the reader's attention to the fact that the choice $2C_{1}=C_{2}$ would give us back the standard GR result. Indeed, we will be forced to make this choice as explained below.
In order to obtain some interesting results given by models with no singularity and with an inflationary stage, we should put certain conditions upon these new
parameters $C_{i}$ and thus, upon $b_{2}$ and $b_{3}$. With this in mind let us again define dimensionless variables in order to work out the evolution of
the scale factor and the density. First, inspired by a class of non-singular models we note a special value of $\rho$ which makes $H^{2}$ go to zero. We denote this density by
$\bar{\rho}_{\max}$ which is given as
\begin{equation}\label{rhomaxAL}
\bar{\rho}_{\max}^{1/2}=-\frac{1}{C_{1}\kappa B_{\frac{1}{3}}}, \quad C_{1}<0.
\end{equation}
Using (\ref{rhomaxAL}) and defining dimensionless variables
\begin{equation}\label{DimensionlessB}
\bar{\sigma}\equiv \frac{\rho}{\bar{\rho}_{\max}}, \mbox{ } \mbox{ } \mbox{ }\mbox{  } \mbox{ } \mbox{ } \mbox{ } \bar{\eta}\equiv t\sqrt{\kappa \bar{\rho}_{\max}}.
\end{equation}
allows us to write the first Friedmann equation as
\begin{equation}
\frac{1}{a}\frac{da}{d\bar{\eta}}=\pm \sqrt{\frac{\bar{\sigma}}{3}}\sqrt{1-\bar{\sigma}^{\frac{1}{2}}}.
\end{equation}
On the other hand, the continuity equation reads 
\begin{equation}
\frac{d\bar{\sigma}}{d\eta}\left(1-\frac{3}{2}\bar{\sigma}^{\frac{1}{2}} \right)=-4\frac{1}{a}\frac{da}{d\bar{\eta}}\bar{\sigma}\left(1-\frac{3}{4}\frac{C_{2}}{C_{1}}\bar{\sigma}^{\frac{1}{2}} \right),
\end{equation}
\begin{equation}\label{DEB}
\frac{d\bar{\sigma}}{d\bar{\eta}}\left(1-\frac{3}{2}\bar{\sigma}^{\frac{1}{2}} \right)=\mp \frac{4}{\sqrt{3}}\bar{\sigma}^{\frac{3}{2}}\left(1-\frac{3}{4}\frac{C_{2}}{C_{1}}\bar{\sigma}^{\frac{1}{2}} \right)\sqrt{1-\bar{\sigma}^{\frac{1}{2}}}.
\end{equation}
It is evident that three critical values are present in this equation. A quick analysis reveals that
\begin{itemize}
\item{$\bar{\sigma}^{\frac{1}{2}}=1$ corresponds to $\rho=\bar{\rho}_{\max}$. This implies $\frac{d\bar{\sigma}}{d\bar{\eta}}=0$ and hence, a critical (extremal) value for $\bar{\sigma}$.}
\item{$\bar{\sigma}^{\frac{1}{2}}=\frac{2}{3}$ corresponds to $\rho=\frac{4}{9}\bar{\rho}_{\max}$. This implies that the 
l.h.s. of (\ref{DEB}) is zero which is a contradiction unless we choose $C_{2}=2C_{1}$. % since the right hand side will never be zero for this value of $\bar{\sigma}$. 
This means that this particular value
for the density is not a critical one as the equation becomes an identity in this case.}
\item{The third case happens when $\bar{\sigma}^{\frac{1}{2}}=\frac{4}{3}\frac{C_{1}}{C_{2}}$ 
corresponding to $\rho=\frac{16}{9}\frac{C_{1}^{2}}{C_{2}^{2}}\bar{\rho}_{\max}$. 
This implies that the r.h.s. of (\ref{DEB}) is zero which is consistent only if $\bar{\sigma}$ has more than one extremal value, or in the particular case when $4C_{1}/3C_{2}=2/3$. We will discuss this choice in more detail below.
Of course, this case
can happen only assuming that $C_2 < 0$. Hence, if $C_2 > 0$ or if $ |C_1|/|C_2| > 3/4$ (the latter case is equivalent to choosing 
the critical density bigger than $\bar{\rho}_{\max}$ which is impossible) there is no physical third critical value for the density. The allowed parameter space which allows us to avoid this third critical value is depicted in Figure \ref{fig:Regionc2}.}
\end{itemize}

\begin{figure}[h] %  figure placement: here, top, bottom, or page
   \centering
   \includegraphics[width=4in]{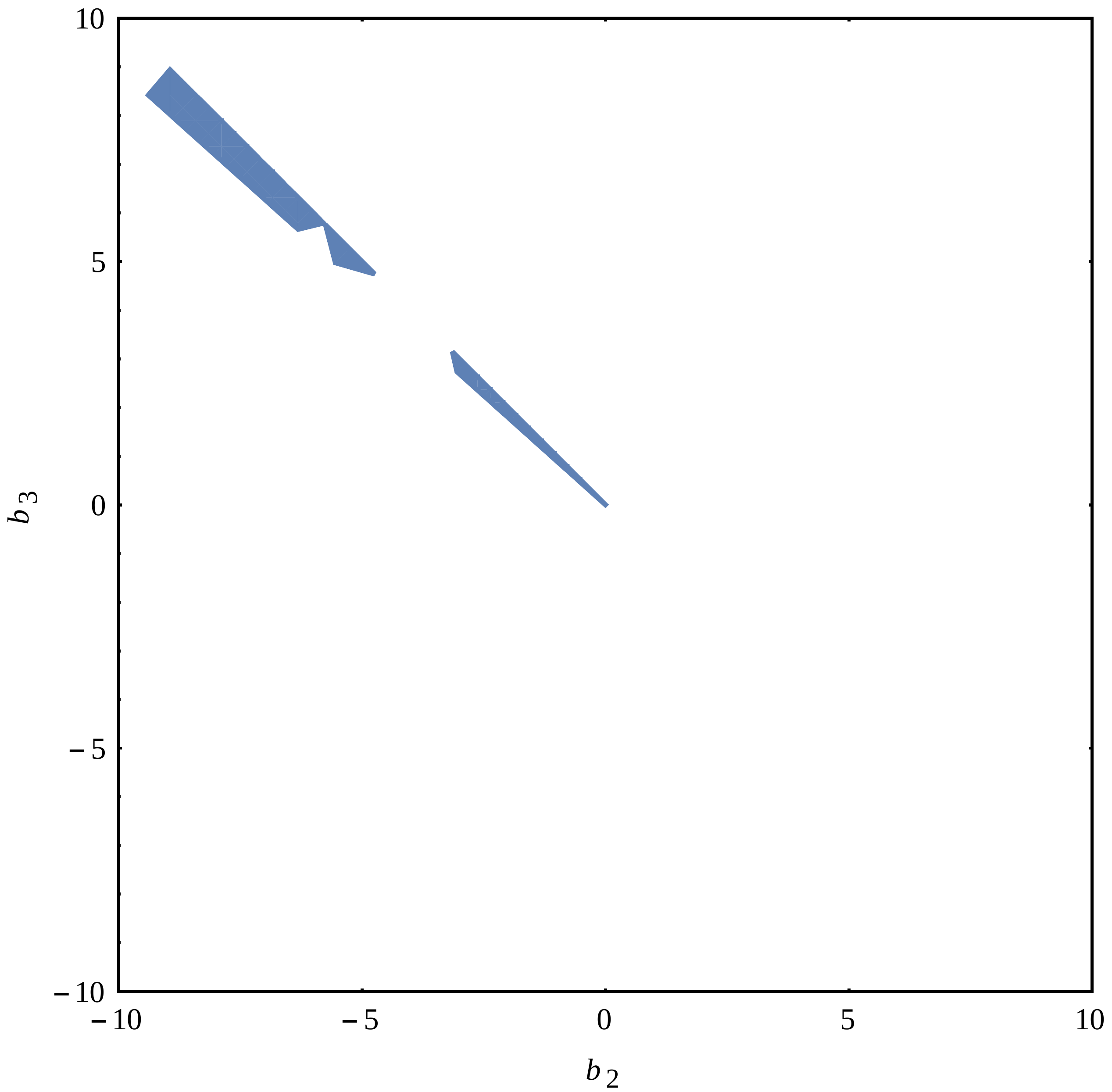} 
   \caption{Region of parameters $b_{2}$ and $b_{3}$ for a non-singular and accelerated inflationary solution.}
   \label{fig:region}
\end{figure}

\begin{figure}[h] %  figure placement: here, top, bottom, or page
   \centering
   \includegraphics[width=4in]{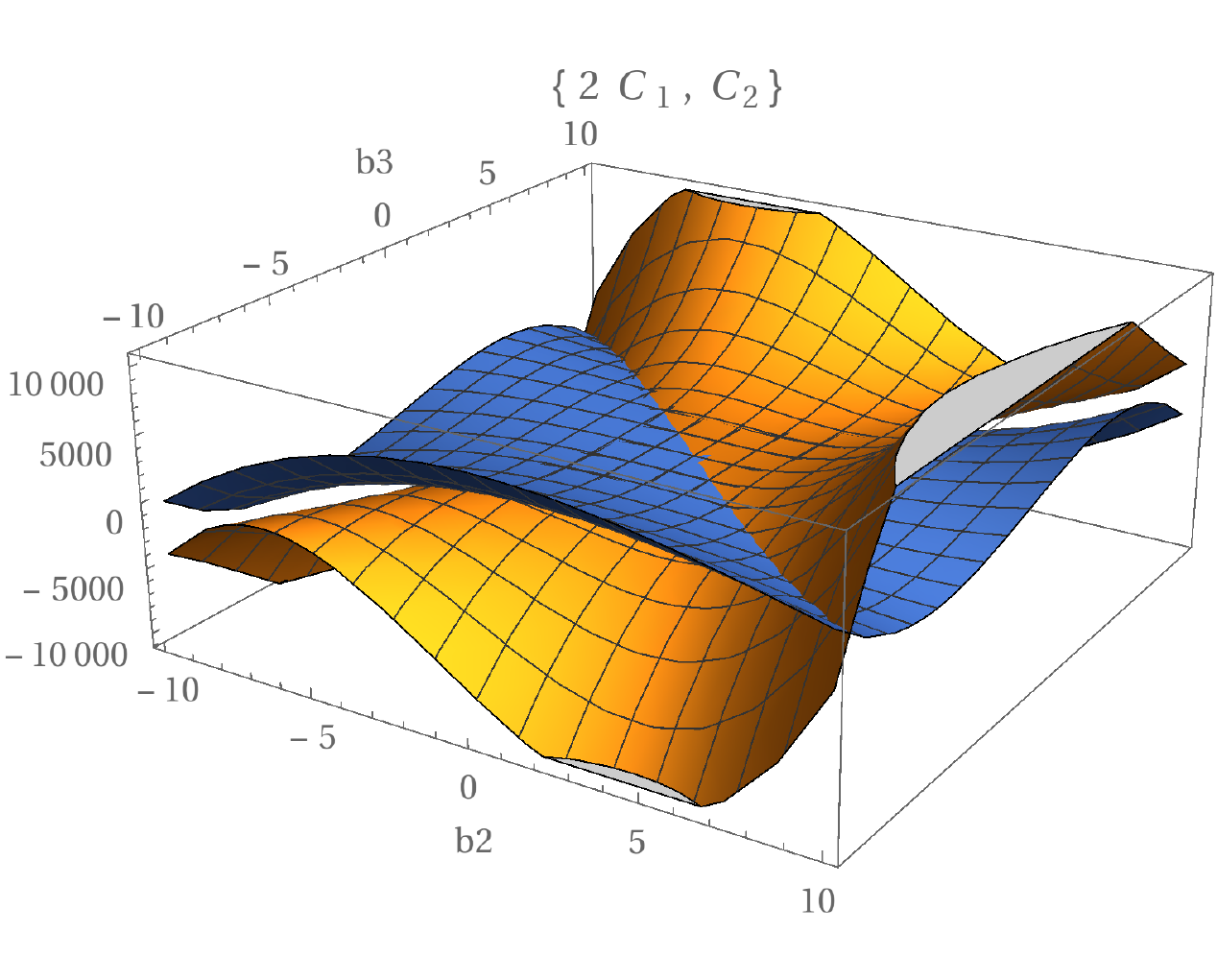} 
   \caption{We plot consistent rescalings of $2C_1$ (yellow) and $C_2$ (blue) as functions of $b_2$ and $b_3$ to show that the intersection between the two resembles a line. We show this in Fig. \ref{fig:lineplot}.}
   \label{fig:intersection}
\end{figure}

\begin{figure}
\centering
\begin{minipage}{.45\textwidth}
\centering
   \includegraphics[width=3.2in]{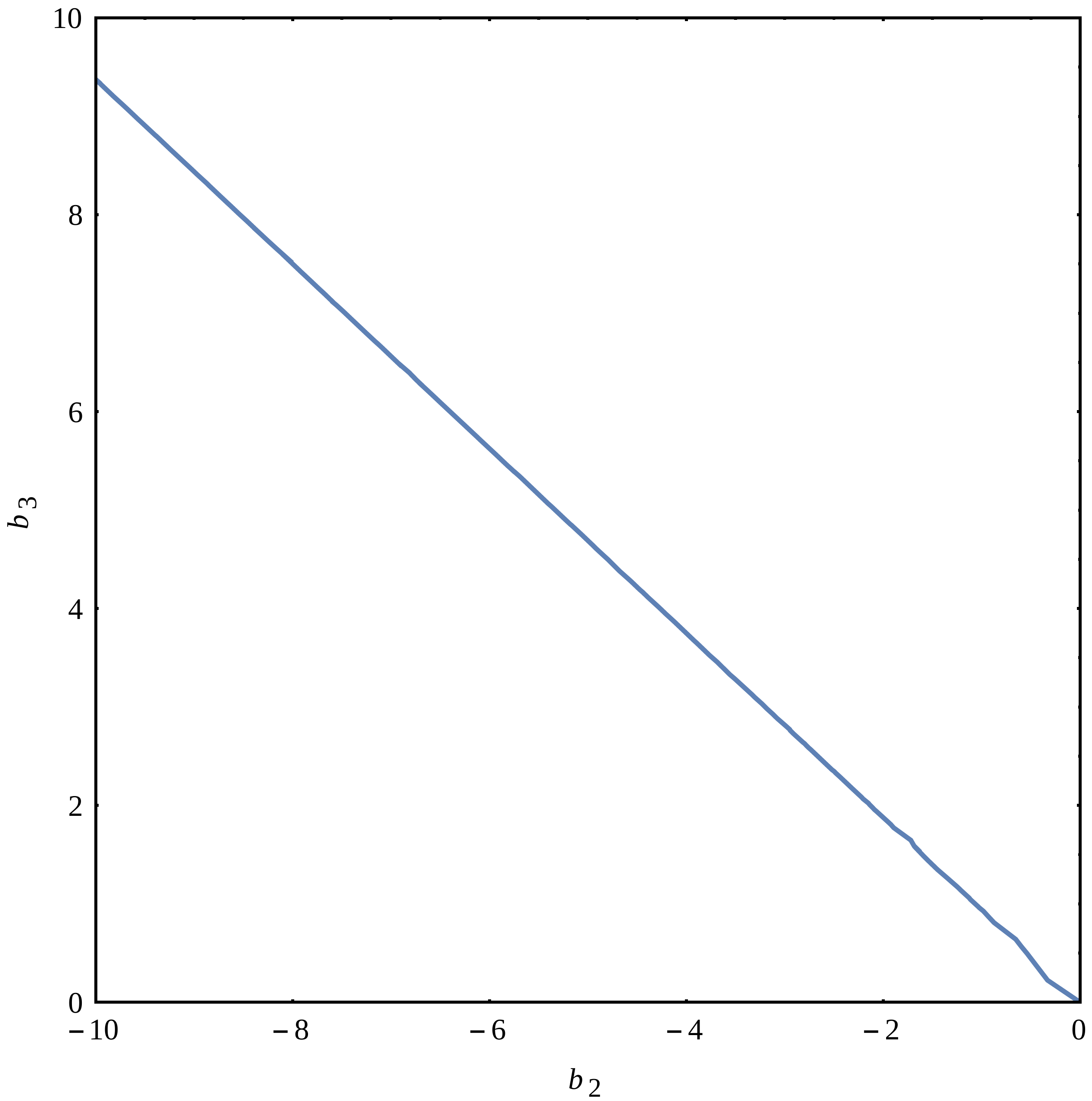} 
   \caption{Plot of the values $b_{2}$ and $b_{3}$ which make the continuity equation consistent, namely the condition in equation (\ref{2C1C2}). This corresponds to the intersection of the surfaces in Figure \ref{fig:intersection}.}
   \label{fig:lineplot}
\end{minipage}\hfill
\begin{minipage}{.45\textwidth}
\centering
   \includegraphics[width=3.2in]{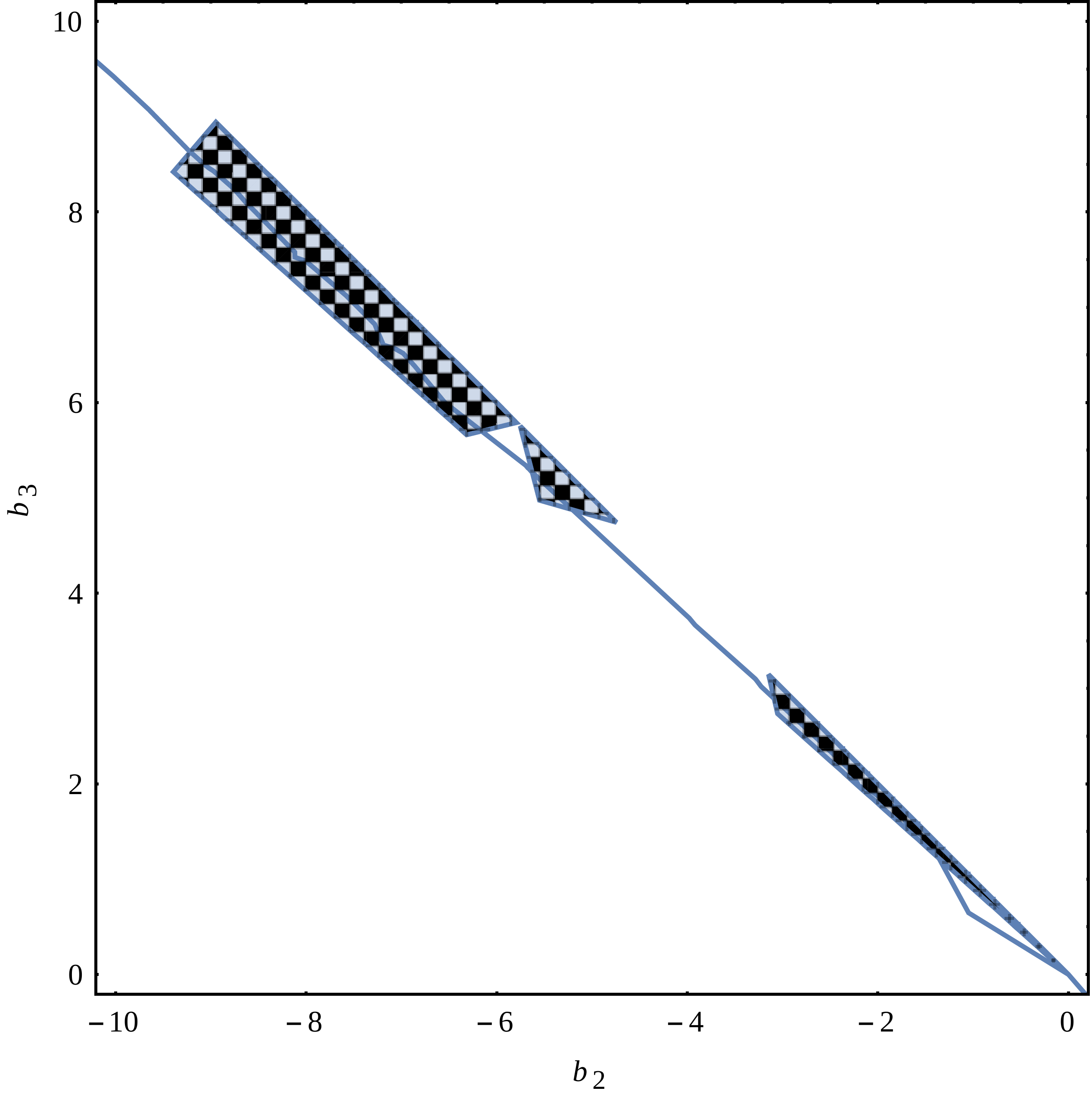} 
   \caption{Combined plot of the values $b_{2}$ and $b_{3}$ which make the continuity equation `consistent' (line plot) and those which yield also a non-singular and accelerated early universe (squared region).}
   \label{fig:RegionPlotBAll}
\end{minipage}\hfill
\end{figure}

In order to impose on the model a non-singular behaviour at the Big Bang (BB), we concentrate on the first Friedmann equation (\ref{FirstFriedmannALA}) and require $ C_{1}<0$ which means
\begin{equation}\label{NSBB}
3b_{3}(A)^{2}+2b_{2}AB+(3b_{2}-2b_{3})(B)^{2}<0,
\end{equation}
so that we may write the first Friedmann equation in the following way
\begin{equation}
H^{2}=\frac{\kappa}{3}\rho\left[1-\left(\frac{\rho}{\rho_{\max}} \right)^{\frac{1}{2}} \right],
\end{equation}
where $\bar{\rho}_{max}$ will play the role of a maximum density as we know already from our analysis above.
%\begin{equation}
%\rho_{c}=\frac{16(b_{3}-b_{2})^{4}}{\left|3b_{2}(B)^{2}+3b_{3}(A)^{2}+2b_{2}AB\right|^{2}}\frac{1}{\kappa^{2}B_{\frac{1}{3}}^{2}}.
%\end{equation}
To guarantee
that indeed $a$ has a minimum at the time when the Universe reaches $\rho_{\max}$, we must require $\ddot{a}(\rho=\bar{\rho}_{\max})>0$.
We write this condition explicitly for the $w=1/3$ case
\begin{equation}
\ddot{a}(\rho=\rho_{\max})=-\frac{\kappa}{3}\rho_{\max}\left(2-\frac{\kappa B_{\frac{1}{3}}}{2}\frac{1}{4(b_{3}-b_{2})^{2}}\left[4b_{2}AB+8b_{3}B^{2} \right]\rho_{\max}^{\frac{1}{2}} \right)a
\end{equation}
or equivalently,
\begin{equation}
\ddot{a}(\rho=\rho_{\max})=-\frac{\kappa}{3}\rho_{\max}\left(1-C_{3}\frac{\kappa B_{\frac{1}{3}}}{2}\rho_{\max}^{\frac{1}{2}} \right)a.
\end{equation}
Since $\rho_{\max}>0$, the condition of $\ddot{a}(\rho=\rho_{\max})>0$ taken together with the inequality (\ref{NSBB}) can be translated into
%\begin{equation}
%-(3b_{2}(B)^{2}+3b_{3}(A)^{2}+2b_{2}AB)<2(4b_{2}AB+3b_{3}B^{2}) \mbox{ } \mbox{ } \Rightarrow \mbox{ } \mbox{ } 36b_{2}^{3}-59b_{2}^{2}b_{3}-8b_{2}b_{3}^{2}+36b_{3}^{2}>0
%\end{equation}
\begin{equation}
\rho_{\max}^{\frac{1}{2}}>\frac{1}{\kappa B_{\frac{1}{3}}}\frac{2}{C_{3}}\equiv \rho_{inf}^{1/2}.
\end{equation}
On the other hand, for an accelerated evolution of the early universe we must look at the second Friedmann equation and require $\ddot{a}(\rho)>0$, %this condition 
%first of all requires the coefficient of $\kappa s^{2}$ to be negative, yielding
at least for certain values of $\rho$. For this to be possible, we see from equation (\ref{SecondFriedmannALA}) written now as 
\begin{equation}
\ddot{a}(\rho)=-\frac{\kappa}{3}\rho\left(1-\frac{\kappa B_{\frac{1}{2}}}{2}C_{3}\rho^{\frac{1}{2}} \right)a>0
\end{equation}
that the coefficient $C_{3}$ needs to be positive leading to the inequality
\begin{equation}
14b_{2}AB+8b_{3}(B)^{2}>0.
\end{equation}
and constraining the density by
\begin{equation}
\rho^{\frac{1}{2}}>\frac{1}{\kappa B_{\frac{1}{3}}}\frac{2}{C_{3}}=\rho_{inf}^{\frac{1}{2}}.%\frac{8(b_{3}-b_{2})^{2}}{[4b_{2}AB+3b_{3}B^{2}]}\frac{1}{\kappa B_{\frac{1}{3}}}
\end{equation}
Note that for an accelerated expansion to occur, $\rho_{inf}$ should be smaller than $\rho_{\max}$.
The above means that for an accelerated expansion to occur, $\rho$ must satisfy the inequality
\begin{equation}
\rho_{\max}^{\frac{1}{2}}>\rho^{\frac{1}{2}}>\rho_{inf}^{\frac{1}{2}}.
\end{equation}
We now have a full set of conditions for both the non-singular behaviour as well as an early accelerated expansion. This may be
interpreted in the following way: at a very early stage post Big Bang the density should lie in the region $\rho_{\max}>\rho>\rho_{inf}$ so that
it has an accelerated expansion. Later on, it will be in the region $\rho_{\max}>\rho_{inf}>\rho$ where there will be no more accelerated expansion
driven by spin terms.
Summarizing the scenario in this model we have
\begin{itemize}
\item{for a non-singular behaviour ($C_{1}<0$)
\begin{equation}\label{NonSingularBBAL1}
3b_{3}(A)^{2}+2b_{2}AB+(3b_{2}-2b_{3})(B)^{2}<0,
\end{equation}
\begin{equation}\label{NonSingularBBAL2}
-\frac{1}{C_{1}}>\frac{2}{C_{3}} \mbox{ } \mbox{ } \Rightarrow \mbox{ } \mbox{ } -\frac{1}{3b_{3}A^{2}+2b_{2}AB+(3b_{2}-2b_{3})B^{2}}>\frac{2}{4b_{2}AB+8b_{3}B^{2}}.
\end{equation}}
\item{For a consistent accelerating stage ($C_{3}>0$)
\begin{equation}\label{AcceleratedAL1}
14b_{2}AB+8b_{3}(B)^{2}>0, \quad \rho_{\max}^{\frac{1}{2}}>\rho^{\frac{1}{2}}>\rho_{inf}^{\frac{1}{2}}
\end{equation}
}
\end{itemize}
In Figure \ref{fig:region} we have plotted the values of parameters $b_{2}$ and $b_{3}$ which fulfill all conditions (\ref{NonSingularBBAL1}), (\ref{NonSingularBBAL2}) and (\ref{AcceleratedAL1}) simultaneously in the shaded region.
If we proceed to solve this model by using the continuity equation
%
%\begin{equation}
%\begin{split}
%\left(P-  \frac{\kappa s^{2}}{4(b_{3}-b_{2})^{2}}\left[b_{3}(A)^{2}+2b_{2}AB+(b_{2}+2b_{3})B^{2} \right]\right)\frac{d}{dt}(a^{3}) & \\ +\frac{d}{dt}\left[ \left(\rho +\frac{\kappa}{4(b_{3}-b_{2})^{2}}\left[3b_{2}(B)^{2}+3b_{3}(A)^{2}+2b_{2}AB \right] s^{2} \right)a^{3}\right] & =0
%\end{split}
%\end{equation}
%
%\begin{equation}
%\begin{split}
%\frac{d}{dt}\left( \rho +\frac{\kappa}{4(b_{3}-b_{2})^{2}}\left[3b_{2}(B)^{2}+3b_{3}(A)^{2}+2b_{2}AB \right] s^{2} \right)=\\ -3H\left(\rho+P+\frac{\kappa s^{2}}{4(b_{3}-b_{2})^{2}}\left[2b_{3}(A)^{2}+(2b_{2}-b_{3})(B)^{2} \right] \right)
%\end{split}
%\end{equation}
%[CHECK AGAIN!!!] $(B)^{2}$ term
%\\
%Taking $s^{2}=B_{\frac{1}{3}}\rho^{\frac{3}{2}}$ the above can be turned into a differential equation for $\rho$, namely
$
\frac{d\rho}{dt}\left(1+\frac{3}{2}C_{1}\kappa B_{\frac{1}{3}}\rho^{\frac{1}{2}} \right)=-4H\rho \left( 1+ \frac{3}{4}C_{2}\kappa B_{\frac{1}{3}}\rho^{\frac{1}{2}}\right),
$
we might introduce a singularity as it is evident by the following integral
\begin{equation}
\int d\rho\frac{\left(1+\frac{3}{2}C_{1}\kappa B_{\frac{1}{3}}\rho^{\frac{1}{2}}\right)}{\rho\left(1+\frac{3}{4}C_{2}\kappa B_{\frac{1}{3}}\rho^{\frac{1}{2}} \right)}=-4\int \frac{da}{a}
\end{equation}
which gives
\begin{equation}
\frac{1}{C_{2}}\left((4C_{1}-2C_{2})\log\left[\frac{ 4+3C_{2}\kappa B_{\frac{1}{3}}\sqrt{\rho}}{4+3C_{2}\kappa B_{\frac{1}{3}}\sqrt{\rho_{0}}}\right] +C_{2}\log\left(\sqrt{\frac{\rho}{\rho_{0}}}\right)\right)=-4\ln \left(a \right).
\end{equation}
This has to do with the third possible critical value of the density as shown previously, which might occur
if we do not put any further constraints on $C_2$. 
To avoid this possibility (and the possibility that the density will have several extrema) 
we insist on the consistency of (\ref{DEB}), which implies $2C_{1}=C_{2}$. Since $C_{1}$ and $C_{2}$ are functions of $b_{2}$ and $b_{3}$, then $2C_{1}=C_{2}$ means that the surfaces described by $C_{i}=C_{i}(b_{2},b_{3})$ intersect, as shown in Figure \ref{fig:intersection}. The path of intersection may be written as
\begin{equation}\label{2C1C2}
4b_{3}A^{2}+4b_{2}AB+4b_{2}B^{2}=0,
\end{equation}
which is essentially a cubic equation in both $b_{2}$ and $b_{3}$. 
The region in which this equation is true turns out to resemble a line in the space parameter defined by $b_{2}$ and $b_{3}$. This is shown in Figure \ref{fig:lineplot} with the help of MATHEMATICA.

\begin{figure}
\centering
\begin{minipage}{.45\textwidth}
\centering
   \includegraphics[width=3.5in]{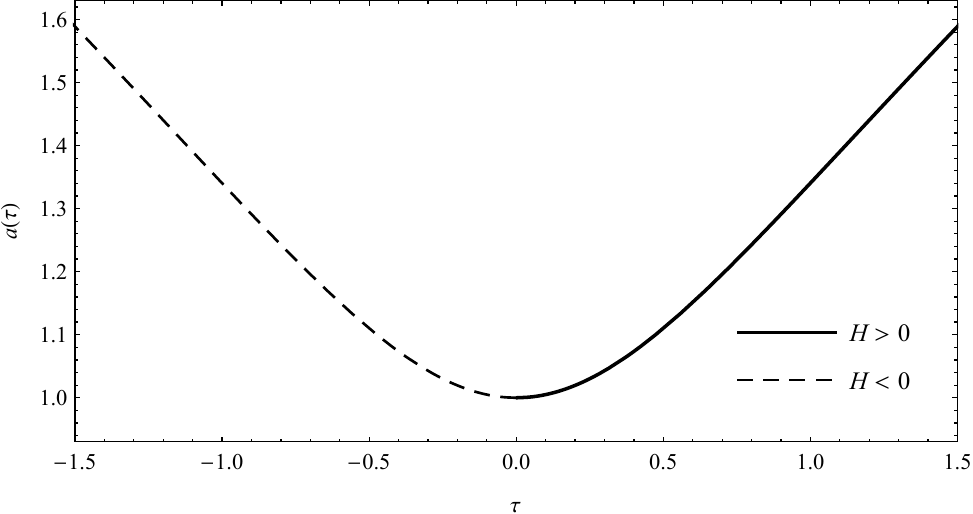} 
   \caption{Plot of both signs of equation (\ref{TauA}) where it is clear to see that $a$ never reaches zero and that at its minimum value it has a smooth bounce.}
   \label{fig:AGeneralCase}
\end{minipage}\hfill
\begin{minipage}{.45\textwidth}
\centering
   \includegraphics[width=3.5in]{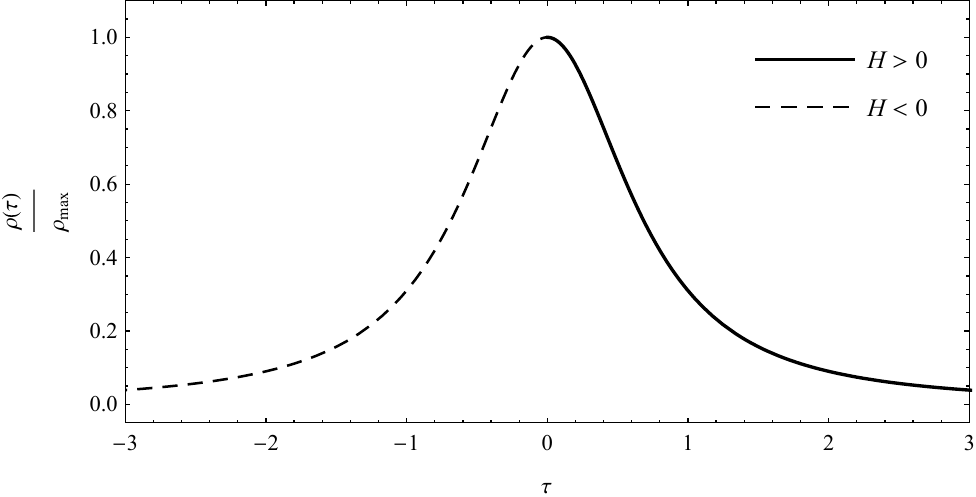} 
   \caption{Plot of both signs of equation (\ref{RhoGeneral}) where it is clear to see that $\rho$ has a maximum value corresponding to $\rho_{\max}$, and that the branches seem to coincide smoothly at this value.}
   \label{fig:RhoGeneral}
\end{minipage}\hfill
\end{figure}

%\begin{figure}[htbp] %  figure placement: here, top, bottom, or page
%   \centering
%   \includegraphics[width=3.5in]{LinePlot.eps} 
%   \caption{Plot of (a line of) values $b_{2}$ and $b_{3}$ which make the continuity equation consistent.}
%   \label{fig:lineplot}
%\end{figure}
We are left with the interesting possibility of exploiting certain values of $b_{2}$ and $b_{3}$ 
in order to avoid initial singularity and have an inflationary phase. We plot these possible values in the combined Figure \ref{fig:RegionPlotBAll}.
We mention that $b_3=1$ and $b_2=2$, which would bring the Lagrangian into the 
canonical form of Ricci scalar, is not a solution here. This is due to the fact that our model even for this particular choice of the
parameters differs from the standard Einstein--Cartan model in the identification, and hence choice, of the energy-momentum tensor.
The cosmology in the Einstein--Cartan theory based on the Lagrangian 
proportional to the Ricci scalar but without the $\accentset{\star}{\nabla}$-terms will be singular as we have shown at the 
end of section IV.
%\begin{figure}[htbp] %  figure placement: here, top, bottom, or page
%   \centering
%   \includegraphics[width=3.5in]{RegionPlotBAll.eps} 
%   \caption{Combined plot of the values $b_{2}$ and $b_{3}$ which make the continuity equation `consistent' and which yield also a non-singular and accelerated early universe.}
%   \label{fig:RegionPlotBAll}
%\end{figure}
\\
\\
The scenario under discussion gets simplified since the continuity equation (\ref{ContALB}) may be written in a way analogous to GR, namely
\begin{equation}
\frac{d\rho}{dt}=-4H\rho,
\end{equation}
\begin{equation}
\int_{\bar{\rho}_{\max}}^{\rho}\frac{d\rho'}{\rho'}=-4\int_{a_{0}=1}^{a}\frac{da'}{a'},
\end{equation}
which implies the usual GR result
\begin{equation}
\rho(a)=\bar{\rho}_{\max}\left(\frac{1}{a} \right)^{4}.
\end{equation}
This result can now be put into the first Friedmann equation to obtain the following integral
\begin{equation}
\int_{a(T_{0})}^{a(t)}da' \frac{a'}{\sqrt{1-\frac{1}{(a')^{2}}}}=\pm \sqrt{\frac{\kappa \bar{\rho}_{\max}}{3}}\int_{T_{0}}^{t}d t'
\end{equation}
which is solved by 
\begin{equation}\label{TauA}
\frac{1}{2}a^{2}\sqrt{1-\frac{1}{a^{2}}}+\frac{1}{2}\log \left[a+\sqrt{a^{2}-1} \right]=\pm \sqrt{\frac{\kappa\bar{\rho}_{\max}}{3}}(t-T_{0}).
\end{equation}
%$\tau$ where $D$ is given by
%
%\begin{equation}
%D=\frac{1}{2}\frac{a^{2}(T_{0})}{a_{0}^{2}}\sqrt{1-\frac{a_{0}^{2}\sigma_{0}}{a^{2}(T_{0})}}+\frac{a_{0}\sigma_{0}}{2}\log \left[a(T_{0})+\sqrt{a^{2}(T_{0})-a_{0}^{2}\sigma_{0}} \right]
%\end{equation}
The plot of (\ref{TauA}) is displayed in Figure \ref{fig:AGeneralCase} and it shows that it 
is possible to reach a non-singular bounce-like behaviour for $a(\tau)$ with $\tau\equiv\sqrt{\frac{\kappa\bar{\rho}_{\max}}{3}}(t-T_{0})$.
%\begin{figure}[htbp] %  figure placement: here, top, bottom, or page
%   \centering
%   \includegraphics[width=5in]{AGeneralCase.eps} 
%   \caption{Plot of both signs of equation (\ref{TauA}) where it is clear to see that $a$ never reaches zero and that at its minimum value it has a smooth bounce.}
%   \label{fig:AGeneralCase}
%\end{figure}
Using this solution for $a(\tau)$ the form of $\rho(\tau)$ can be obtained by replacing $\rho$ in terms of $a$. 
This gives
\begin{equation}\label{RhoGeneral}
\frac{1}{2}\sqrt{\frac{\bar{\rho}_{\max}}{\rho}}\sqrt{1-\sqrt{\frac{\rho}{\bar{\rho}_{\max}}}}+\frac{1}{2}\log \left[\left(\frac{\bar{\rho}_{\max}}{\rho}\right)^{\frac{1}{4}}+\sqrt{\left(\frac{\bar{\rho}_{\max}}{\rho} \right)^{\frac{1}{2}}-1} \right]=\pm \sqrt{\frac{\kappa \bar{\rho}_{\max}}{3}}(t-T_{0}).
\end{equation}
We plot this in Figure \ref{fig:RhoGeneral}.
%\begin{figure}[htbp] %  figure placement: here, top, bottom, or page
%   \centering
%   \includegraphics[width=5in]{RhoGeneral.eps} 
%   \caption{Plot of both signs of equation (\ref{RhoGeneral}) where it is clear to see that $\rho$ has a maximum value corresponding to $\rho_{\max}$, and that the branches seem to coincide smoothly at this value.}
%   \label{fig:RhoGeneral}
%\end{figure}
Alternatively, we could also solve numerically the equation  
\begin{equation}
\frac{d\bar{\sigma}}{d\eta}=\mp \frac{4}{\sqrt{3}}\bar{\sigma}^{\frac{3}{2}}\sqrt{1-\bar{\sigma}^{\frac{1}{2}}},
\end{equation}
%from which we obtain a numerical solution plotted in figure \ref{fig:SigmaGeneral} for both signs, note that when a value close to $\bar{\sigma}=1$ is chosen as
%an initial value, the solutions appear to smoothly transition from one branch to the other.
%\begin{figure}
%\centering
%\begin{minipage}{.45\textwidth}
%\centering
%   \includegraphics[width=3.3in]{SigmaGeneral-New.eps} 
%   \caption{Plot of $\bar{\sigma}$ in terms of $\bar{\eta}$ for both signs with $\bar{\sigma}(0)\simeq 1$.}
%   \label{fig:SigmaGeneral}
%\end{minipage}\hfill
%\begin{minipage}{.45\textwidth}
%\centering
%   \includegraphics[width=3.3in]{AGeneralNum-New.eps} 
%   \caption{Plot of the numerical solution for $a$ in terms of $\bar{\eta}$ for both signs with $\bar{\sigma}(0)\simeq 1$.}
%   \label{fig:AGeneralNum}
%\end{minipage}\hfill
%\end{figure}
%\begin{figure}[htbp] %  figure placement: here, top, bottom, or page
%   \centering
%   \includegraphics[width=5in]{SigmaGeneral.eps} 
%   \caption{Plot of $\bar{\sigma}$ in terms of $\bar{\eta}$ for both signs with $\bar{\sigma}(0)\simeq 1$.}
%   \label{fig:SigmaGeneral}
%\end{figure}
from which
\begin{equation}
a=\exp\left[\pm \int_{0}^{\bar{\eta}} d\bar{\eta}'\sqrt{\frac{\bar{\sigma}}{3}}\sqrt{1-\bar{\sigma}^{\frac{1}{2}}}  \right].
\end{equation}
The results agree with our previous method.
%What we obtain numerically we also plot in figure \ref{fig:AGeneralNum}, where we see indeed the bounce behaviour as $a(\eta)$ reaches a minimum value.

%\begin{figure}[htbp] %  figure placement: here, top, bottom, or page
%   \centering
%   \includegraphics[width=5in]{AGeneralNum.eps} 
%   \caption{Plot of the numerical solution for $a$ in terms of $\bar{\eta}$ for both signs with $\bar{\sigma}(0)\simeq 1$.}
%   \label{fig:AGeneralNum}
%\end{figure}

%\begin{figure}[htbp] %  figure placement: here, top, bottom, or page
%   \centering
%   \includegraphics[width=3.5in]{RegionPlotBAll.eps} 
%   \caption{Plot of (a line of) values $b_{2}$ and $b_{3}$ which make the continuity equation consistent, and the region which allows non-singular type of solutions as well as an inflationary behaviour.}
%   \label{fig:regionplotalb}
%\end{figure}
%Sergio: 1.we will have to leave out figs 14 and 15 (NOT DONE - just commenting)
% 2. Instead we can plot in 12 and 13 more solutions withe different initial values for the density 
% 3. this brings me to the quesstion: wha is the actaul initail value for density in 12/13? 0.999 as in 14/15? (It is done by plotting rho/rho_max...)
% 4. we will have to explore more the other case. either C_2 > 0 or |C_1/C_2| > 3/4. In both cases there is no third critical value
%we can discuss it when u come.
\section{Different Averaging}
A recent work \cite{Berredo} on the issue of averaging the spin terms $\langle s\indices{_{\mu}^{\lambda}}s_{\nu\lambda} \rangle$ 
claims a result different from the one we have discussed above. 
 The claim amounts to
%
%\begin{equation}
%\langle s\indices{_{\mu}^{\lambda}}s_{\nu\lambda} \rangle=\frac{2}{3}\left(g_{\mu\nu}-u_{\mu}u_{\nu} \right)\sigma^{2}
%\end{equation}
%\begin{equation}
%\langle s_{\alpha\beta}s^{\alpha\beta}\rangle = 2\sigma^{2}
%\end{equation}
%
%so in our notation they claim
%
\begin{equation}\label{DifferentAverage}
\langle s\indices{_{\mu}^{\lambda}}s_{\nu\lambda} \rangle=\frac{2}{12}\left(g_{\mu\nu}-u_{\mu}u_{\nu} \right)s^{2}.
\end{equation}
We do not agree that this is the result obtained by Gasperini in \cite{Gasperini} and stress that Gasperini has implicitly used the condition
$\langle s\indices{_{\mu}^{\lambda}}s_{\nu\lambda} \rangle=0$. We think, however, that it may be worthwhile to check the implications of such an averaging on each
of the models presented before. Note that in the standard model discussed in section IV such a term cancels out so this model will not be modified by this change.
%Sergio: can u substantiate this more?

\subsection{Weyssenhoff perfect fluid with spin}
We discuss below the alternative averaging for the model we outlined at the end of section IV where we dropped all terms proportional to $\accentset{\star}{\nabla}$.
Using the result given in equation (\ref{EffectiveEM1}) and considering 
the modification given by the averaging results in equation (\ref{DifferentAverage}), the new effective energy--momentum tensor is 
\begin{equation}
\tilde{\sigma}_{\mu\nu}=\left(\rho+p+\frac{5}{6}\kappa s^{2} \right)u_{\mu}u_{\nu}-\left(p+\frac{1}{12}\kappa s^{2} \right)g_{\mu\nu}.
\end{equation}
This change modifies only the second Friedmann equation, resulting in the following set of equations
\begin{eqnarray}
H^{2}&=&\frac{\kappa}{3}\left(\rho+\frac{3}{4}\kappa s^{2} \right),\\
3H^{2}+2\dot{H}=-\kappa\left( p+\frac{1}{12}\kappa s^{2}\right) &\Rightarrow& \dot{H}+H^{2}=-\frac{\kappa}{6}\left(3p+\rho+\kappa s^{2} \right),\\
\dot{H}&=&-\frac{\kappa}{2}\left(p+\rho+\frac{5}{6}\kappa s^{2}\right).
\end{eqnarray}
The continuity equation for the radiation case takes the form
\begin{equation}
\dot{\rho}\left(1+\frac{9}{8}\kappa B_{\frac{1}{3}}\rho^{\frac{1}{2}} \right)=-4H\rho\left(1+\frac{5}{8}\kappa B_{\frac{1}{3}}\rho^{\frac{1}{2}} \right).
\end{equation}
Note that no critical densities are present. Solving for $a$ in terms of $\rho$ gives
\begin{equation}\label{aDifferentAverage1}
a=\left(\frac{8+5\kappa B_{\frac{1}{3}}\sqrt{\rho_{0}}}{8+5\kappa B_{\frac{1}{3}}\sqrt{\rho}} \right)^{\frac{2}{5}}\left(\frac{\rho_{0}}{\rho} \right)^{\frac{1}{4}}.
\end{equation}
The plot of (\ref{aDifferentAverage1}) suggests a slightly modified behaviour with respect to standard General Relativity 
but it still exhibits the standard behaviour $a\rightarrow 0$ as $\rho\rightarrow \infty$.

\subsection{Simplified alternative Lagrangian}
In the case of the invariants $\mathcal{A}'$, taking the different average will yield the following effective energy-momentum tensor in an analogous way to the previous
model, namely
\begin{equation}
\tilde{\sigma}_{\mu\nu}=\left(\rho+p+\frac{5}{6a_{1}}\kappa s^{2} \right)u_{\mu}u_{\nu}-\left(p+\frac{1}{12a_{1}}\kappa s^{2} \right)g_{\mu\nu},
\end{equation}
which gives the Friedmann equations
\begin{equation}
H^{2}=\frac{\kappa}{3}\left( \rho + \frac{3}{4a_{1}}\kappa s^{2}\right),
\end{equation}
\begin{equation}
\dot{H}=-\frac{\kappa}{2}\left(p+\rho+\frac{5}{6a_{1}}\kappa s^{2} \right).
\end{equation}
The continuity equation for the radiation case is
\begin{equation}
\dot{\rho}\left(1+\frac{9}{8a_{1}}\kappa B_{\frac{1}{3}}\rho^{\frac{1}{2}} \right)=-4H\rho\left(1-\frac{5}{8a_{1}}\kappa B_{\frac{1}{3}}\rho^{\frac{1}{2}} \right)
\end{equation}
We see again that the case with $a_{1}>0$ is rather uninteresting since it is a small generalization upon the first model. On the other hand, as we had before,
the case $a_{1}<0$ is indeed promising. For this case let us write this continuity equation in the form
\begin{equation}
\dot{\rho}\left[1-\left(\frac{\rho}{\rho_{c1}} \right)^{\frac{1}{2}} \right]=-4H\rho \left[1-\left(\frac{\rho}{\rho_{c2}} \right)^{\frac{1}{2}} \right]
\end{equation}
with
\begin{equation}
\rho_{c1}^{1/2}\equiv \frac{8|a_{1}|}{9}\frac{1}{\kappa B_{\frac{1}{3}}},\mbox{ } \mbox{ } \mbox{ } \mbox{ } \mbox{ } \mbox{ } \rho_{c2}^{1/2}\equiv \frac{8|a_{1}|}{5}\frac{1}{\kappa B_{\frac{1}{3}}}.
\end{equation}
We observe that again the first Friedmann equation defines a maximum density, namely
\begin{equation*}
\rho_{\max 1}^{1/2}\equiv \frac{4|a_{1}|}{3}\frac{1}{\kappa B_{\frac{1}{3}}}.
\end{equation*}
Thus, one of the effects of taking a different averaging is just to modify the numerical values of the
critical densities which satisfy the inequality
\begin{equation}
\rho_{c1}^{1/2}<\rho_{\max 1}^{1/2}<\rho_{c2}^{1/2}.
\end{equation}
A similar relation was found before but now
\begin{equation}
\rho_{c1}^{1/2}=\frac{2}{3}\rho_{\max}^{1/2}, \mbox{ } \mbox{ } \mbox{ } \mbox{ } \mbox{  } \mbox{ } \mbox{ } \mbox{ } \mbox{ } \rho_{c_{2}}^{1/2}=\frac{6}{5}\rho_{\max}.
\end{equation}
At first glance it appears that $\ddot{a}$ may be positive. Taking $a_{1}<0$ yields for the radiation case  
\begin{equation}
\ddot{a}=-\frac{\kappa}{3}\rho \left(1-\frac{\kappa B}{2|a_{1}|}\rho^{\frac{1}{2}} \right)a.
\end{equation}
Hence, $\ddot{a}$ will be positive only if $\rho^{1/2}>\frac{2|a_{1}|}{\kappa B_{\frac{1}{3}}}=\frac{3}{2}\rho_{\max 1}^{1/2}$. The restrictions set by the
first Friedmann equation are sufficient. The different averaging only reflects in the change of $\rho_{c1}$ and $\rho_{c2}$.

\subsection{General alternative Lagrangian}
Using the set of invariants $\mathcal{A}$ and its general Lagrangian, the different averaging
of equation (\ref{DifferentAverage}) leads to the following effective energy--momentum tensor
\begin{equation}
\begin{split}
\tilde{\sigma}_{\mu\nu}=&\left(\rho+p+\frac{\kappa s^{2}}{2(b_{3}-b_{2})^{2}}\left[\frac{5}{3}b_{3}A^{2}+\frac{2}{3}b_{2}AB+(b_{2}-2b_{3})B^{2} \right] \right)u_{\mu}u_{\nu}\\
&-\left(p-\frac{\kappa s^{2}}{4(b_{3}-b_{2})^{2}}\left[-\frac{1}{3}b_{3}A^{2}+\frac{2}{3}b_{2}AB+(b_{2}+2b_{3})B^{2} \right] \right)g_{\mu\nu},
\end{split}
\end{equation}
which results in the following Friedmann equations
\begin{equation}
H^{2}=\frac{\kappa}{3}\left(\rho+\frac{\kappa s^{2}}{4(b_{3}-b_{2}^{1})}\left[3b_{3}A^{2}+2b_{2}AB+(3b_{2}-2b_{3})B^{2} \right] \right),
\end{equation}
\begin{equation}
H^{2}+\dot{H}=-\frac{\kappa}{6}\left(\rho+3p+\frac{\kappa s^{2}}{4(b_{3}-b_{2})^{2}}\left[4b_{3}A^{2}-8b_{3}B^{2} \right] \right),
\end{equation}
\begin{equation}
\dot{H}=-\frac{\kappa}{2}\left(\rho+p+\frac{\kappa s^{2}}{4(b_{3}-b_{2})^{2}}\left[\frac{10}{3}b_{3}A^{2}+\frac{4}{3}b_{2}AB+(2b_{2}-4b_{3})B^{2} \right] \right),
\end{equation}
Note that as expected, the first Friedmann equation remains unchanged with respect to the Gasperini averaging \cite{Gasperini}. This means that we can use the same
$C_{1}$ as defined in equation (\ref{c1def}) but we need to define the following new constants
\begin{equation}
\tilde{C}_{2}\equiv \frac{1}{4(b_{3}-b_{2})^{2}}\left\{ \frac{10}{3}b_{3}A^{2}+\frac{4}{3}b_{2}AB+(2b_{2}-4b_{3})B^{2}\right\},
\end{equation}
\begin{equation}
\tilde{C}_{3}\equiv \frac{1}{4(b_{3}-b_{2})^{2}}\left[4b_{3}A^{2}-8b_{3}B^{2} \right].
\end{equation}
With these new constants the Friedmann equations will simplify as follows
\begin{equation}\label{FFtilde}
H^{2}=\frac{\kappa}{3}(\rho+C_{1}\kappa s^{2}),
\end{equation}
\begin{equation}\label{SFtilde}
\ddot{a}=-\frac{\kappa}{6}(\rho+3p-\tilde{C}_{3}\kappa s^{2})a, \mbox{ } \mbox{ } \mbox{ }  \mbox{ } \mbox{ } \mbox{ } \dot{H}=-\frac{\kappa}{2}(\rho+p+\tilde{C}_{2}\kappa s^{2}).
\end{equation}
We see that the same conditions apply for these coefficients as before, particularly the critical density is the same when $w=1/3$, namely $\bar{\rho}_{\max}^{1/2}=-\frac{1}{C_{1}\kappa B_{\frac{1}{3}}}$. The differential equation may be written in an analogous way using the dimensionless variables of equation (\ref{DimensionlessB}) and one finds
\begin{equation}
\frac{d\bar{\sigma}}{d\eta}\left(1-\frac{3}{2}\bar{\sigma}^{\frac{1}{2}}\right)=\mp \frac{4}{\sqrt{3}}\bar{\sigma}^{\frac{3}{2}}\left(1-\frac{3}{4}\frac{\tilde{C_{2}}}{C_{1}}\bar{\sigma}^{\frac{1}{2}} \right)\sqrt{1-\bar{\sigma}^{\frac{1}{2}}}.
\end{equation}
The difference to the model we discussed before lies in the different values for $\tilde{C}_i$.
\section{Conclusions and Outlook}
In this paper we have examined in some detail the cosmological aspects of Einstein--Cartan theories
which allow an anti-symmetric part of the connection (contortion). Starting from the beginning, i.e. introducing the contortion
and relevant identities connected to it, we proceeded to the action principle based on the Lagrangian proportional to $R$. We presented this
variation in two different manners to exhibit the one freedom which we have: the choice of the physical source. We argued that the latter can be
either the metric or the canonical energy-momentum tensor. We briefly discussed the standard approach with the canonical energy-momentum tensor
which essentially is imposed by us if we wish to have Einstein-like ($R_{\mu \nu}-1/2g_{\mu \nu}R=\kappa \Sigma_{\mu \nu}$) equations for the metric.
The results are a bouncing universe with an inflationary expansion.
This approach is contrasted to a theory with the metric energy-momentum tensor where the determining equations are not canonical, i.e. not of the Einstein form.
In this set-up the theory resembles more the standard cosmology where the initial singularity cannot be avoided.
As a next step we have generalized the Lagrangian allowing for an arbitrary linear combination of diffeomorphic invariants. Based on these we constructed   
two different cosmological scenarios. One with a finite non-zero bounce to be identified with a Big Bang and the other one describing a universe
born at finite non-zero value of the scale factor. We indicated possible acceleration of the expansion by a proper choice of the parameter space.
This shows that the cosmologies of the Einstein--Cartan type can have different cosmological consequences, but many of the versions share
the desirable feature of avoiding the initial singularity and possible having an accelerated early expansion. We find this a remarkable aspect of the
Einstein--Cartan cosmologies. More so as the extension of Einstein's gravity to achieve this goal is rather mild and in many ways unconstrained as, for instance,
it does not alter any vacuum solutions. Notably the $s_{\mu \nu}$ part of the torsion source $s_{\mu \nu}u^{\alpha}$ can be identified with one of the 
generators of the Poincar\'e--Lie algebra which is connected to the spin observable in the operator language. Here, obviously we can only talk about expectations values
and connect the spin expectation value to the expectation value of $s_{\mu \nu}$. Such an interpretation connects the source of the torsion directly with
quantum mechanics.

As we have demonstrated not all cosmological models based on the Einstein--Cartan theory lead to a singularity-free 
cosmology (in the form a bounce). A thorough discussion of the connection between the bounce induced
by an anti-symmetric connection and the singularity theorems \cite{Chaubey, Senovilla1}
would be welcome.  These singularity theorems are based on three assumptions: energy conditions (week, null, strong or
their averaged versions), causality condition and an initial boundary condition. In the past cosmologists 
probing into bounces concentrated on the violation of one of the energy conditions, but the example of Senovilla's
discovery \cite{Senovilla2} of a singularity-free solution with no violation of the energy and causality conditions
showed us that a more careful treatment is necessary. One of the first papers making a connection
between the absence of an initial cosmological singularity in theories with torsion and the singularity theorems
\cite{Kerlick} also notices that in models with Dirac fields one finds an 'enhancement' of the singularity formation. 
In \cite{Pesmatsiou} the authors point out that the weak energy condition is not always satisfied in
Einstein--Cartan spacetimes. Based on generalization of the singularity theorems \cite{Esposito} reaches the conclusion
that in Einstein--Cartan spacetime the occurrence of singularities is less generic as compared to standard
General Relativity.  The energy conditions (in particular the null one) are important for the wormhole solution
in theories with torsion \cite{Grezia}.

Another interesting variant of the Einstein--Cartan theory not studied in this paper 
is to include in the Lagrangian parity violating terms proportional to $\epsilon^{\alpha \beta \gamma \delta}
R_{\alpha \beta \gamma \delta}$ and examine in detail their cosmological implications \cite{ParityViolation} concentrating on the aspects of the early universe.
  
\begin{appendices}
\section*{Appendix A. Derivation of the contortion tensor}

Demanding metricity when torsion is non vanishing fixes the behaviour of the contortion tensor in terms of the torsion 
tensor. To get the contortion tensor we shall use the metricity condition,  namely
\begin{eqnarray}\label{Metricity1}
\nabla_{\mu}g_{\alpha\beta}=0 &\rightarrow& \partial_{\mu}g_{\alpha\beta}=\Gamma_{\mu\alpha}^{\lambda}g_{\lambda\beta}+\Gamma_{\mu\beta}^{\lambda}g_{\alpha\lambda},\\ \label{Metricity2}
\nabla_{\beta}g_{\mu\alpha}=0 &\rightarrow& \partial_{\beta}g_{\mu\alpha}=\Gamma_{\beta\mu}^{\lambda}g_{\lambda\alpha}+\Gamma_{\beta\alpha}^{\lambda}g_{\mu\lambda},\\\label{Metricity3}
\nabla_{\alpha}g_{\mu\beta}= 0 &\rightarrow& \partial_{\alpha}g_{\mu\beta}=\Gamma_{\alpha\mu}^{\lambda}g_{\lambda\beta}+\Gamma_{\alpha\beta}^{\lambda}g_{\mu\lambda}.
\end{eqnarray}

Following \cite{Inverno}, we consider the combination (\ref{Metricity1})+(\ref{Metricity2})-(\ref{Metricity3}) and contract them with $g^{\alpha\lambda}$.
Furthermore, using $\Gamma_{\mu\nu}^{\alpha}=\accentset{\circ}{\Gamma}_{\mu\nu}^{\alpha}-K\indices{_{\mu\nu}^{\alpha}}$ leads to 
\begin{equation}
\frac{1}{2}g^{\nu\alpha}\left(\partial_{\mu}g_{\alpha\beta}+\partial_{\beta}g_{\mu\alpha}-\partial_{\alpha}g_{\mu\beta} \right)=\accentset{\circ}{\Gamma}_{\mu\beta}^{\nu}+2K\indices{_{[\alpha\mu]}^{\lambda}}g_{\lambda\beta}g^{\nu\alpha}+2K\indices{_{[\alpha\beta]}^{\lambda}}g_{\mu\lambda}g^{\nu\alpha}-2K\indices{_{(\mu\beta)}^{\lambda}}g_{\lambda\alpha}g^{\nu\alpha},
\end{equation}
\begin{equation}
\accentset{\circ}{\Gamma}_{\mu\beta}^{\nu}=\accentset{\circ}{\Gamma}_{\mu\beta}^{\nu}-2S\indices{_{\alpha\mu}^{\lambda}}g_{\lambda\beta}g^{\nu\alpha}-2S\indices{_{\alpha\beta}^{\lambda}}g_{\mu\lambda}g^{\nu\alpha}-2K\indices{_{(\mu\beta)}^{\lambda}}\delta_{\lambda}^{\nu},
\end{equation}
where we have used $K\indices{_{[\mu\nu]}^{\lambda}}=\Gamma_{[\mu\nu]}^{\lambda}=-S\indices{_{\mu\nu}^{\lambda}}$. 
Obviously this gives the symmetric part of the contortion tensor as
\begin{equation}
K\indices{_{(\mu\beta)}^{\nu}}=-S\indices{^{\nu}_{\mu\beta}}-S\indices{^{\nu}_{\beta\mu}}.
\end{equation}
Adding its antisymmetric part $K\indices{_{[\mu\beta]}^{\nu}}=-S\indices{_{\mu\beta}^{\nu}}$  the rearranged contortion tensor turns out to be
\begin{equation}
K\indices{_{\mu\beta}^{\nu}}=-S\indices{_{\mu\beta}^{\nu}}+S\indices{_{\beta}^{\alpha}_{\mu}}-S\indices{^{\alpha}_{\mu\beta}}
\end{equation}

\section*{Appendix B. Standard FRLW cosmology approach}
\begin{figure}
\centering
\begin{minipage}{.45\textwidth}
\centering
   \includegraphics[width=3.1in]{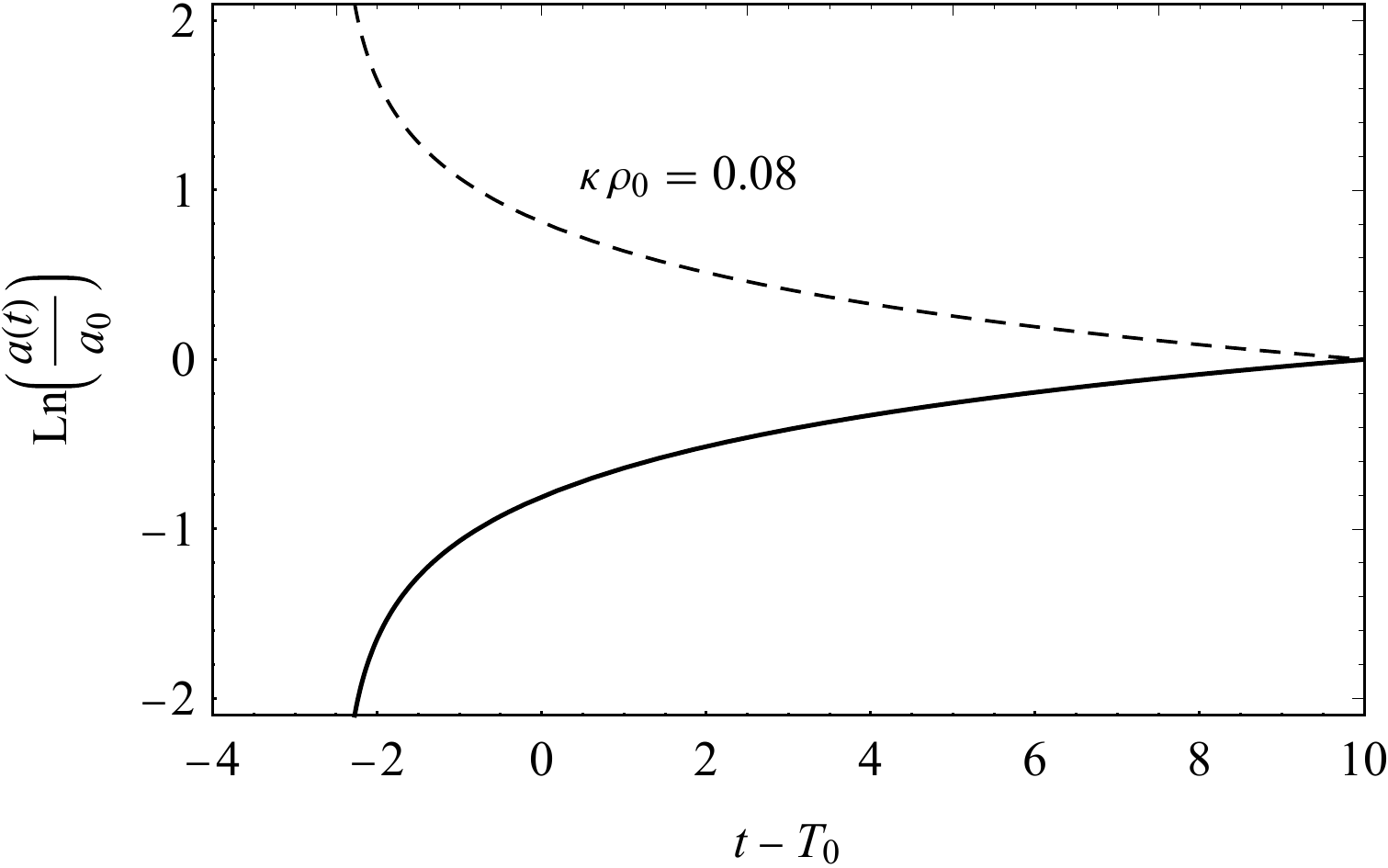} 
   \caption{Plot of the solution to equation (\ref{ln}). Observe that for a certain value of $t$, $\ln(a/a_{0})$ goes to $\pm \infty$.}
   \label{fig:ln}
\end{minipage}\hfill
\begin{minipage}{.45\textwidth}
\centering
   \includegraphics[width=3.1in]{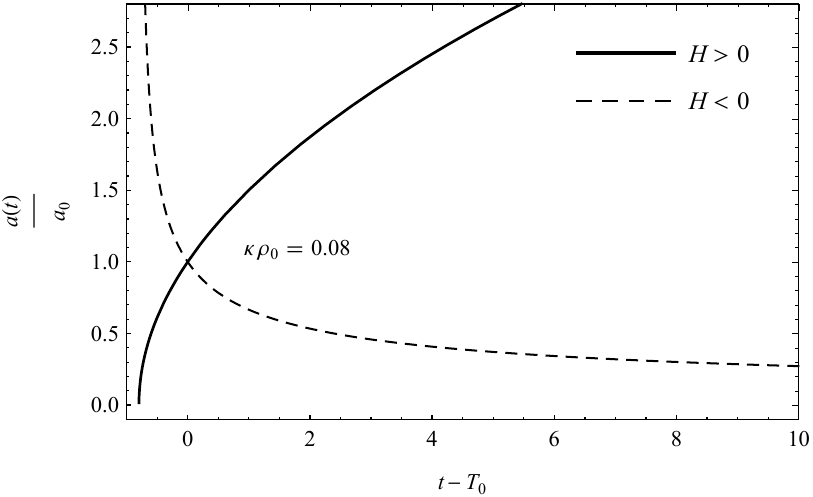} 
   \caption{Plot of the solution to equation (\ref{a1}). Note that for a certain value of $t$, $a(t)$ goes either to zero or $\infty$.}
   \label{fig:a1}
\end{minipage}\hfill
\end{figure}

\begin{figure}
\centering
\begin{minipage}{.45\textwidth}
\centering
   \includegraphics[width=3.1in]{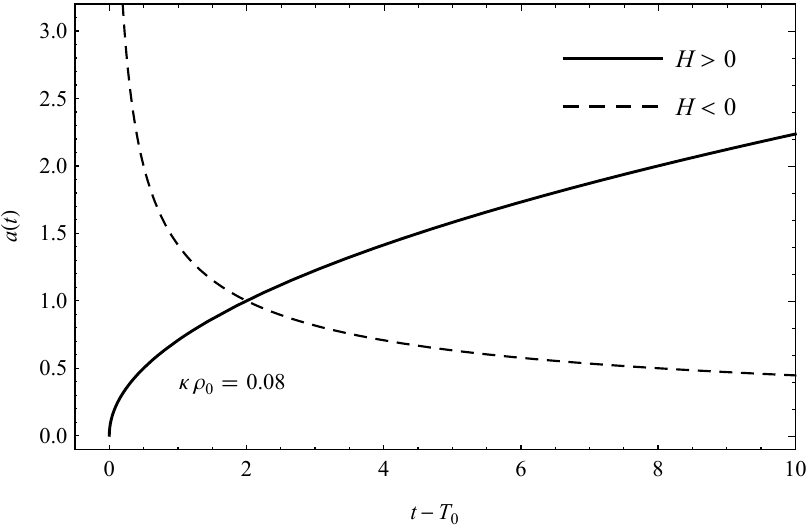} 
   \caption{Plot the solution to equation (\ref{a1}). Observe that for $t=0$, $a(t)\rightarrow 0$ or to $\infty$. Note that the positive value of the exponent corresponds to $a(t=0)=0$.}
   \label{fig:a2}
\end{minipage}\hfill
\begin{minipage}{.45\textwidth}
\centering
%   \includegraphics[width=3.1in]{rho_Plot1-New.eps} 
%   \caption{Sketch of equation (\ref{rho1}) where it is clear that for a certain value of $t$, $%\rho(t)$ tends to $\infty$.}
%   \label{fig:rho1}
%\end{minipage}\hfill
%\end{figure}

%\begin{figure}[htbp] %  figure placement: here, top, bottom, or page
%   \centering
%   \includegraphics[width=3.5in]{a_Plot2.eps} 
%   \caption{Sketch of equation (\ref{a2}) where it is clear that for $t=0$, $a(t)$ goes to zero or to $\infty$. Note that the positive value of the exponent corresponds to $a(t=0)=0$}
%   \label{fig:a2}
%\end{figure}

%\begin{figure}[htbp] %  figure placement: here, top, bottom, or page
%   \centering
%   \includegraphics[width=3.5in]{rho_Plot1.eps} 
%   \caption{Sketch of equation (\ref{rho1}) where it is clear that for a certain value of $t$, $\rho(t)$ tends to $\infty$.}
%   \label{fig:rho1}
%\end{figure}

%\begin{figure}
%\centering
%\begin{minipage}{.45\textwidth}
%\centering
  \includegraphics[width=3.1in]{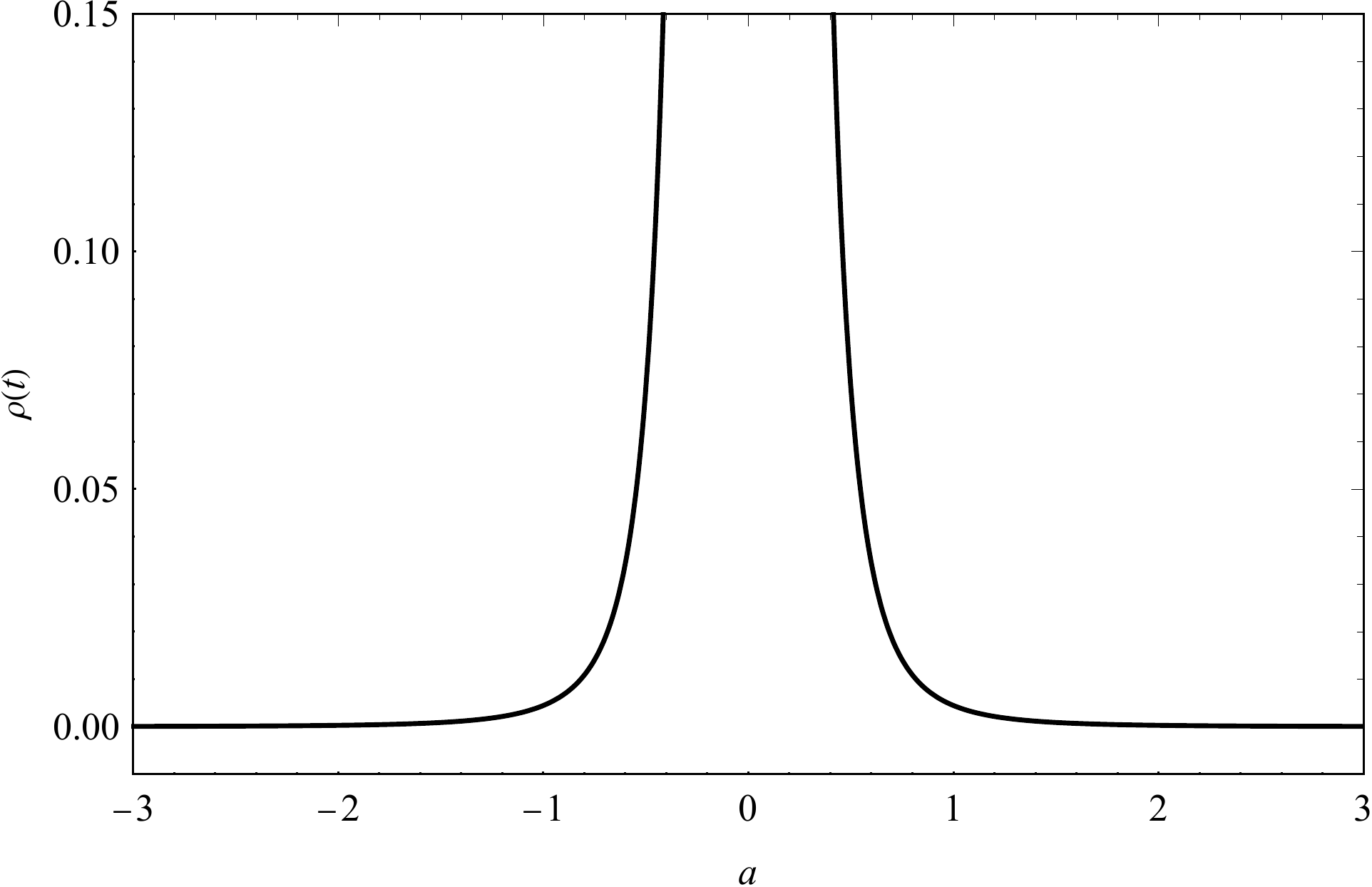} 
   \caption{Plot of the solution to equation (\ref{rho2}). Here, $\rho\rightarrow\infty$ as $t\rightarrow 0$.}
   \label{fig:rho2}
\end{minipage}\hfill
%\begin{minipage}{.45\textwidth}
%\centering
%   \includegraphics[width=3.1in]{rho_Plot_a-New.eps} 
%   \caption{Sketch of equation (\ref{rhoa}) where it is clear that $a=0$, corresponds to $\rho$ going to $\infty$.}
%   \label{fig:rhoa}
%\end{minipage}\hfill
\end{figure}

%\subsection{}
It is instructive to outline explicitly the differences of the models discussed above as compared to
the standard cosmological model which we briefly describe here.  
We start from the following Friedman equation and the usual continuity equation
\begin{equation}
H^{2}=\frac{\kappa}{3}\rho,\quad
\dot{\rho}=-3H(\rho+p).
\end{equation}
Now, in order to solve for $\rho$ in terms of $t$ we will first take the equation of state in the form $p=w\rho$ and choose $w=\frac{1}{3}$ since we are
interested in the radiation dominated universe. From this we obtain
\begin{equation}\label{rhodotstandard}
\dot{\rho}=\mp 4\sqrt{\frac{\kappa}{3}}\rho^{\frac{3}{2}}.
\end{equation}
Integrating (\ref{rhodotstandard}) leads to
\begin{equation}
\left(\frac{1}{\sqrt{\rho}}-\frac{1}{\sqrt{\rho_{0}}} \right)=\pm 2 \sqrt{\frac{\kappa}{3}}\left(t-T_{0} \right)
\end{equation}
which solved for $\rho(t)$ gives
\begin{equation}\label{rho1}
\rho(t)=\frac{\rho_{0}}{\left[2\sqrt{\frac{\kappa\rho_{0}}{3}}(t-T_{0})+1 \right]^{2}}.
\end{equation}
The result above implies that $\rho(t=T_{0})=\rho_{0}$. Furthermore if 
$(t-T_{0})=-\frac{1}{2}\sqrt{\frac{3}{\kappa\rho_{0}}}$, then
$\rho$ blows up. Thus, $\rho$ grows faster as $t \rightarrow T_{0}-\frac{1}{2}\sqrt{\frac{3}{\kappa\rho_{0}}}$. We solve for
$a(t)$ by using the first Friedmann equation we have written above. Since we have
\begin{equation}
\int_{1}^{a(t)} \frac{da'}{a'}=\pm \sqrt{\frac{\kappa\rho_{0}}{3}}\int_{t_{0}}^{t} \frac{dt'}{2\sqrt{\frac{\kappa\rho_{0}}{3}}(t'-T_{0})+1},
\end{equation}
then
\begin{equation}\label{ln}
\ln a(t) =\pm \frac{1}{2}\ln \left[\frac{1+2(t-T_{0})\sqrt{\frac{\kappa\rho_{0}}{3}}}{1+2(t_{0}-T_{0})\sqrt{\frac{\kappa\rho_{0}}{3}}} \right],
\end{equation}
the plots for these solutions are plotted in Figures \ref{fig:ln} and \ref{fig:a1}. 
Indeed, we see that if $t=T_{0}-\frac{1}{2}\sqrt{\frac{3}{\kappa\rho_{0}}}$ or $t_{0}=T_{0}-\frac{1}{2}\sqrt{\frac{3}{\kappa\rho_{0}}}$ 
we obtain a divergent r.h.s. which implies a divergent
corresponding left hand side suggesting that either $a\rightarrow 0$ or $a\rightarrow\infty$. Since we have a freedom of choice, let us
assume that $t_{0}$ corresponds to the present epoch. This sets $t_{0}\geq T_{0}$. It follows that the only value at which the
expression on the r.h.s. is divergent is $t=T_{0}-\frac{1}{2}\sqrt{\frac{3}{\kappa\rho_{0}}}$ taking $a\rightarrow 0$ on the l.h.s. 
When solving for $a(t)$, we find
\begin{equation}\label{a1}
a(t)=\left(\frac{1+2(t-T_{0})\sqrt{\frac{\kappa\rho_{0}}{3}}}{1+2(t_{0}-T_{0})\sqrt{\frac{\kappa\rho_{0}}{3}}} \right)^{\pm \frac{1}{2}}.
\end{equation}
From equation (\ref{a1})
we see that $a\rightarrow 0$ as $t\rightarrow T_{0}-\frac{1}{2}\sqrt{\frac{3}{\kappa \rho_{0}}}$, we plot both signs of this equation in Figure \ref{fig:a2}. This could in principle be associated with the
Big Bang time, so that the singularity can be associated to the `beginning' of the Universe. To emphasize this, we take $T_{0} = \frac{1}{2}\sqrt{\frac{3}{\kappa\rho_{0}}}$
and the solution can be written as
$
a(t)=\left(\frac{t}{t_{0}} \right)^{\pm\frac{1}{2}}.
$
Moreover, given the expected Big Bang behaviour of the universe, it is evident we should assume an expanding early Universe, and thus, the $\pm$ in the solution
can be neglected by choosing the monotonically growing solution
\begin{equation}
a(t)=\left(\frac{t}{t_{0}} \right)^{\frac{1}{2}}.
\end{equation}
This has also some implication upon $\rho(t)$ since 
\begin{equation}\label{rho2}
\rho(t)=\frac{3}{4\kappa}\frac{1}{t^{2}}.
\end{equation}
Hence, the divergence of $\rho$ corresponds to the value $t=0$.
The continuity equation allows to obtain $\rho$ in terms of $a$ by writing 
\begin{equation}
\frac{d\rho}{dt}=-4\frac{da}{dt}\frac{1}{a}\rho.
\end{equation}
This leads to 
\begin{equation}
\rho({a})=\frac{\rho_{\rm cr}}{a^{4}}
\end{equation}
where $\rho_{\rm cr}$ is taken as the present value of the density of the universe, we plot this 
behaviour in Figure \ref{fig:rho2}. 
%Sergio: Take only figure 18 and 20! Delete the other.
The two different branches of the solution for $a$ cannot be glued together to arrive at a differentiable result. 

%\begin{figure}[htbp] %  figure placement: here, top, bottom, or page
%   \centering
%   \includegraphics[width=3.5in]{Ln_Plot.eps} 
%   \caption{Sketch of equation (\ref{ln}) where it is clear that for a certain value of $t$, $\ln(a/a_{0})$ goes to $\pm \infty$.}
%   \label{fig:ln}
%\end{figure}

%\begin{figure}[htbp] %  figure placement: here, top, bottom, or page
%   \centering
%   \includegraphics[width=3.5in]{a_Plot1.eps} 
%   \caption{Sketch of equation (\ref{a1}) where it is clear that for a certain value of $t$, $a(t)$ goes either to zero or $\infty$.}
%   \label{fig:a1}
%\end{figure}

%\begin{figure}[htbp] %  figure placement: here, top, bottom, or page
%   \centering
%   \includegraphics[width=3.5in]{rho_Plot2.eps} 
%   \caption{Sketch of equation (\ref{rho2}) where it is clear that for $t=0$, $\rho$ blows up to $\infty$.}
%   \label{fig:rho2}
%\end{figure}

%\begin{figure}[htbp] %  figure placement: here, top, bottom, or page
%   \centering
%   \includegraphics[width=3.5in]{rho_Plot_a.eps} 
%   \caption{Sketch of equation (\ref{rhoa}) where it is clear that $a=0$, corresponds to $\rho$ going to $\infty$.}
%   \label{fig:rhoa}
%\end{figure}

\section*{Appendix C. Conservation laws}
%Sergio: this is completly unclear. I mean what u wriote below is not clear (the whole appendix C)
One can obtain conservation laws from identities obtained in ECKS theory. Let us start from equation (\ref{BianchiTorsion})
\begin{equation*}
\accentset{\star}{\nabla}_{\nu}\Sigma\indices{_{\mu}^{\nu}}=\tau\indices{_{\nu\lambda}^{\gamma}}R\indices{_{\mu\gamma}^{\nu\lambda}}+2\Sigma\indices{_{\gamma}^{\nu}}S\indices{_{\mu\nu}^{\gamma}}.
\end{equation*}
When choosing $\mu=0$ in GR theory with the FLRW metric, one obtains the continuity equation $\dot{\rho}+3H(\rho+p)=0$. We expect an equivalent result considering the
energy--momentum identification as done in \cite{Hehl}. This should match what we obtained in equation (\ref{ContinuityHehl1}). We start with the r.h.s. of equation (\ref{BianchiTorsion}) for $\mu=0$
\begin{equation}\label{starsigma1}
\accentset{\star}{\nabla}_{\nu}\Sigma\indices{_{0}^{\nu}}=\nabla_{\nu}\Sigma\indices{_{0}^{\nu}}+2S\indices{_{\nu\alpha}^{\alpha}}\Sigma\indices{_{0}^{\nu}}=\nabla_{\nu}\Sigma\indices{_{0}^{\nu}}
\end{equation}
since $S\indices{_{\nu\alpha}^{\alpha}}=\frac{\kappa}{2}\tau\indices{_{\alpha\lambda}^{\lambda}}=s_{\alpha\lambda}u^{\lambda}=0$. Calculating the covariant derivative on the r.h.s. of (\ref{starsigma1}) gives
\begin{eqnarray}
\nabla_{\nu}\Sigma\indices{_{0}^{\nu}}&=&\accentset{\circ}{\nabla}_{\nu}\Sigma\indices{_{0}^{\nu}}+K\indices{_{\nu 0}^{\alpha}}\Sigma\indices{_{\alpha}^{\nu}}-K\indices{_{\nu\alpha}^{\nu}}\Sigma\indices{_{0}^{\alpha}}, \nonumber \\
&=&\accentset{\circ}{\nabla}_{\nu}\Sigma\indices{_{0}^{\nu}}+\kappa\left(-\tau\indices{_{\nu 0}^{\alpha}}+\tau\indices{_{0}^{\alpha}_{\nu}}-\tau\indices{^{\alpha}_{\nu 0}} \right)\Sigma\indices{_{\alpha}^{\nu}},\nonumber\\ \label{CovDerivativeSigma}
&=&\accentset{\circ}{\nabla}_{\nu}\Sigma\indices{_{0}^{\nu}}-\kappa s\indices{^{\alpha}_{\nu}}u_{0}\Sigma\indices{_{\alpha}^{\nu}},
\end{eqnarray}
where we have used the conditions $s_{i0}=0$ and $\tau\indices{_{\alpha\lambda}^{\lambda}}=0$. We see that given the antisymmetry of $s_{\mu\nu}$ only the antisymmetric 
part of $\Sigma\indices{_{\alpha}^{\nu}}$ in the last term survives, that is $\Sigma_{[\alpha\nu]}=\accentset{\circ}{\nabla}_{\lambda}(s_{\alpha\nu}u^{\lambda})$.
On the other hand it is useful to write the expression $\accentset{\circ}{\nabla}_{\nu}\Sigma\indices{_{0}^{\nu}}$ as follows
\begin{equation}
\begin{split}
\accentset{\circ}{\nabla}_{\nu}\Sigma\indices{_{0}^{\nu}}&=\partial_{\nu}\Sigma\indices{_{0}^{\nu}}-\accentset{\circ}{\Gamma}_{\nu 0}^{\alpha}\Sigma\indices{_{\alpha}^{\nu}}+\accentset{\circ}{\Gamma}_{\nu\alpha}^{\nu}\Sigma\indices{_{0}^{\alpha}},\\
&=\dot{\rho}+3Hp+3H\rho,
\end{split}
\end{equation}
where we have used the Christoffel Symbols $\accentset{\circ}{\Gamma}_{\mu\nu}^{\alpha}$ for the FLRW metric.
Furthermore, the second term on the r.h.s. of (\ref{CovDerivativeSigma}) may be written as
\begin{equation}
-\kappa s\indices{^{\alpha}_{\nu}}u_{0}\Sigma\indices{_{\alpha}^{\nu}}=-\kappa s^{\alpha\nu}u_{0}\Sigma_{\alpha\nu}.
\end{equation}
We explore what happens to the Weyssenhoff term in $\Sigma$ 
\begin{equation}
-\kappa s^{\alpha\nu}u_{0}\accentset{\circ}{\nabla}_{\lambda}(u^{\lambda}s_{\alpha\nu})=-\kappa s^{\alpha\nu}u_{0}\left(\partial_{\lambda}(u^{\lambda}s_{\alpha\nu})+\accentset{\circ}{\Gamma}_{\lambda\beta}^{\lambda}u^{\beta}s_{\alpha\nu}-\accentset{\circ}{\Gamma}_{\lambda\alpha}^{\beta}u^{\lambda}s_{\beta\nu}-\accentset{\circ}{\Gamma}_{\lambda\nu}^{\beta}u^{\lambda}s_{\alpha\beta} \right).
\end{equation}
We note that the first term on the r.h.s. may be rewritten using
\begin{equation}
\partial_{0}(s_{\alpha\nu}s^{\alpha\nu})=\partial_{0}(s_{\alpha\nu})s^{\alpha\nu}+s_{\alpha\nu}\partial_{0}(s^{\alpha\nu})=2\partial_{0}(s_{\alpha\nu})s^{\alpha\nu}=\frac{d}{dt}\left(\frac{1}{2}s^{2}\right)
\end{equation}
so that we finally obtain for the Weyssenhoff term
\begin{equation}
-\kappa s^{\alpha\nu}u_{0}\accentset{\circ}{\nabla}_{\lambda}(u^{\lambda}s_{\alpha\nu})=-\frac{d}{dt}\left(\frac{1}{4}\kappa s^{2} \right) - 3H\left(\frac{1}{2}\kappa s^{2} \right)+3H\left(\frac{1}{2}\kappa s^{2} \right)+3H\left(\frac{1}{2}\kappa s^{2} \right).
\end{equation}
Finally, for the l.h.s. of equation (\ref{BianchiTorsion}) we get
\begin{equation}
\accentset{\star}{\nabla}_{\nu}\Sigma\indices{_{0}^{\nu}}=\frac{d}{dt}\left(\rho - \frac{1}{4}\kappa s^{2} \right)+3H\left(\rho+p-\frac{1}{2}\kappa s^{2} \right).
\end{equation}
Now, the r.h.s. can be shown to be zero for the $\mu=0$ equation. Let us first observe that
\begin{equation}
\tau\indices{_{\nu\lambda}^{\gamma}}R\indices{_{0 \gamma}^{\nu\lambda}}=s_{\nu\lambda}u^{\gamma}R\indices{_{0\gamma}^{\nu\lambda}}=s_{\nu\lambda}u^{0}R\indices{_{00}^{\nu\lambda}}=0
\end{equation}
because only $u^{0}$ does not vanish. For the other term, we have
\begin{equation}
2\Sigma\indices{_{\gamma}^{\nu}}S\indices{_{0\nu}^{\gamma}}=2\Sigma\indices{_{\gamma}^{\nu}}\kappa \left(\tau\indices{_{0\nu}^{\gamma}} \right)=2\kappa \Sigma\indices{_{\gamma}^{\nu}}s_{0\nu}u^{\gamma}=0,
\end{equation}
because $s_{i0}=-s_{0i}=0$. Indeed, we get the conservation law in the standard case as in
equation (\ref{ContinuityHehl1}).

\end{appendices}

\end{document}